\documentclass[a4paper,fleqn]{cas-sc}

\usepackage[numbers,sort&compress]{natbib}
\usepackage{amsmath,amsfonts}
\usepackage{array}
\usepackage{setspace}
\usepackage{textcomp}
\usepackage{multirow}
\usepackage{stfloats}
\usepackage{url}
\usepackage{verbatim}
\usepackage{graphicx}
\usepackage{textcomp}
\usepackage{subfigure}
\usepackage{amsmath}
\usepackage{amsthm}
\usepackage{diagbox}
\newtheorem{definition}{Definition}
\newtheorem{lemma}{Lemma}
\newtheorem{theorem}{Theorem}

\begin{document}
\let\WriteBookmarks\relax
\def\floatpagepagefraction{1}
\def\textpagefraction{.001}

\shorttitle{Temporal Network with Online and Hidden Vertices}

\shortauthors{Ziyan Zeng et~al.}

\title [mode = title]{Temporal network modeling with online and hidden vertices based on the birth and death process}                      
\author[1]{Ziyan Zeng}[style=chinese]
\credit{Conceptualization of this study, Methodology, Simulations}

\affiliation[1]{organization={The College of Artificial Intelligence, Southwest University},
    addressline={No.2 Tiansheng Road, Beibei}, 
    city={Chongqing},
    postcode={400715}, 
    country={China}}

\author[1]{Minyu Feng}[style=chinese]
\credit{Conceptualization of this study, Methodology, Simulations}
\cormark[1]
\cortext[cor1]{Minyu Feng}
\ead{myfeng@swu.edu.cn}
\author[2,3]{J\"{u}rgen Kurths}
\affiliation[2]{organization={Potsdam Institute for Climate Impact Research, 14437 Potsdam},
    country={Germany}}
\affiliation[3]{organization={Department of Physics, Humboldt University, 12489 Berlin},
countru={Germany}}
\credit{Conceptualization of this study, Methodology, Simulations}

\nonumnote{This work was partly supported by the Humanities and Social Science Fund of Ministry of Education of the People’s Republic of China under Grant No.21YJCZH028 and partly by the National Nature Science Foundation of China (NSFC) under Grant No.62206230.  }
\begin{abstract}
Complex networks have played an important role in describing real complex systems since the end of the last century. Recently, research on real-world data sets reports intermittent interaction among social individuals. In this paper, we pay attention to this typical phenomenon of intermittent interaction by considering the state transition of network vertices between online and hidden based on the birth and death process. By continuous-time Markov theory, we show that both the number of each vertex's online neighbors and the online network size are stable and follow the homogeneous probability distribution in a similar form, inducing similar statistics as well. In addition, all propositions are verified via simulations. Moreover, we also present the degree distributions based on small-world and scale-free networks and find some regular patterns by simulations. The application in fitting real networks is discussed. 
\end{abstract}

\begin{highlights}
\item We propose a temporal network model considering the stochastic phase transition of vertices. 
\item Theoretical analysis is derived based on the continuous Markov chain method and confirmed by simulations. 
\item Application in fitting the real network is discussed.
\end{highlights}

\begin{keywords}
Complex Networks \sep Birth-death Process\sep Stationary Distribution \sep Hidden Vertices
\end{keywords}
\maketitle

\section{Introduction}
\label{sec:introduction}
Network science has been playing a significant role in studies of social structures and dynamics. In 1998, the proposition of the small-world network (SW) model \cite{watts1998collective} successfully reveals the internal mechanism of the six degrees of separation \cite{guare1990six} by rewiring regular networks to introduce increasing amounts of disorder. In 1999, Barab\'{a}si and Albert first noticed the network expansion and preferential attachment in real systems and proposed the scale-free network (SF) model \cite{barabasi1999emergence}. Based on these two studies, in the past two decades, studies of complex network modeling and statistical mechanisms have sprung up \cite{albert2002statistical}. By the end of the last century, Kleinberg modified the formation rules of SWs and first studied the navigation on SWs \cite{kleinberg2000navigation}\cite{kleinberg2000small}. Later, Barab\'{a}si and Albert proposed an extended SF model, which allows reconnection among edges\cite{albert2000topology}. In 2001, Bianconi and Barab\'{a}si noticed that networks could follow Bose statistics and can undergo Bose-Einstein condensation, revealing the "first-mover-advantage," "fit-get-rich," and "winner-takes-all" phenomena in complex networks \cite{bianconi2001bose}. In the study of the world trade web, Li et al. found that the global preferential attachment is not always applicable and proposed the local-world evolving network model \cite{li2003local}\cite{li2003complexity}. Topological properties of complex networks \cite{albert2000topology}, which present stability in continuously evolving, have been widely studied as well. Topological properties of SWs, including the clustering coefficient, the average path length, and the degree distribution, have been studied by Barrat et al.\cite{barrat2000properties} and Newman et al.\cite{newman2000mean}. Additionally, the degree distribution \cite{barabasi1999mean} and the clustering coefficient \cite{fronczak2003mean} have been studied via the mean-field theory as well. Furthermore, dynamical behaviors on complex networks, including epidemic spreadings \cite{castellano2009statistical}, evolutionary games \cite{perc2017statistical}, and synchronization \cite{arenas2008synchronization}, have been revealing social dynamics in real systems. Recently, the higher-order structures of complex networks are attracting research's attention \cite{majhi2022dynamics}, focusing in particular on the novel aspects of the dynamics that emerge on higher-order networks. In addition to network models and dynamics, the algorithms in complex networks are widely studied as well and shown to be effective in real social systems, e.g., optimal estimation \cite{li2020optimal} and community detection \cite{li2020optimization}. 

Nevertheless, most networked systems of scientific interest are characterized by temporal links, meaning the network's structure changes over time \cite{li2017fundamental}\cite{holme2012temporal}, such as neural and brain networks \cite{sporns2004organization}\cite{bullmore2009complex}, ecological networks \cite{de2005dynamic}\cite{bajardi2011dynamical} and some other systems \cite{dagum1992dynamic}\cite{medo2011temporal}. Recently, the concept of temporal networks has emerged as the times require, on which connections of networks are time-varying. From the perspective of network modeling, the activity-driven network \cite{perra2012activity} has formed a helpful framework for a dynamical analysis of social systems, e.g., the citation networks. Besides, researchers noticed that vertices in networks are not only growing but also decreasing at the same time. To this end, Zhang et al. proposed a network model considering the random birth and death of complex networks \cite{zhang2016random}. Furthermore, based on previous works, Feng et al. proposed a model to describe the growth and decline of vertices in continuous time \cite{feng2018evolving}\cite{feng2023heritable}, where the evolving network model is regarded as a queueing system. This framework provides a special research idea for the social dynamics, including the epidemic propagation \cite{li2021protection}, the evolutionary game \cite{zeng2022spatial}, and belief dynamics \cite{li2022fast}. Li et al. considered the birth-death process in a specific area around each vertex and proposed the network model with the degree increase and decrease mechanism \cite{li2023evolving}. In addition to temporal links, the higher-order and multilayer structures of complex networks provide researchers with fundamental advantages in the study of social physics \cite{jusup2022social}. 

However, the aforementioned network models neglect the typical social phenomenon of temporarily offline vertices. In other words, once a vertex is removed from the network, it will not appear again in the network. In social and communication networks, individuals often leave the network temporarily, but go online again to chat with each other instead of being removed permanently. Besides, there is research noticing the existence of hidden vertices \cite{smith2019hidden,mocanu2018decentralized,li2019dynamical,li2021measuring} and trying to establish network models to describe them. In this paper, we focus on network modeling with the vertex phase switching between online and hidden states. Based on the birth and death process, we propose a novel network model which allows vertices to be online or hidden with two independent exponential rates. If vertices become hidden, they disconnect from their neighbors temporarily, and once hidden vertices become online, they reconnect to their neighbors again. This forms the mentioned vertex phase transition between two different states. In the modeling part, we introduce our model and theoretically analyze the properties of the model based on the birth and death process. In detailed simulations, we compare our theoretical analysis and simulation results. Additionally, we present degree distributions of our network model under different pairs of parameters. 

Generally, the contributions of the proposed model are as follows.

\begin{enumerate}
  \item We propose the network model considering the vertex phase transitions between online and hidden states. This intermittent phenomenon of individual interaction is widely reported before. Based on continuous Markov chain theory, we perform the theoretical analysis on the probability model of online network size. 
  \item We verify the mentioned theoretical analysis and validate the effectiveness in fitting real-world network data sets by simulation. By showing the degree distributions, we present that the phase switching of individuals between online and hidden states maintains the homogeneity of SWs but reduces the heterogeneity of SFs. 
  \item We provide a new network framework for the study of social dynamics, such as epidemic propagation\cite{wang2019coevolution}, evolutionary games \cite{zeng2022spatial2}, and synchronization \cite{pan2015synchronization}\cite{li2023new} in the complex systems. 
\end{enumerate}

A detailed organization of this paper is as follows: In Sec. \ref{sec:Complex Networks With Online and Hidden Vertices}, we introduce our model in detail and perform a theoretical analysis based on the birth and death process. In Sec. \ref{sec:simulation}, we present simulation results and demonstrate the accuracy of our theoretical analysis. In Sec. \ref{sec:conclusion}, we conclude our work and give some outlook.

\section{Complex Networks With Online and Hidden Vertices}
\label{sec:Complex Networks With Online and Hidden Vertices}

In this section, we introduce a novel network model with online and hidden vertices (NOH) based on the birth and death process and related analysis of network properties. We pay attention to the ubiquitous phenomenon that an individual may be temporarily "turned off" and "divorced" from the system it belongs to. For example, considering an electric light in daily life, we usually do not keep it on all the time, instead, we turn it on when we need it and turn it off when we do not. In social media, a person can only contact others when it is online. Besides, if it goes offline or hidden, it is not able to interact with its neighbors or friends, and once it is online again, it reconnects the network and contacts its existing friends, which can be described as the vertex phase transitions between online and hidden states. Under the above circumstances, an individual stays detectable or online for a certain duration, and once it goes offline or hidden, there is also a duration until it goes online again, which can be regarded as a cyclic but not simply periodic process. The individuals won't be online or offline permanently but can undergo the state transition to be online and offline alternatively. 

To this end, we establish the NOH model to describe this phenomenon. We use the exponential distribution and Poisson statistics to model the temporal network because the Poisson statistics have been widely applied to the model of human behaviors and confirmed \cite{katti1968handbook,reynolds2003call,chakrabarti2020statistical}, although there is a study showing that there are also heavy tail characteristics in human activate time \cite{barabasi2005origin}. Take a specific example, Ref. \cite{miritello2011dynamical} analyzed the mobile phone activity of 20 million people in one country, showing that the probability distribution of interevent time is almost exponential as in a Poisson process. Additionally, using the exponential interevent time helps to simplify our analysis and induce understandable conclusions. If there are $n$ online vertices in the network, the hidden (online) vertices transform their states into online (hidden) at an exponential rate $\lambda_{n}$ ($\mu_{n}$), also interpreted as the online rate (hidden rate), i.e., an online (hidden) vertex stays visible (invisible) in the network for a duration that follows an exponential distribution, and if the duration ends, it becomes hidden (online). Once a vertex becomes hidden, it disconnects from its neighbors temporarily, and once a vertex becomes online, it reconnects to its online and original neighbors. 

The ideas of common methods to simulate a complex network system are mostly based on the SW and SF models, and we now describe some models briefly. The construction of an SW starts from a regular graph where each vertex has $K/2$ neighbors on its left and right respectively \cite{watts1998collective}. Then, the networked system undergoes the random connection process, leading to a short average path length and high clustering coefficient. The SF model considers the growth and preferential attachment \cite{barabasi1999emergence}. The network keeps growing, and each vertex obtains a new neighbor with the probability to its degree. The network model with variable elements (NVE) extends the new connection number in the SF to random variables \cite{feng2020accumulative}. Additionally, the activity-driven network model allows the vertices to activate with a certain probability and then connect to others \cite{perra2012activity}. Refs. \cite{feng2020accumulative} and \cite{perra2012activity} present a good performance in the paper citation network. 

\subsection{Model}
\begin{figure*}[htbp]
\centering
\includegraphics[scale=0.55]{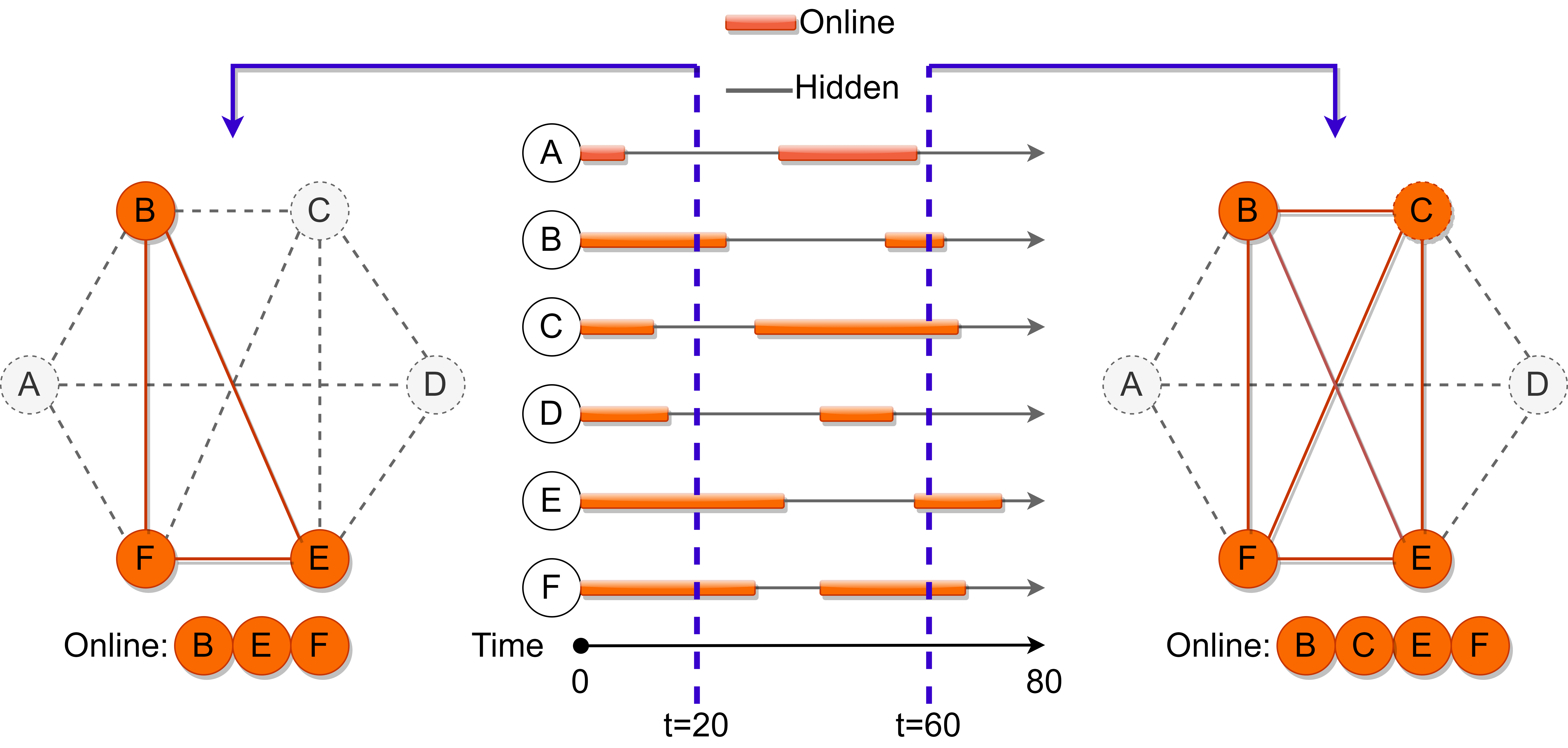}
\caption{\textbf{An example of the NOH model. }This figure presents an evolution example of the network model with online and hidden vertices. The example network is composed of 6 vertices. Orange and grey durations denote the online and hidden duration of each vertex respectively. The time range is set as $[0,80]$, and we observe the network at $t=$20 and 60. When $t=20$, the vertices $B$, $E$, $F$ are online (colored in orange), and $A$, $C$, $D$ are hidden. When $t=80$, the vertices $B$, $C$, $E$ and $F$ are online (colored in orange), and $A$, $D$ are hidden. Connections between two online vertices are marked in orange solid lines. Connections that are temporarily cut are marked in grey dotted lines. (color online)}
\label{fig:example}
\end{figure*}
We now introduce the process of constructing the NOH model, including initialization and the state transitions of vertices.
\label{subsec:model}

\textbf{Initialization. }To start with, there are $n=N(0)$ online vertices in the network.

\textbf{State transitions of vertices. }Provided a vertex becomes hidden at time $t$, it then becomes online by an exponential rate $\lambda$, i.e., it stays hidden for a time duration that follows an exponential distribution with the parameter $\lambda$ and then becomes online. If a vertex becomes online at time $t$, it then becomes hidden by an exponential rate $\mu$, i.e., it stays online for a time duration that follows an exponential distribution with the parameter $\mu$. Besides, each hidden vertex disconnects from its neighbors temporarily until it becomes online. Once it is online, it reconnects to its online and original neighbors. For each vertex, its hidden duration $T_{h}(t)$ follows
\begin{equation}\label{eq:hidden duration}
T_{h}(t)=1-\exp(-\lambda t),
\end{equation}
and its online duration $T_{o}(t)$ follows
\begin{equation}\label{eq:online duration}
T_{o}(t)=1-\exp(-\mu t). 
\end{equation}

The above Eqs. \ref{eq:hidden duration} and \ref{eq:online duration} are the cumulative form of exponential distributions, denoting the probability distribution of online and hidden time respectively. Note that an online vertex may be isolated and have no neighbor. We assume that, in this case, the isolated vertex does not have to do a random search for new neighbors to get rid of the isolated state as Ref. \cite{dos2020generative} from the data set \cite{sociopatterns} indicates. Besides, this assumption is reasonable because, in the real data, this random research process is rarely observed. Additionally, the potential links among individuals in our NOH model are assumed to be statically determined by the initial network as the above real data suggests as well. Dynamics on both nodes and edges can be applied in the models of temporal networks \cite{holme2012temporal}. In this article, we establish our model based on vertex dynamics, i.e., we focus on the stochastic process of nodes in a network. Additionally, previous studies have shown the effectiveness of modeling the node dynamics in empirical social dynamics, e.g., epidemic spreading \cite{ruan2021role} and active dynamic \cite{dos2020generative} in real networks. 

\begin{figure}[htbp]
\centering
\includegraphics[scale=0.1]{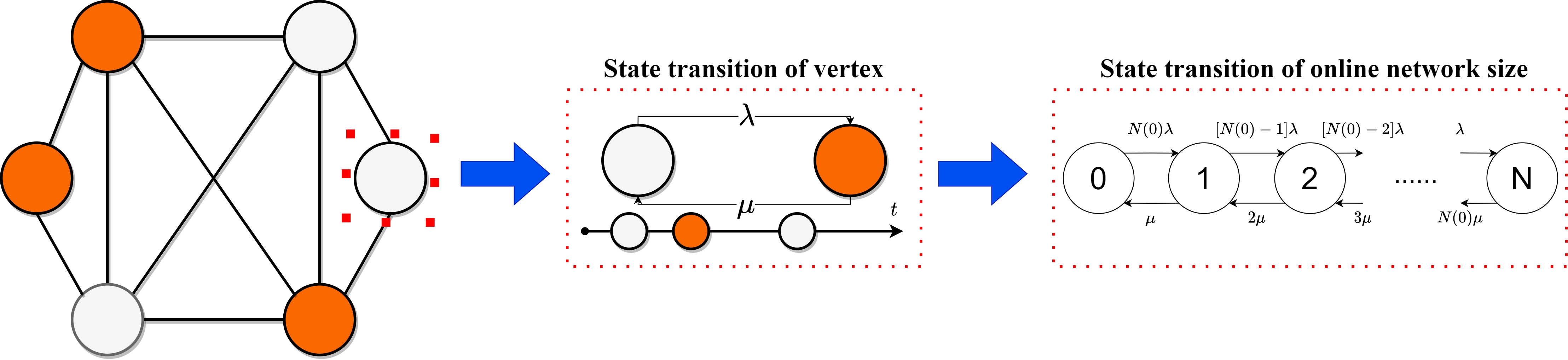}
\caption{\textbf{System diagram of the proposed model.} This figure presents an illustration of one vertex's state transition and online network size transition. The state transition of a large number of vertices constitutes our model. (color online)}
\label{fig:StateTransition}
\end{figure}

The initial network can be an existing model with $N$ vertices, e.g., SF, ER, or SW, where we assume that all vertices are online at time $t=0$. In Fig. \ref{fig:example}, we show an example of the NOH model, which contains 6 vertices. On the time axis, each vertex's online duration is colored in orange. For a better illustration, in the beginning, all vertices are online and possess an online duration. When $t=20$, the online vertices are $B$, $E$, $F$ (colored in orange), and the hidden ones are $A$, $C$, and $D$. According to our model, hidden vertices disconnect their neighbors temporarily, hence there are inactive edges (grey dotted lines) beside $A$, $C$ and $D$. When $t=60$, the online vertices are $B$, $C$, $E$, $F$, and the hidden ones are $A$, $D$. Compared to $t=20$, the vertex $C$ becomes online. Therefore, the vertex $C$ reconnects to its original neighbors. However, the vertex $D$ is hidden, hence the vertex $C$ cannot reconnect to $D$ at $t=60$. In Fig. \ref{fig:StateTransition}, we show a system diagram of the state transitions of vertex and online network size. 
\subsection{Definitions and Theoretical Analysis}

As illustrated in the modeling part, the NOH model makes vertices hidden (online) and temporarily breaks (reconnects) edges, which can be regarded as a homogeneous continuous and irreducible Markov chain. In this subsection, we introduce some theoretical analysis on NOHs principally based on the birth and death process.

To explain our theory more clearly, we first introduce some required definitions based on Ref. \cite{ross2014introduction}. 

\begin{definition}\label{def:1}
Let $\{k_i(t), t\geq 0\}$ denote a stochastic process of the number of the vertex $i$'s online neighbors with the state space $\Omega_{i}=\{0, 1, 2, \ldots, k_i(0)\}$, where $k_i(0)$ indicates the initial degree of vertex $i$.
\end{definition}
\begin{definition}\label{def:2}
Let $\{N(t), t\geq 0\}$ denote a stochastic process of the network size with the state space $\Omega=\{0, 1, 2, \ldots, N(0)\}$, where $N(0)$ indicates the initial size of the network.
\end{definition}

Defs. \ref{def:1} and \ref{def:2} define the stochastic processes of the number of each vertex's online neighbors and the network size respectively. Visually, $k_i(t)$ and $N(t)$ are the degree of vertex $i$ and online network size at time $t$. As introduced previously, vertices of the network do not grow a larger size than the initial state but become hidden and online alternatingly based on the birth and death process. Therefore, in Def. \ref{def:1}, the maximum value for the vertex $i$'s state space $\Omega_i$ is its initial degree $k_{i}(0)$. If the vertex $i$ is online at time $t$, $k_i(t)$ denotes the vertex $i$'s degree. In a similar manner, in Def. \ref{def:2}, the maximum value in the state space of the network size $\Omega$ is the initial network size $N(0)$.
\begin{definition}\label{def:3}
Let $p_{i(m,n)}(\Delta t)$ be the probability that the number of the vertex $i$'s online neighbors is $m$ and will next transfer to $n$ in the time interval $\Delta t$, where
\begin{equation}\label{eq:transition probability of degree}
p_{i(m,n)}(\Delta t)=P\{k_i(t+\Delta t)=n\vert k_i(t)=m\}, m,n\in \Omega_{i}
\end{equation}
and $q_{i(m,n)}$ be the corresponding probability transition rate, where
\begin{equation}\label{eq:transition rate of degree}
q_{i(m,n)}=\left\{
\begin{array}{lr}
\lim_{t\rightarrow 0^{+}}\frac{p_{i(m,n)}(t)}{t}, & m\neq n\\
\lim_{t\rightarrow 0^{+}}\frac{1-p_{i(m,n)}(t)}{t}, & m=n \\
\end{array}
\right.m,n\in\Omega_i.
\end{equation}
\end{definition}
\begin{definition}\label{def:4}
Let $p_{m,n}(\Delta t)$ denote the probability that the network size is $m$ and will next transfer to $n$ in the time interval $\Delta t$, where
\begin{equation}\label{eq:transition probability of size}
p_{m,n}(\Delta t)=P\{N(t+\Delta t)=n\vert N(t)=m\}, m,n\in\Omega,
\end{equation}
\end{definition}
and $q_{m,n}$ be the corresponding probability transition rate, where
\begin{equation}\label{eq:transition rate of size}
q_{m,n}=\left\{
\begin{array}{lr}
\lim_{t\rightarrow 0^{+}}\frac{p_{m,n}(t)}{t}, & m\neq n\\
\lim_{t\rightarrow 0^{+}}\frac{1-p_{m,n}(t)}{t}, & m=n \\
\end{array}
\right.m,n\in\Omega
\end{equation}

Defs. \ref{def:3} and \ref{def:4} define the transition probability in the time interval $\Delta t$ (Eqs. \ref{eq:transition probability of degree} and \ref{eq:transition probability of size}) and the transition rate (Eqs. \ref{eq:transition rate of degree} and \ref{eq:transition rate of size}) of each vertex's degree and the network size separately. $p_{i(m,n)}(\Delta t)$ denotes the probability that the vertex $i$'s online neighbor number changes from $m$ to $n$ during $\Delta t$, and $p_{m,n}(\Delta t)$ is the probability that the online network size changes from $m$ to $n$ during $\Delta t$. In addition, $q_{i(m,n)}$ and $q_{m,n}$ denote the state transition speeds of the corresponding stochastic process. 
\begin{definition}\label{def:5}
Let $p_{i(n)}=\lim_{t\rightarrow\infty}p_{i(n)}(t)$ denote the limiting probability that the vertex $i$'s degree is $n$, where
\begin{equation}
p_{i(n)}(t)=P\{k_{i}(t)=n\},n\in\Omega_i
\end{equation}
\end{definition}
\begin{definition}\label{def:6}
Let $p_{n}=\lim_{t\rightarrow\infty}p_{n}(t)$ denote the limiting probability that the network size is $n$, where
\begin{equation}
p_{n}(t)=P\{N(t)=n\}, n\in\Omega
\end{equation}
\end{definition}

Defs. \ref{def:5} and \ref{def:6} define limiting probabilities of the stochastic process $\{k_i(t), t\geq 0\}$ and $\{N(t), t\geq 0\}$ under long-range time. Here, $p_{i(n)}$ denotes the probability that the number of vertices $i$'s online neighbors is $n$, and $p_{n}$ denotes the probability that the network size is $n$. Based on the definitions above, it is obvious that the difference between the stochastic process of the vertex $i$'s degree $k_i(t)$ and the network size $N(t)$ is the state space, where the state space of the process $k_i(t)$ and $N(t)$ are the vertex $i$'s possible degree and the possible size of the network respectively. Therefore, if the stationary distribution of $N(t)$ exists, we can regard $k_i(t)$ as the stochastic process of the size of a subnetwork, which contains the vertex $i$ and its neighbors. In addition, the stationary distribution of $k_i(t)$ has a similar form as $N(t)$, where we simplify the subsequent description in a theoretical analysis.

Before we do further theoretical analysis, we note that the aforementioned stochastic processes are all homogeneous, continuous, and irreducible. According to Eqs. \ref{eq:transition probability of degree} and \ref{eq:transition probability of size}, each transition probability is only related to the time interval $\Delta t$ but not to the initial time $t$. Therefore, each mentioned Markov chain is homogeneous. Besides, considering an extremely small time $t$, it is apparent that each chain stays in its state with probability 1, i.e., $\lim _{t\rightarrow0^+}p_{i(m,m)}(t)=1$ for the process $k_i(t)$ and $\lim _{t\rightarrow0^+}p_{m,m}(t)=1$ for the process $N(t)$. Moreover, provided $m\neq n$, we have $\lim _{t\rightarrow0^+}p_{i(m,n)}(t)=0$ for the process $k_i(t)$ and $\lim _{t\rightarrow0^+}p_{m,n}(t)=0$ for the process $N(t)$, which indicates that the probability to change the state of each chain is 0. Therefore, each chain's continuity holds. Based on Defs. \ref{def:1} and \ref{def:2}, as we introduced in Sec. \ref{subsec:model}, all states in the state space $\Omega_i$ are connective, and the same is true in the state space $\Omega$, i.e., the Markov chain $k_i(t)$ and $N(t)$ are irreducible. Therefore, the limiting probability and stationary distribution are equivalent in our model. Then, to study the effect of our proposed mechanisms, we are particularly concerned about the stationary distribution of the two stochastic processes defined in Defs. \ref{def:1} and \ref{def:2}. 

For the stationary distribution of the NOH model, we first give a lemma for the stationary distribution without proof. 
\begin{lemma}\label{lem:PQ=0}
For a homogeneous, continuous and irreducible Markov chain $\{X(t), t\geq 0\}$ with the state space $E$, its stationary distribution is $\{\pi_j, j\in E\}$, where
\begin{equation}\label{eq:kolmogrov}
-\pi_{j}q_{j,j}+\sum_{i\neq j\in E}\pi_{i}q_{ij}=0(\boldsymbol{\Pi Q}=\boldsymbol{0}).
\end{equation}
\end{lemma}

In Eq. \ref{eq:kolmogrov}, $\pi_i$ is the probability to find the state $i$ in a stochastic process, and $q_{i,j}$ is the probability transition rate from state $i$ to $j$. State spaces of the stochastic process of $k_i(t)$ and $N(t)$ are both limited. Therefore, the stationary distributions of these two processes can be obtained by solving the system of equations Eq. \ref{eq:kolmogrov}. 

We next present the transition probability of the stochastic process $k_i(t)$ and $N(t)$ as a lemma for the subsequent use without proof. 
\begin{lemma}\label{theorem:1}
For $m,n\in\Omega_i$, the transition probability of the stochastic process $k_i(t)$ is
\begin{equation}\label{eq:transition probability of kit}
p_{i(m,n)}(\Delta t)=
\left\{
\begin{array}{ll}
[k_i(0)-m]\lambda \Delta t+o(\Delta t),& n=m+1 \\
m\mu \Delta t +o(\Delta t),& n=m-1 \\
1-[k_i(0)-m]\lambda\Delta t\\
- m\mu \Delta t+o(\Delta t),& n=m \\
o(\Delta t),& \vert n-m\vert\geq 2\\
\end{array}.
\right.
\end{equation}
For $m,n\in\Omega$, the transition probability of the stochastic process $N(t)$ is
\begin{equation}\label{eq:transition probability of Nt}
p_{m,n}(\Delta t)=
\left\{
\begin{array}{ll}
[N(0)-m]\lambda \Delta t+o(\Delta t),& n=m+1 \\
m\mu \Delta t +o(\Delta t),& n=m-1 \\
1-[N(0)-m]\lambda\Delta t\\
- m\mu \Delta t+o(\Delta t),& n=m \\
o(\Delta t),& \vert n-m\vert\geq 2\\
\end{array}.
\right.
\end{equation}
\end{lemma}

Next, we carry out the stationary distribution of the process $k_i(t)$ and $N(t)$ based on Lemmas \ref{lem:PQ=0} and \ref{theorem:1}.
\begin{theorem}
\label{theorem:2}
For the stochastic process $k_i(t)$, the stationary distribution is
\begin{equation}\label{eq:stationary distribution kit}
\pi_{i(n)}=\frac{C_{k_i(0)}^{n}(\lambda\mu^{-1})^n}{(1+\lambda\mu^{-1})^{k_i(0)}}.
\end{equation}
For the stochastic process $N(t)$, the stationary distribution is
\begin{equation}\label{eq:stationary distribution Nt}
\pi_{n}=\frac{C_{N(0)}^{n}(\lambda\mu^{-1})^n}{(1+\lambda\mu^{-1})^{N(0)}}.
\end{equation}
\end{theorem}

The proof of Theorem 1 can be found in Appendix \ref{sec: appendix-proof1}. In Theorem 1, the stationary distribution presents the probability that the stochastic process stays in a certain state under a long-range time. According to Eq. \ref{eq:stationary distribution kit}, the probability that the individual $i$ has no neighbor is denoted as
\begin{equation}\label{eq: isolate}
    \pi_{i(0)}=\frac{1}{(1+\lambda\mu^{-1})^{k_i(0)}}\neq 0.
\end{equation}
Therefore, it is possible for an individual to be both online and isolated. It is worth noting that the probability distribution is only determined by $\lambda\mu^{-1}$ and $N(0)$. Therefore, even if different $\lambda$s and $\mu$s are given, once $\lambda\mu^{-1}$ is fixed, the probability distribution remains the same. 
We next analyze some statistical characteristics of the NOH model. 
\begin{theorem}\label{theorem:3}
For the vertex $i$, the expected number of online vertices in its neighbors is
\begin{equation}\label{eq:Ekit}
E[k_{i}]=\frac{k_i(0)\lambda\mu^{-1}}{1+\lambda\mu^{-1}},
\end{equation}
and its variance is
\begin{equation}\label{eq:Dki}
D[k_{i}]=\frac{k_i(0)\lambda\mu^{-1}}{(1+\lambda\mu^{-1})^2}.
\end{equation}
The expected online network size is
\begin{equation}\label{eq:EN}
E[N]=\frac{N(0)\lambda\mu^{-1}}{1+\lambda\mu^{-1}},
\end{equation}
and its variance is
\begin{equation}\label{eq:DN}
D[N]=\frac{N(0)\lambda\mu^{-1}}{(1+\lambda\mu^{-1})^2}.
\end{equation}
\end{theorem}

The proof of Theorem 2 can be found in Appendix \ref{sec: appendix-proof2}. In Theorem 2, we obtain the expected number and the variance of each vertex's online neighbors and the online network size. Besides, both expected values are only related to $\lambda\mu^{-1}$ and $N(0)$.

\section{Simulations}
\label{sec:simulation}

We start our simulations as follows: Initial networks, considering SFs and SWs, are generated via the Python package \emph{networkx}, where SFs and SWs are generated by the functions \emph{barabasi\_albert\_graph()} and \emph{watts\_strogatz\_graph()} respectively. To simulate the state transition of hidden and online vertices, we apply the Python function \emph{expovariate()} in the package \emph{random} to generate each vertex's duration in each state, where the parameter is set as $\lambda$ for a hidden vertex and $\mu$ for an online vertex. In following simulations, for parameters in our NOH model, we let the online rate $\lambda=[0.005, 0.010, 0.015, 0.020]$, the hidden rate $\mu=[0.005, 0.010, 0.015]$, $N(0)=[10^3, 2\times10^3, 4\times10^3, 6\times10^3]$, for the parameters of the initial networks, we let $m=[5, 10, 20]$ in SFs and $K=[10, 20, 40, 60]$, $p=[0.20, 0.30]$ in SWs. We note that the networks in all our simulation results are generated randomly and tested for more than 10 times. One can easily repeat our simulations and obtain the same conclusions as ours. 

\subsection{Snapshots}
\label{subsec:snapshots}
\begin{figure*}[htbp]
\centering
  \subfigure[$t=2\times10^3$]{
    \includegraphics[scale=0.12]{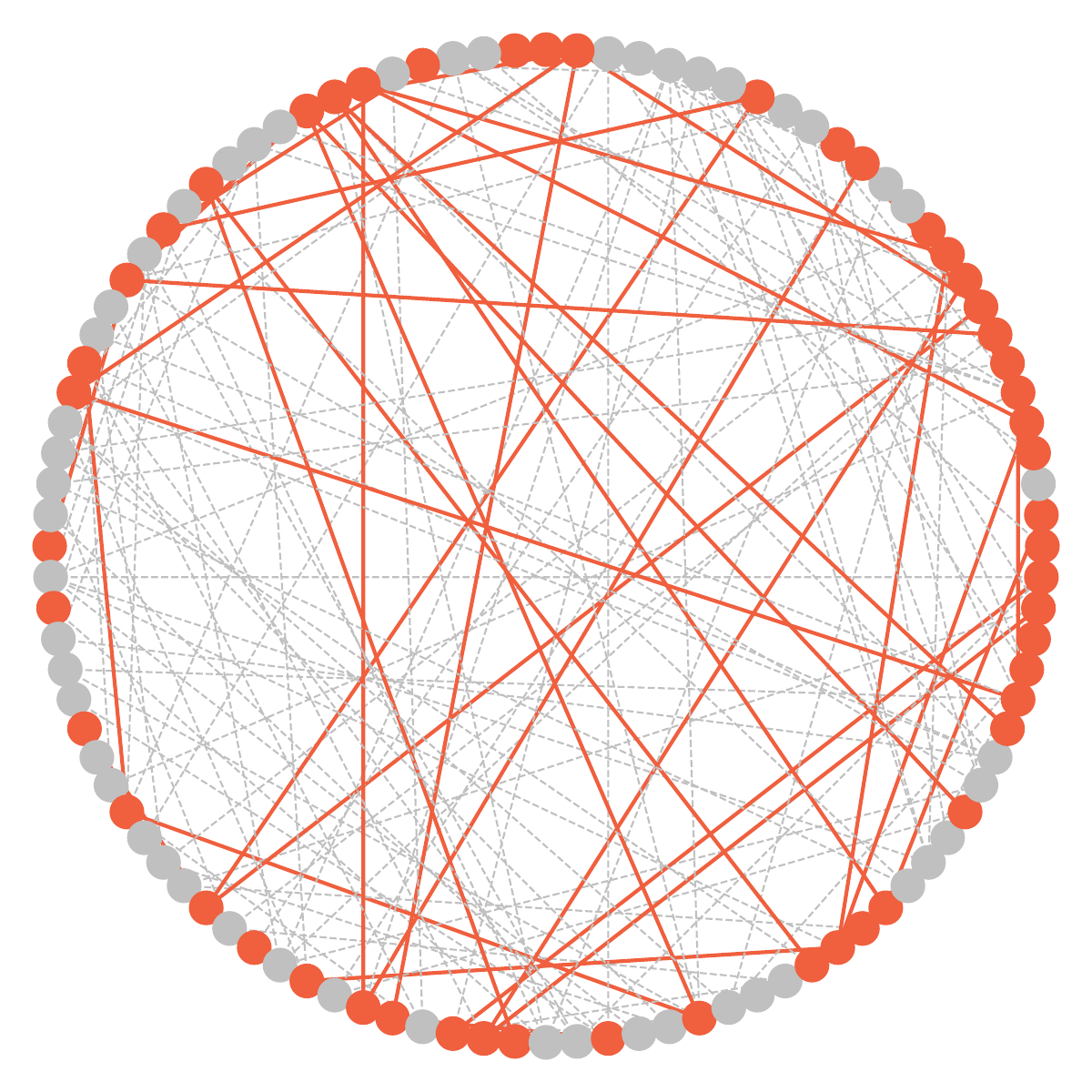}
    \label{fig:snapshots/1_1}
  }
  \subfigure[$t=4\times10^3$]{
    \includegraphics[scale=0.12]{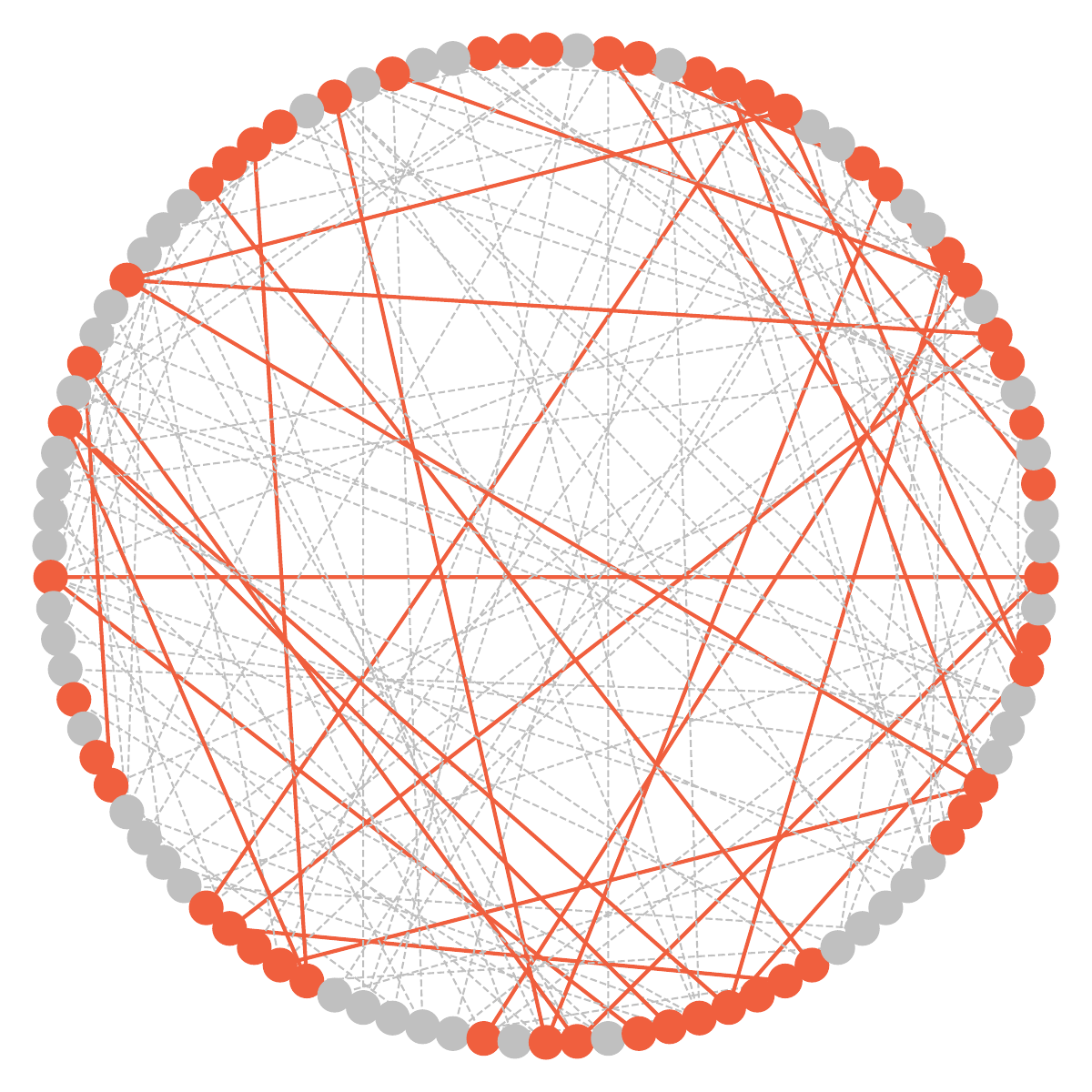}
    \label{fig:snapshots/1_2}
  }
  \subfigure[$t=6\times10^3$]{
    \includegraphics[scale=0.12]{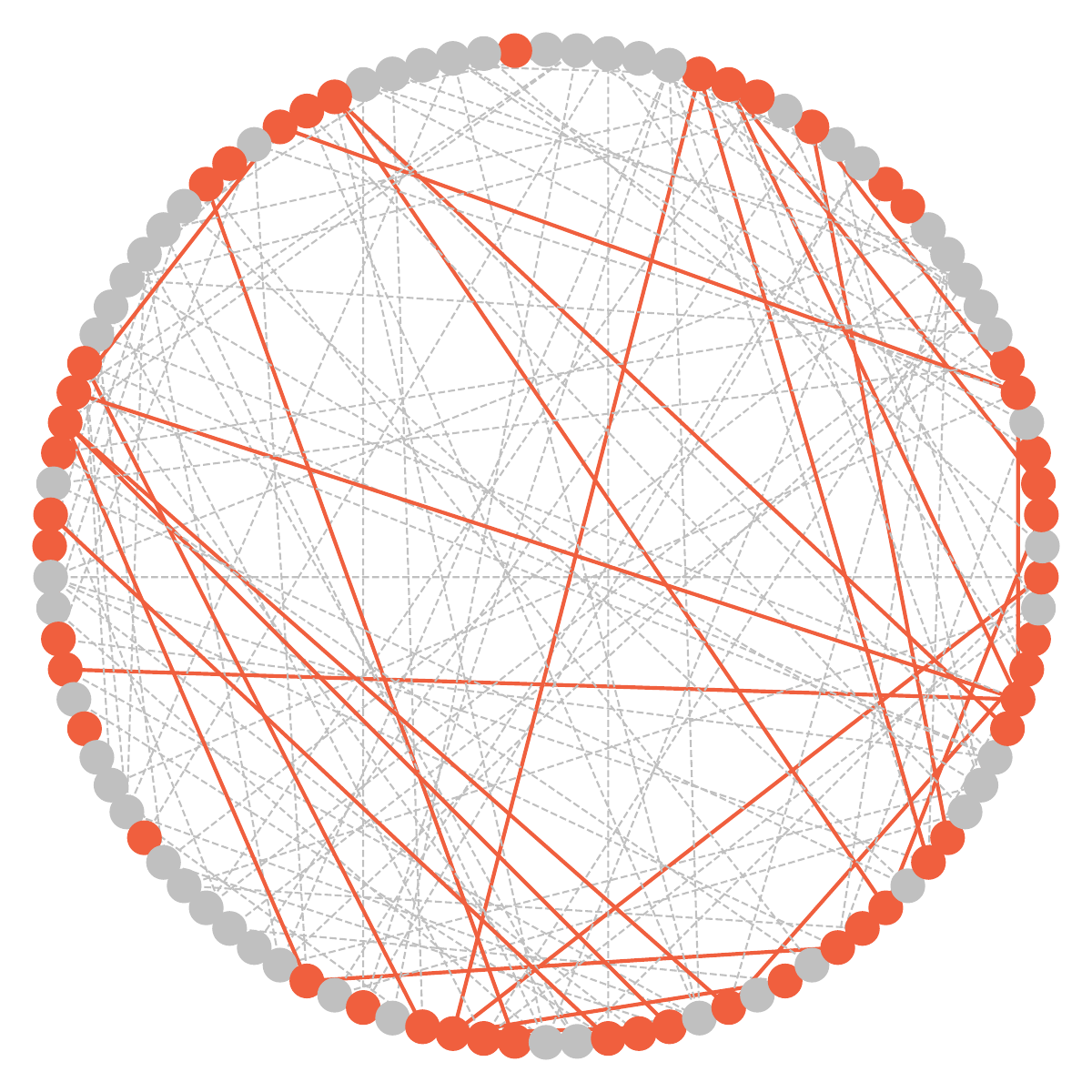}
    \label{fig:snapshots/1_3}
  }
  \subfigure[$t=8\times10^3$]{
    \includegraphics[scale=0.12]{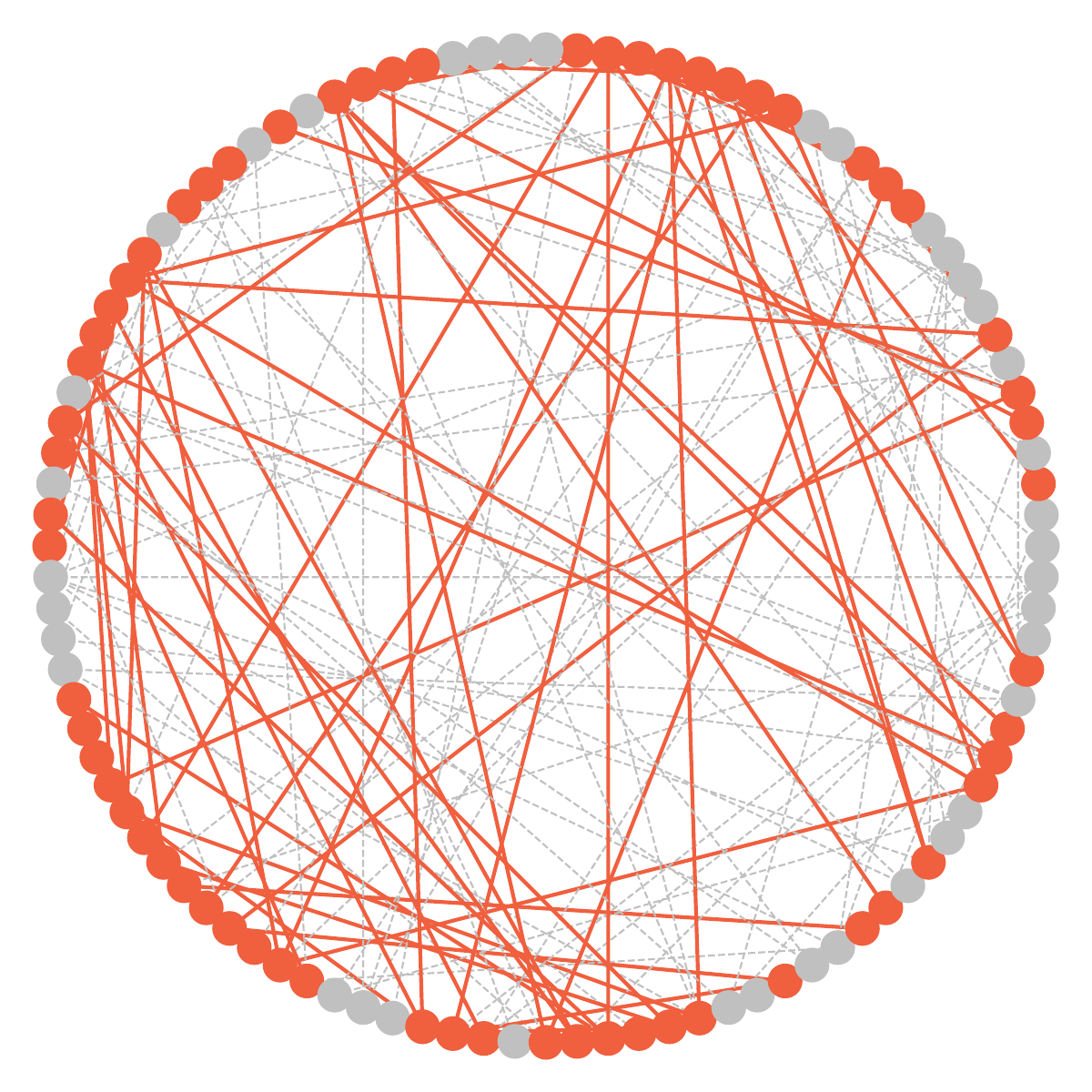}
    \label{fig:snapshots/1_4}
  }
  \subfigure[$t=10^4$]{
    \includegraphics[scale=0.12]{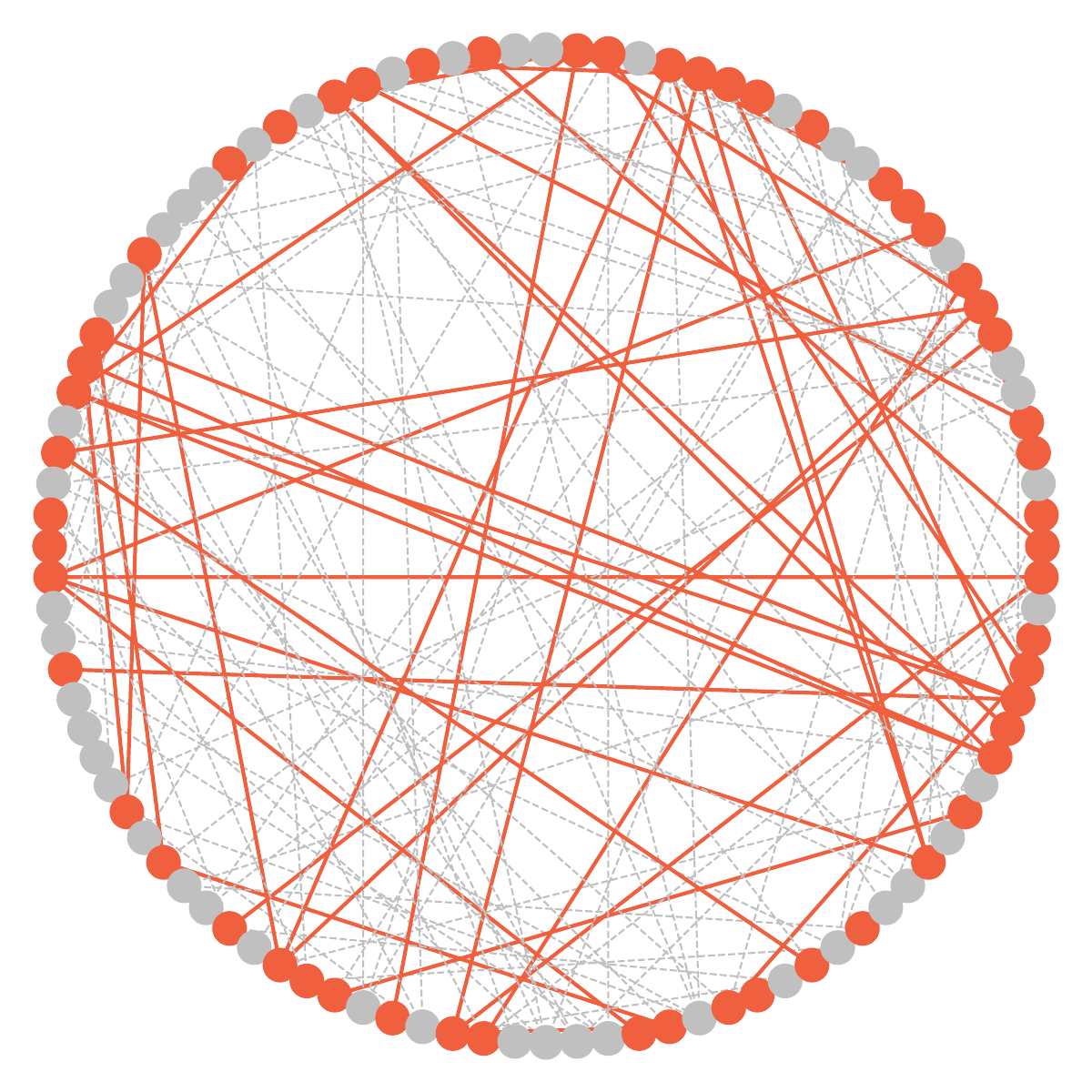}
    \label{fig:snapshots/1_5}
  }
  \vspace{-3.3mm}

  \subfigure[$t=2\times10^3$]{
    \includegraphics[scale=0.12]{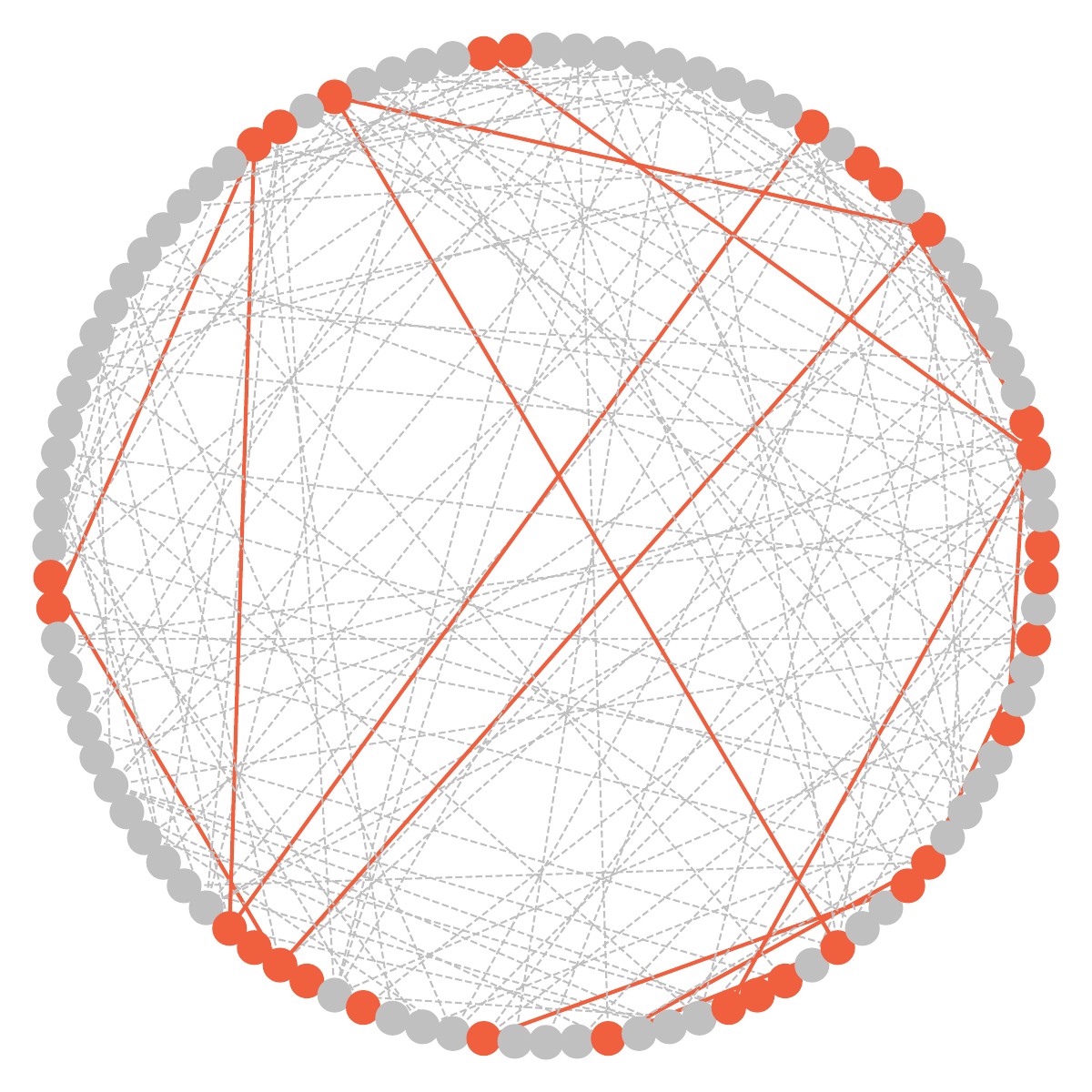}
    \label{fig:snapshots/2_1}
  }
  \subfigure[$t=4\times10^3$]{
    \includegraphics[scale=0.12]{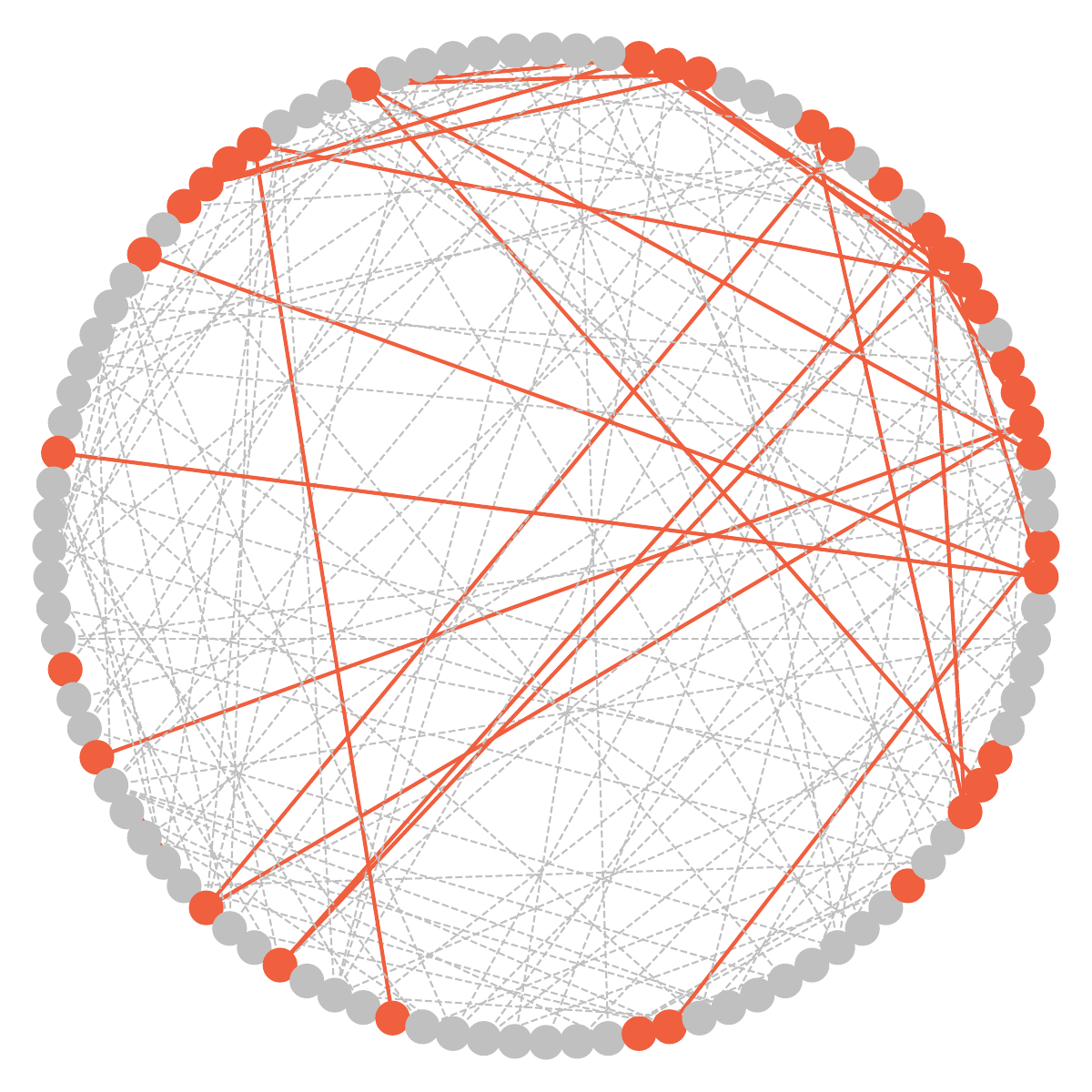}
    \label{fig:snapshots/2_2}
  }
  \subfigure[$t=6\times10^3$]{
    \includegraphics[scale=0.12]{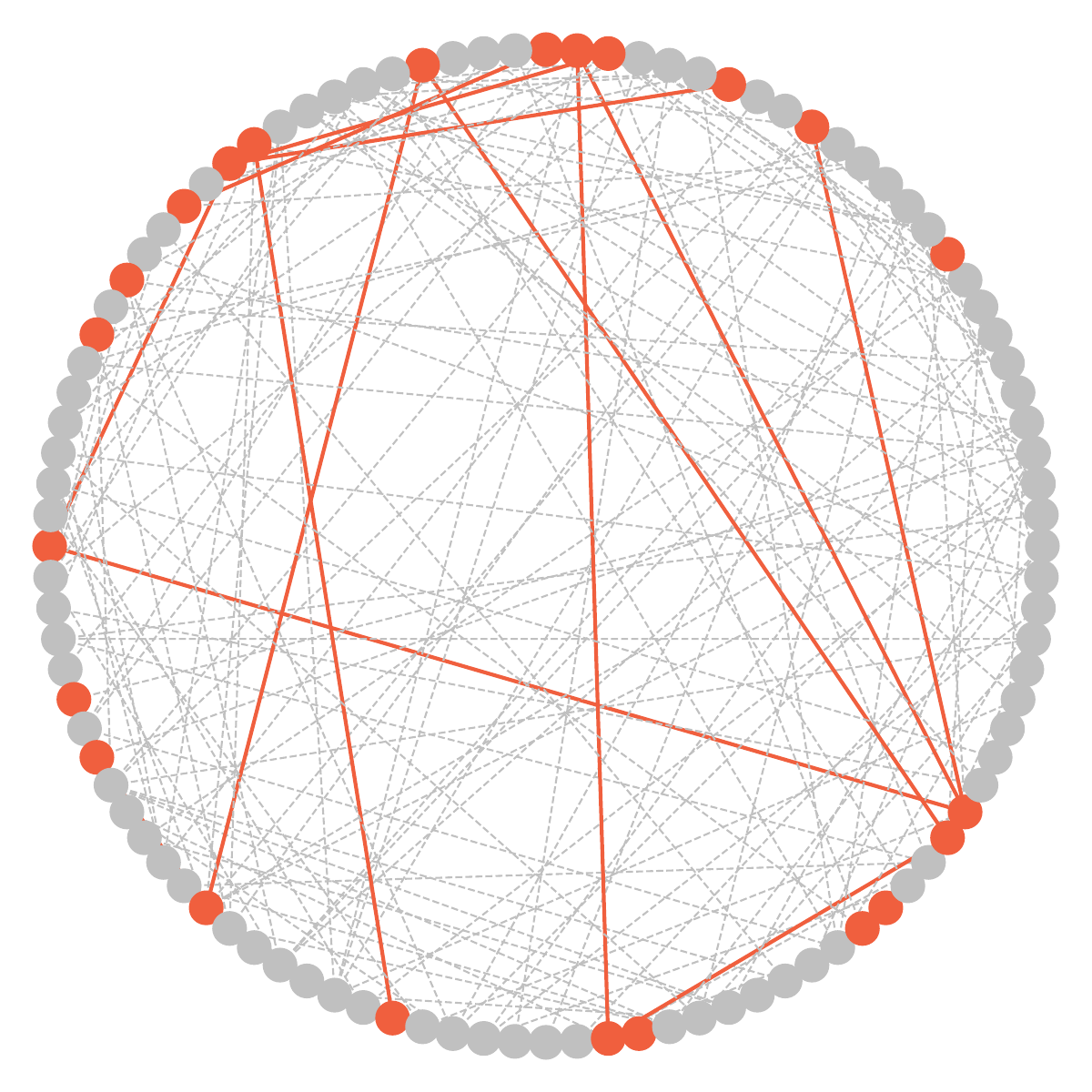}
    \label{fig:snapshots/2_3}
  }
  \subfigure[$t=8\times10^3$]{
    \includegraphics[scale=0.12]{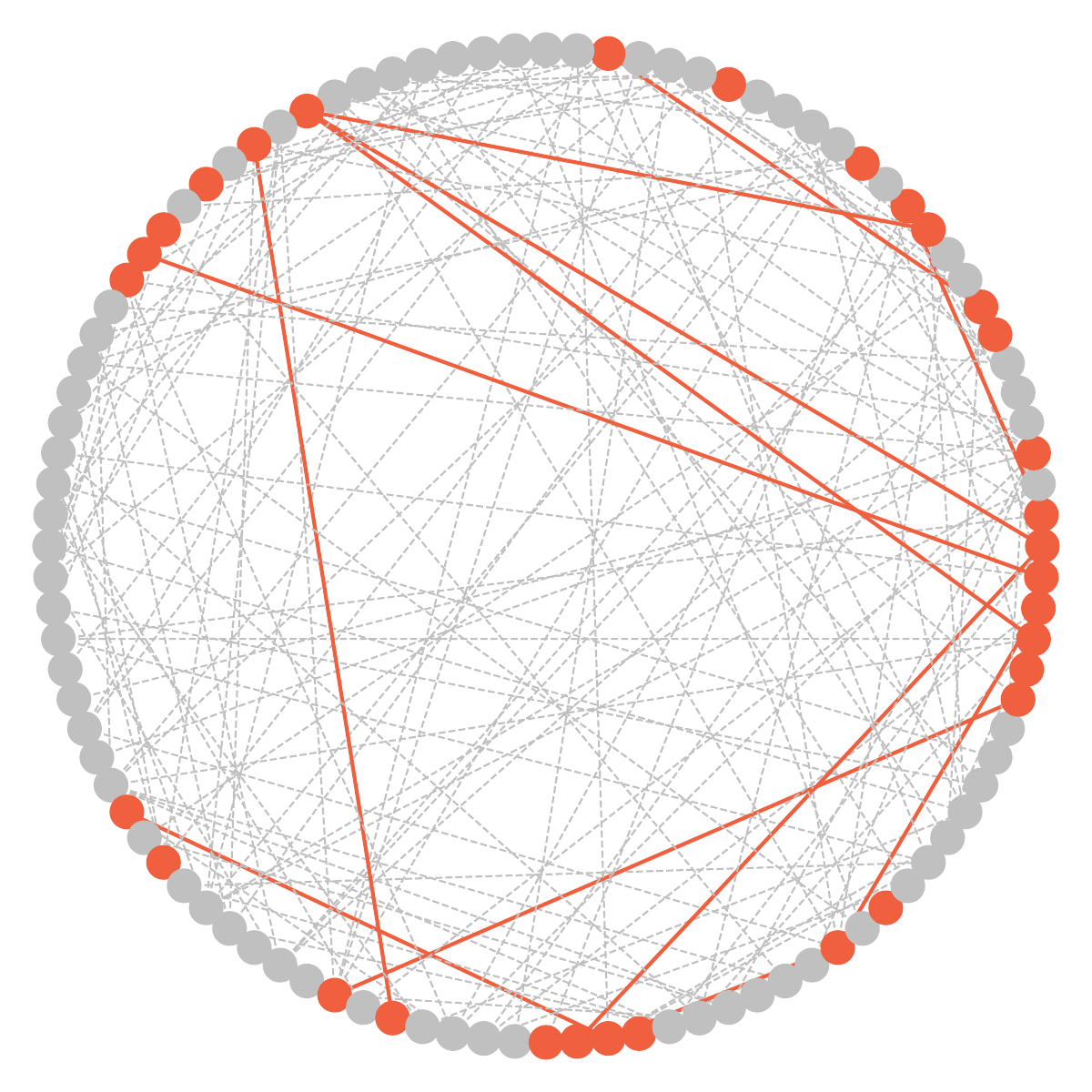}
    \label{fig:snapshots/2_4}
  }
  \subfigure[$t=10^4$]{
    \includegraphics[scale=0.12]{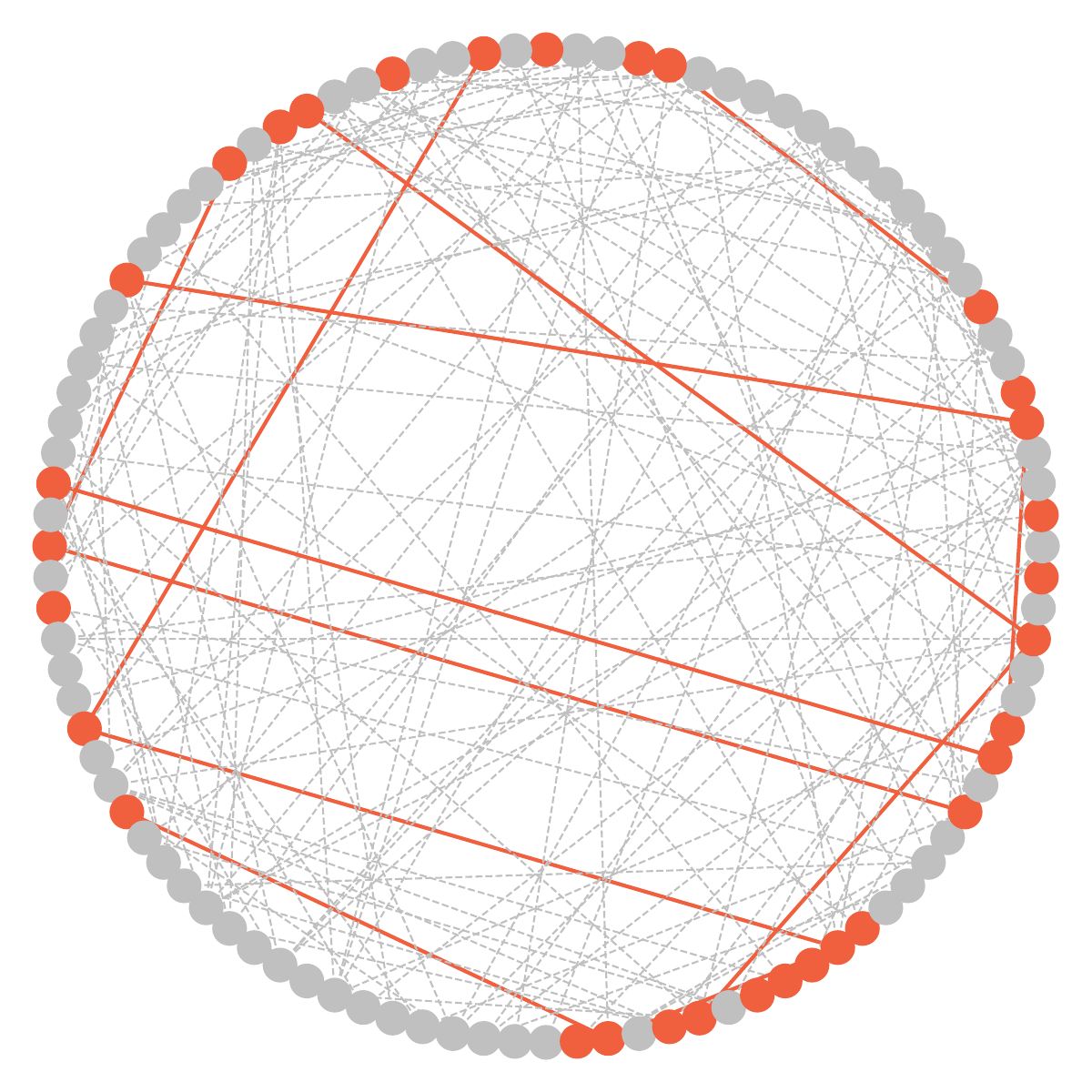}
    \label{fig:snapshots/2_5}
  }
  \vspace{-3.3mm}
\caption{\textbf{Snapshots of NOHs. }The online network structure is stable as time evolves and different as $\lambda$ and $\mu$ change. We present snapshots at time $t=2\times10^3,4\times10^3,6\times10^3,8\times10^3,10^4$ with fixed $\lambda=0.005$ and $\mu=0.005$ for Figs. \ref{fig:snapshots/1_1}-\ref{fig:snapshots/1_5}, and $\mu=0.010$ for Figs. \ref{fig:snapshots/2_1}-\ref{fig:snapshots/2_5}. The initial network is set as SW with 100 vertices (small number for a better presentation), $K=10$ and $p=0.30$. Vertices in orange are online and in grey are hidden. Solid links in orange are for two online individuals, and dashed links in grey are for at least one hidden vertex. (color online)}
\label{fig:snapshots}
\end{figure*}
In this subsection, by setting the initial network containing 100 vertices as the SW with $K=10$, $p=0.30$, we give characteristic snapshots of NOHs. Fixing $\lambda=0.005$, we compare network structures under two sets of parameters $\mu=0.005$ in Figs. \ref{fig:snapshots/1_1}-\ref{fig:snapshots/1_5} and $\mu=0.010$ in Figs. \ref{fig:snapshots/2_1}-\ref{fig:snapshots/2_5}. As shown in Figs. \ref{fig:snapshots/1_1}-\ref{fig:snapshots/1_5}, the online vertices (colored in orange) occupy approximately half of the initial network, and in Figs. \ref{fig:snapshots/2_1}-\ref{fig:snapshots/2_5}, they take up around 1/3, and the amount of online interactions are declined. Although online interaction relationships are varying as time passes, the amount of interaction and the number of online individuals are stable. Besides, each online network above is more sparse than its initial state because a link is detectable (colored in orange) if and only if two vertices at its both ends are online. Additionally, more isolated vertices are observed when the hidden rate $\mu=0.010$ is given, indicating that a vertex is more likely to have no friend to contact with for the reason that its neighbors online duration is relatively small.
\subsection{Numbers of vertices' online neighbors}
\label{subsec:kit}

\begin{figure*}[htbp]
\centering
\vspace{-3.3mm}
  \subfigure[SF, $\lambda=0.010$, $\mu=0.005$]{
    \includegraphics[scale=0.21]{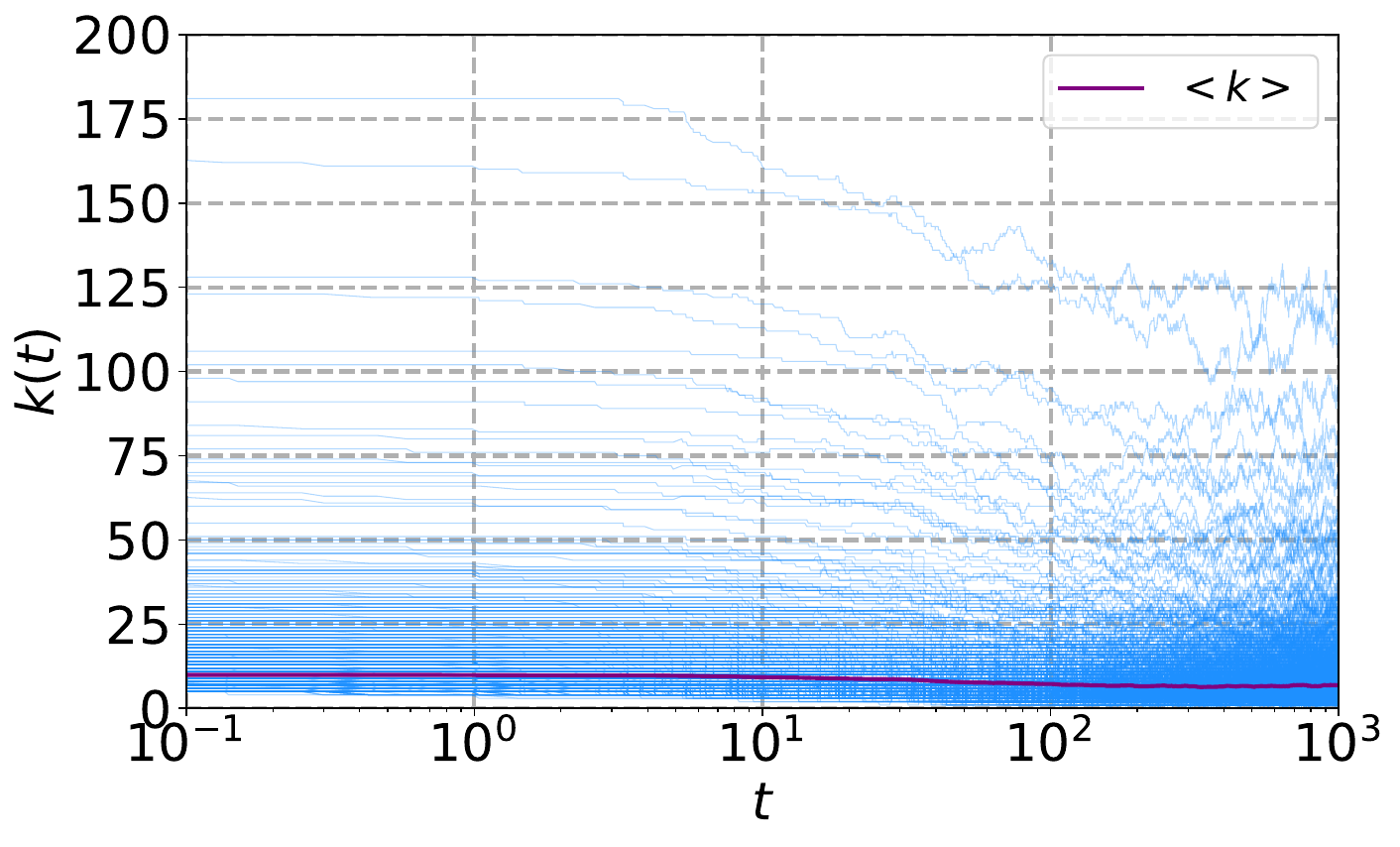}
    \label{fig:1/SF_1.00_0.50}
  }
  \hspace{-5mm}
  \subfigure[SF, $\lambda=0.010$, $\mu=0.010$]{
    \includegraphics[scale=0.21]{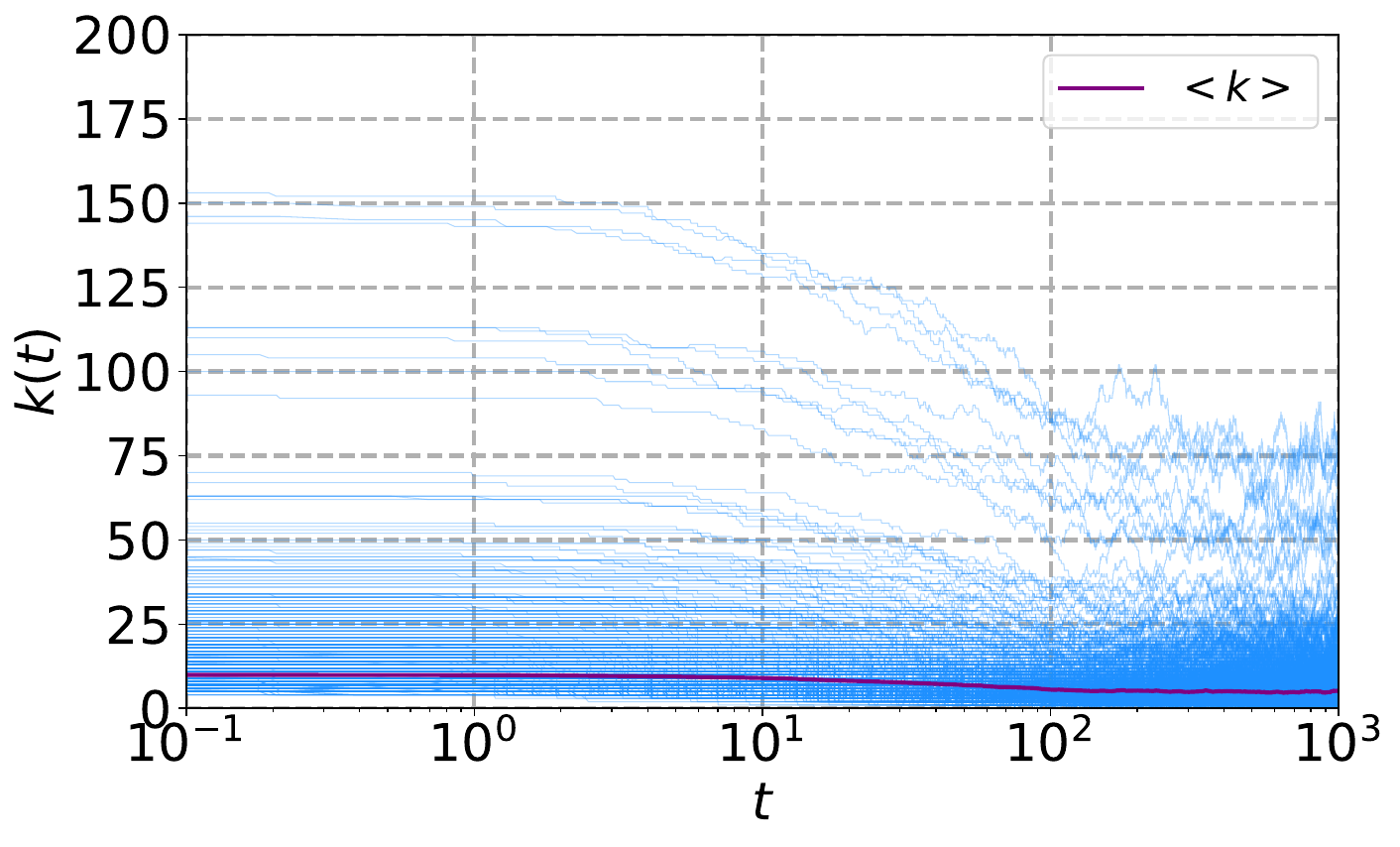}
    \label{fig:1/SF_1.00_1.00}
  }
    \hspace{-5mm}
  \subfigure[SF, $\lambda=0.010$, $\mu=0.015$]{
    \includegraphics[scale=0.21]{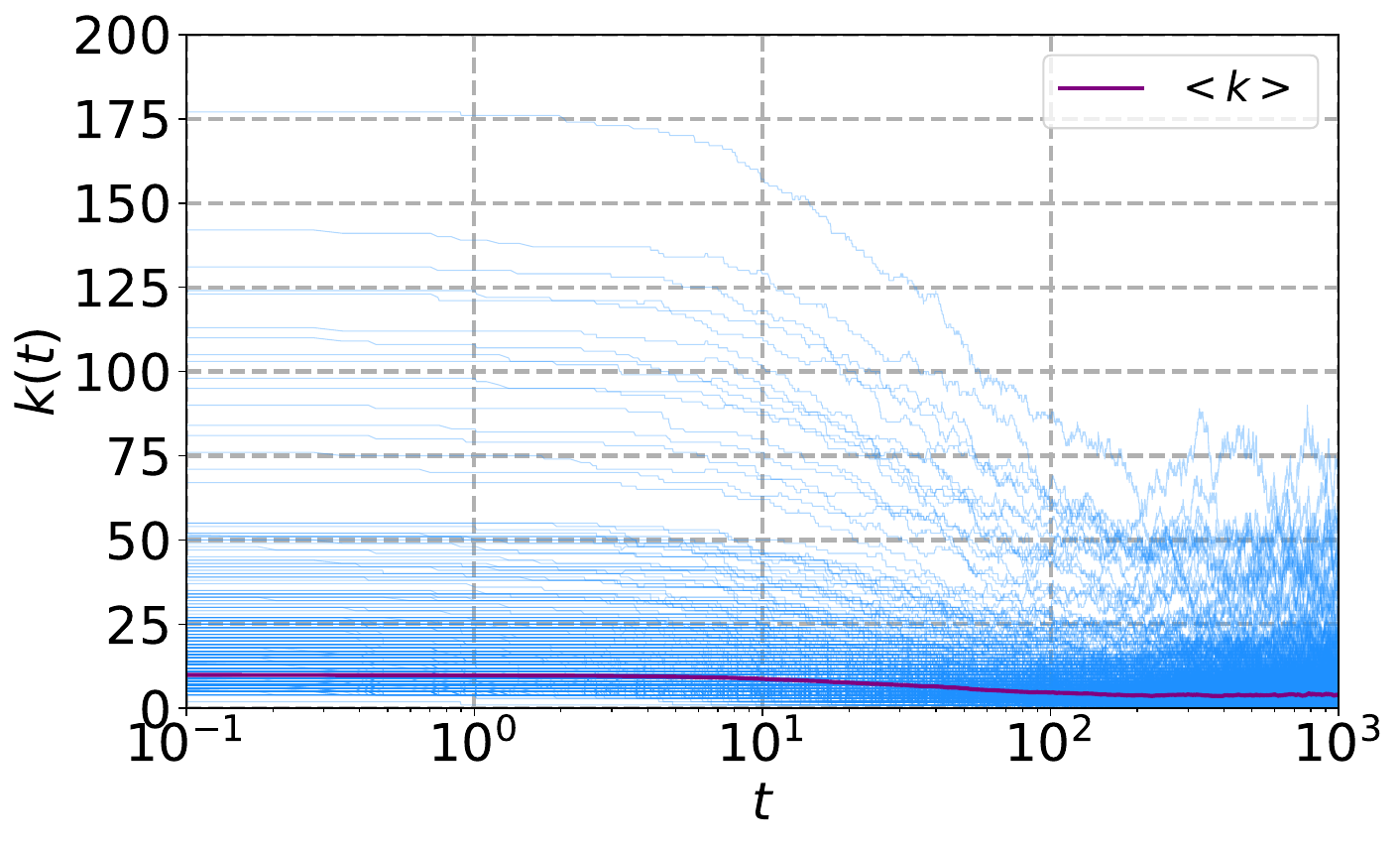}
    \label{fig:1/SF_1.00_1.50}
  }
    \hspace{-5mm}
\vspace{-3.3mm}

  \subfigure[SW, $\lambda=0.010$, $\mu=0.005$]{
    \includegraphics[scale=0.21]{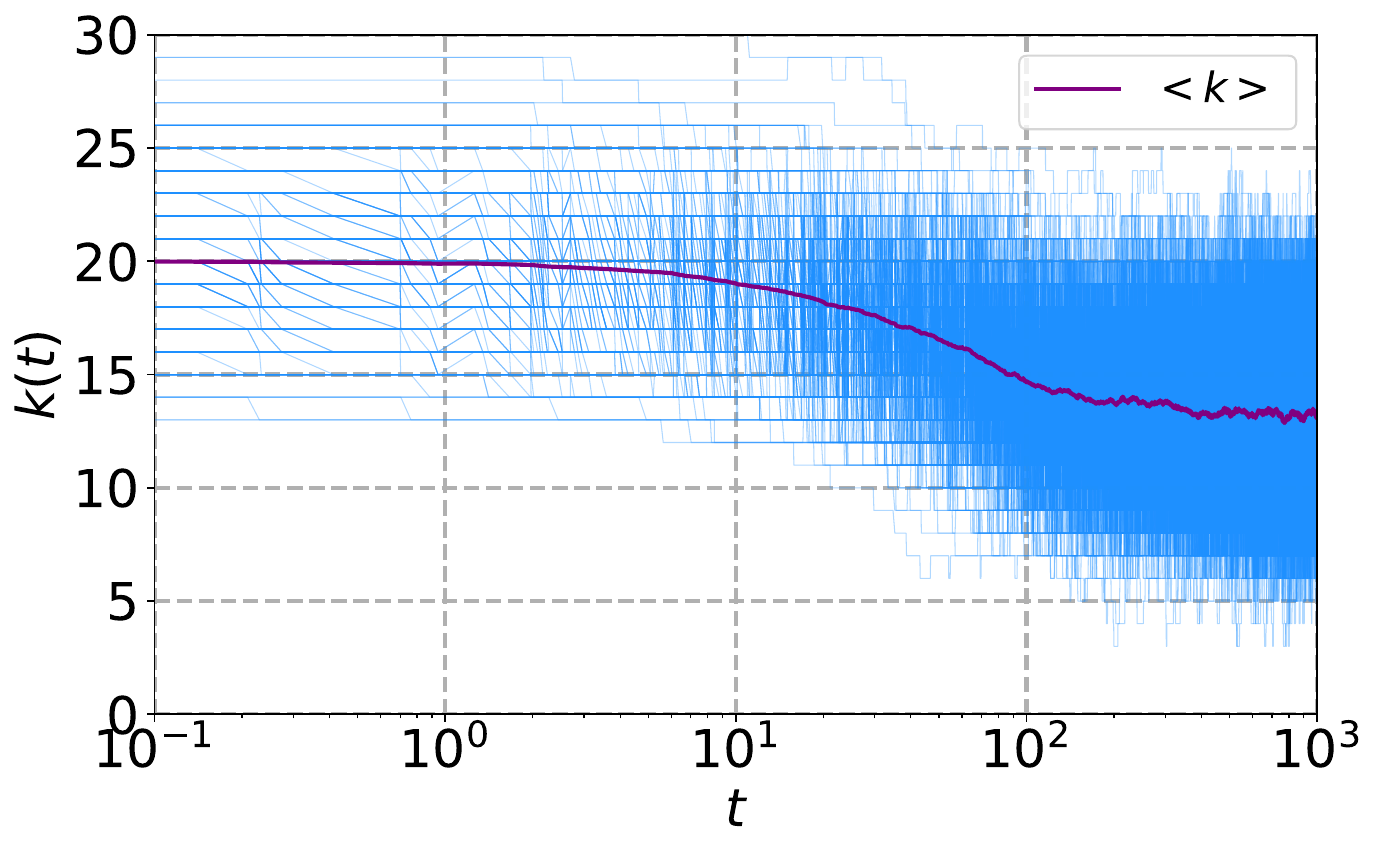}
    \label{fig:1/SW_1.00_0.50}
  }
    \hspace{-5mm}
  \subfigure[SW, $\lambda=0.010$, $\mu=0.010$]{
    \includegraphics[scale=0.21]{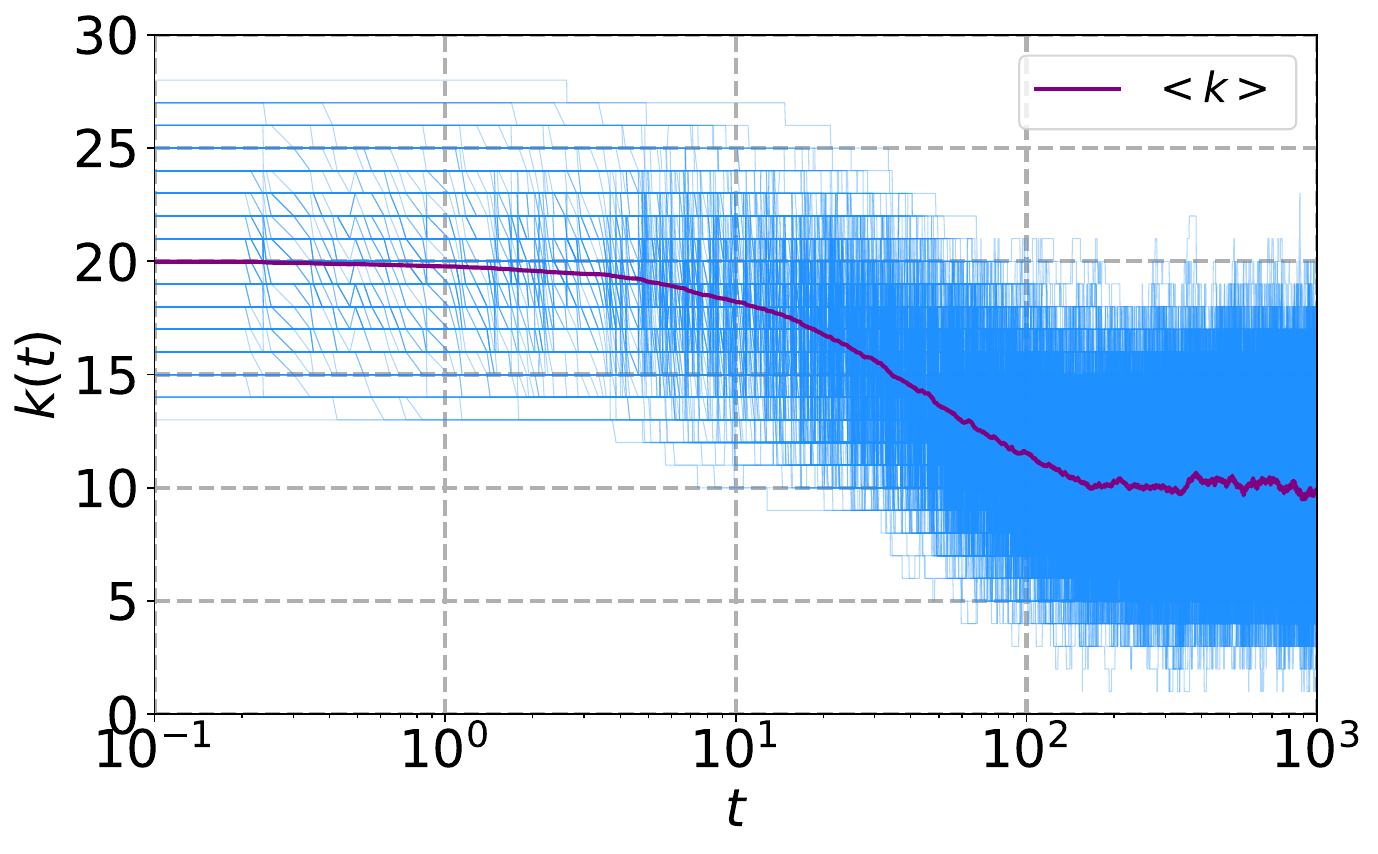}
    \label{fig:1/SW_1.00_1.00}
  }
    \hspace{-5mm}
  \subfigure[SW, $\lambda=0.010$, $\mu=0.015$]{
    \includegraphics[scale=0.21]{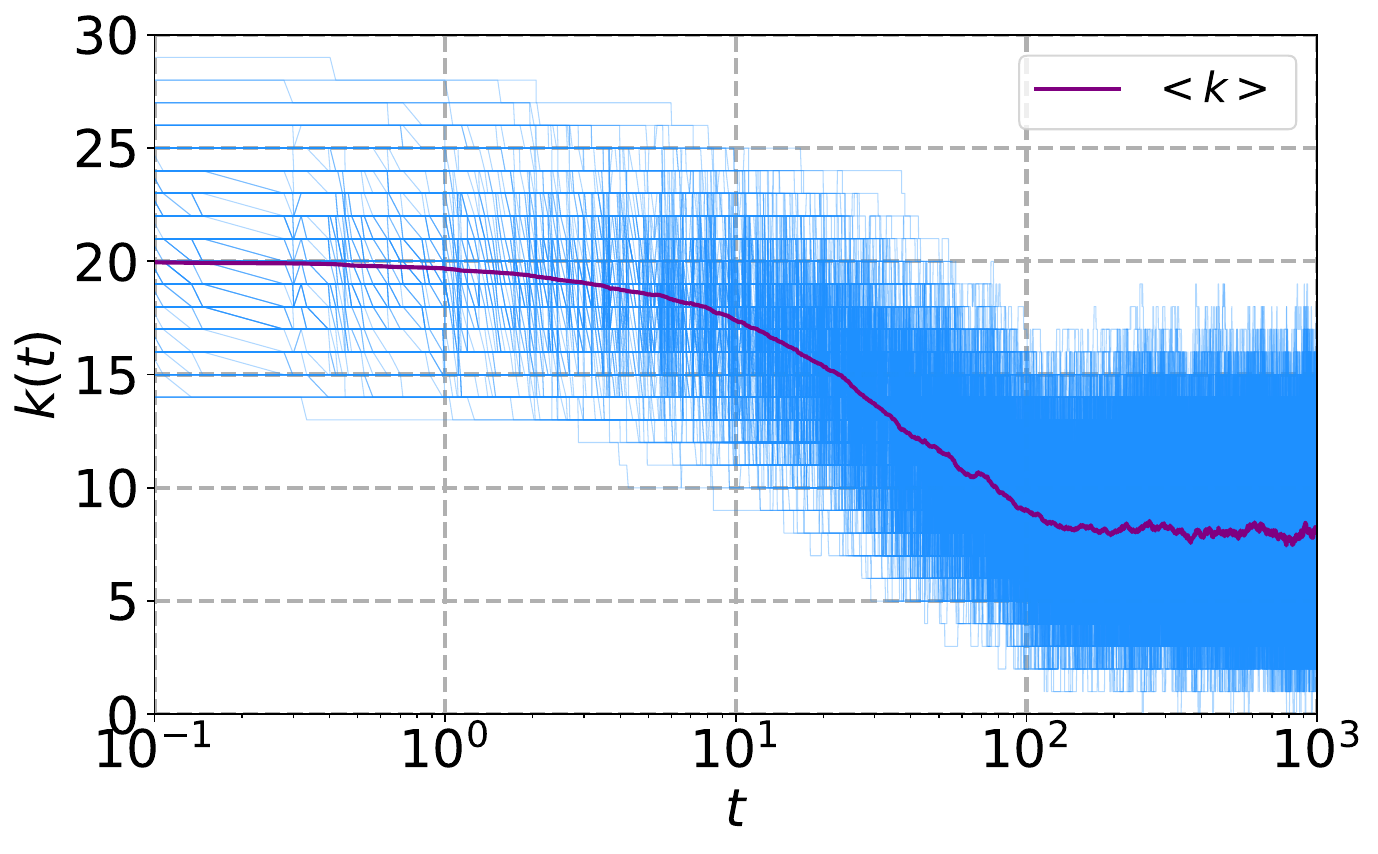}
    \label{fig:1/SW_1.00_1.50}
  }
    \hspace{-5mm}
  \vspace{-3.3mm}
\caption{\textbf{The number of each vertex's online neighbors as functions of time. }The numbers of each vertex's online neighbors along with the expectation are stable as time evolves. Each subplot presents the numbers of all vertices' online neighbors for $t\leq10^3$ (blue plots) and the average number of each vertex's online neighbors (purple plots) with initial sizes $N(0)=2\times10^3$ and $\lambda$s$=0.010$. Network types and $\mu$s are set as \textbf{\ref{fig:1/SF_1.00_0.50}} SF, $\mu=0.005$, \textbf{\ref{fig:1/SF_1.00_1.00}} SF, $\mu=0.010$, \textbf{\ref{fig:1/SF_1.00_1.50}} SF, $\mu=0.015$, \textbf{\ref{fig:1/SW_1.00_0.50}} SW, $\mu=0.005$, \textbf{\ref{fig:1/SW_1.00_1.00}} SW, $\mu=0.010$, \textbf{\ref{fig:1/SW_1.00_1.50}} SW, $\mu=0.015$. The $x$-axis is set as the time $t$ in the range $[10^{-1}, 10^3]$. Besides, for SFs, we set $m=5$ and the $y$-axis as the number of online neighbors $k_t$ and its range as $[0, 200]$. For SWs, we set $K=20$, $p=0.3$ and the $y$-axis range as $[0, 30]$. (color online) }
\label{fig:1}
\end{figure*}

In this subsection, we focus on the number of each vertex's online neighbors, which is interpreted as the stochastic process $k_i(t)$ for the vertex $i$. As stated in Sec. \ref{sec:Complex Networks With Online and Hidden Vertices}, a larger $\lambda$ provides a higher rate for hidden vertices to become online, and a larger $\mu$ displays a higher rate for vertices to hide themselves or go offline. In the first simulation, we set the online rate $\lambda=0.010$, the hidden rate $\mu=0.005, 0.010, 0.015$ and observe the evolution process of the number of each vertex's online neighbors on SFs ($m=5$) and SWs ($K=20, p=0.30$) (Fig. \ref{fig:1}).

Blue plots in Fig. \ref{fig:1} present the evolution process for each vertex, and the purple ones the average situation. Each blue line is set to be transparent, therefore if lines are denser, the frequency is higher. Although the initial state of each same type network is distinguishing as a consequence of the randomness of the network generating algorithm, the characteristic of each network has no essential difference. Apparently, in each plot, the number of each vertex's online neighbors $k_i(t)$ reduces at the beginning and becomes stable at $t=10^3$. Besides, a network with a higher hidden rate $\mu$ leads to fewer online neighbors for each vertex. As given in Theorem \ref{theorem:2}, the number of each vertex's online neighbors follows a distribution describe as Eq. \ref{eq:stationary distribution kit}, which indicates that each $k_i(t)$ steadily floats within a bound as shown in Fig. \ref{fig:1}. Additionally, the variance we obtained in Eq. \ref{eq:Dki} ensures that the stochastic sequence has a value far from the expected value with an extremely low probability. When the initial network is set as SF, in Fig. \ref{fig:1/SF_1.00_0.50}, there are vertices still having more than 100 online neighbors, and plots are sparse, which indicates that the online duration is longer for each vertex's online neighbors. Additionally, with a higher hidden rate $\mu$s, in Fig. \ref{fig:1/SF_1.00_1.00}, there are no vertex having more than 75 online neighbors, and in Fig. \ref{fig:1/SF_1.00_1.50}, each vertex possesses fewer online neighbors than in the situations above. Besides, although the average number of each vertex's online neighbors is not prominent due to the network heterogeneity, to guarantee the integrity of simulations, we plot the whole evolution process. It is worth noting that the average number declines as time passes, and it declines faster with a higher $\mu$.

Provided that the initial network is SW, the difference becomes more significant. In Figs. \ref{fig:1/SW_1.00_0.50}-\ref{fig:1/SW_1.00_1.50}, blue plots decline faster if the hidden rate $\mu$ is larger, which indicates that the number of each vertex's online neighbors grows lower with a larger hidden rate $\mu$. Besides, the purple plots report the average number of each vertex's online neighbors, which declines and becomes stable as well. As a consequence of the homogeneity of SWs, the average number of each vertex's online neighbors is presented clearly in Figs. \ref{fig:1/SW_1.00_0.50}-\ref{fig:1/SW_1.00_1.50}. In Fig. \ref{fig:1/SW_1.00_0.50}, the purple plot becomes stable at $<k>=14$ approximately, and in Figs. \ref{fig:1/SW_1.00_1.00} and \ref{fig:1/SW_1.00_1.50}, it becomes stable around $<k>=10$ and $<k>=7$ respectively.

Note that it is possible for an individual to be isolated according to our analysis in Eq. \ref{eq: isolate}, which is shown in Fig. \ref{fig:1} as well. A blue plot may decrease to zero for a certain time, but increase as the network evolves. That is, an individual may be isolated at a time. As the time goes, its offline neighbors will be online again and help this node out of isolation. 

\subsection{Online network sizes}
\begin{figure*}[htbp]
\centering
\vspace{-3.3mm}
  \subfigure[$\lambda=0.010, \mu=0.005$]{
    \includegraphics[scale=0.21]{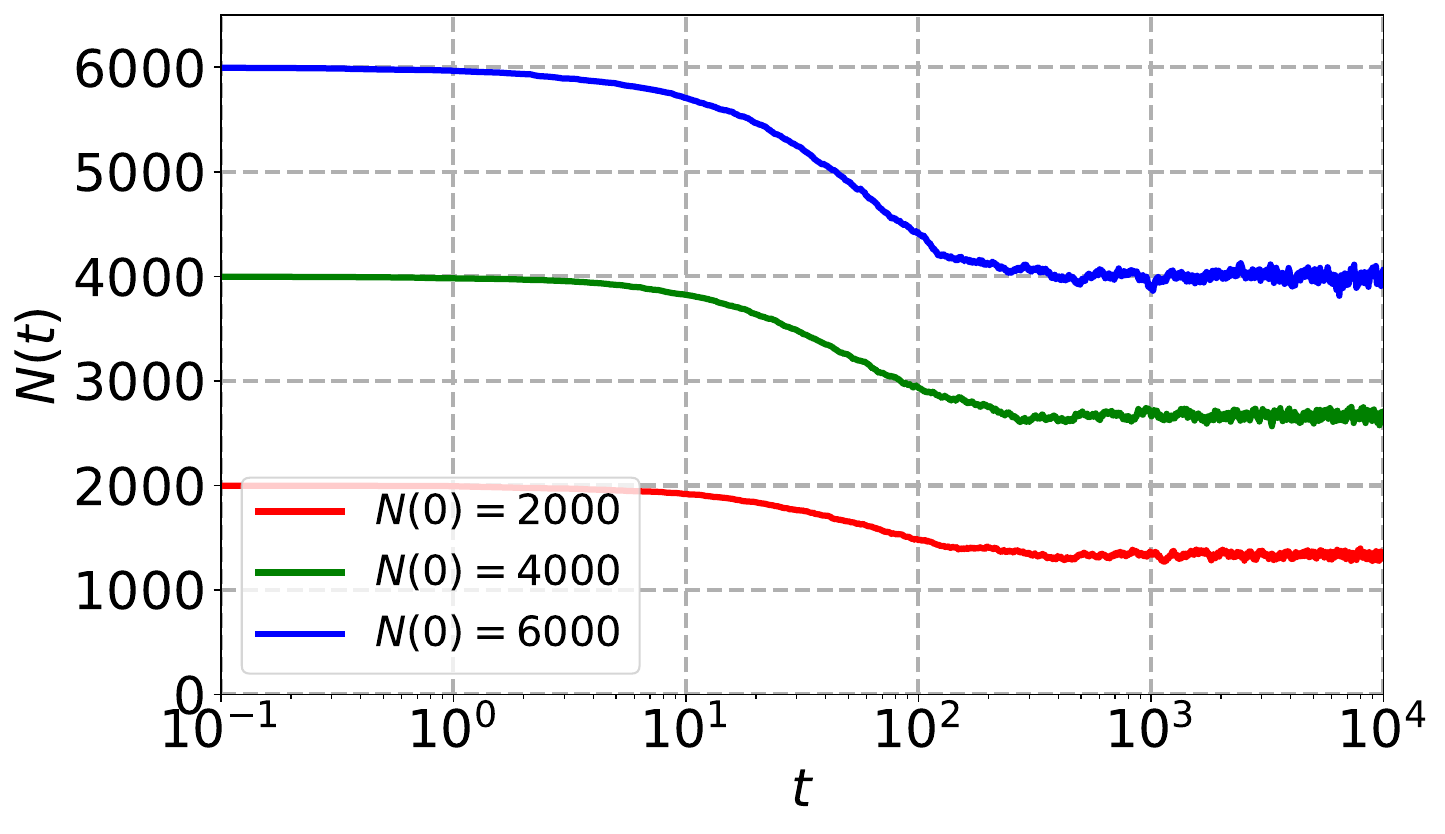}
    \label{fig:2/1}
  }
    \hspace{-5mm}
  \subfigure[$\lambda=0.010, \mu=0.010$]{
    \includegraphics[scale=0.21]{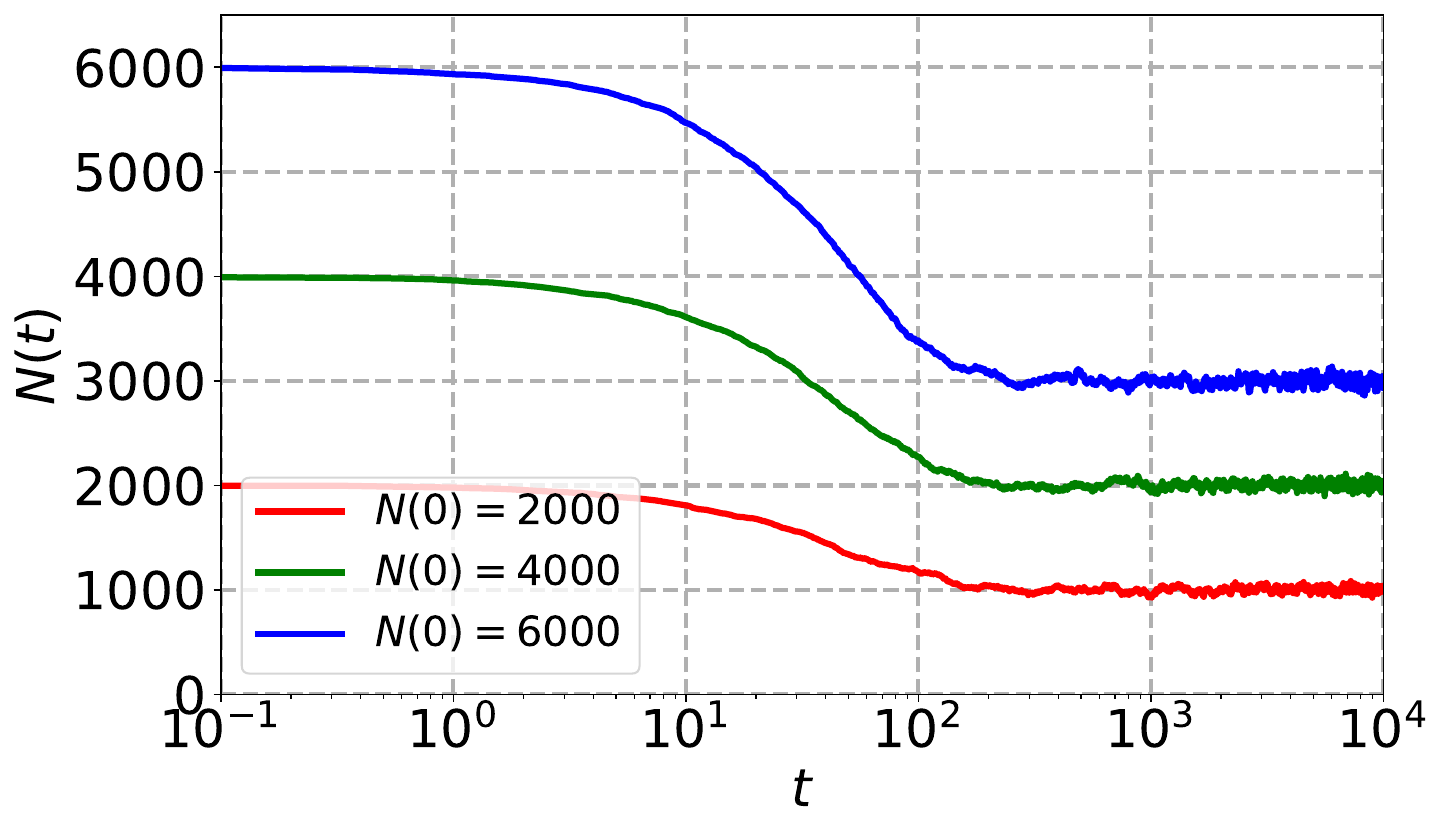}
    \label{fig:2/2}
  }
    \hspace{-5mm}
  \subfigure[$\lambda=0.010, \mu=0.015$]{
    \includegraphics[scale=0.21]{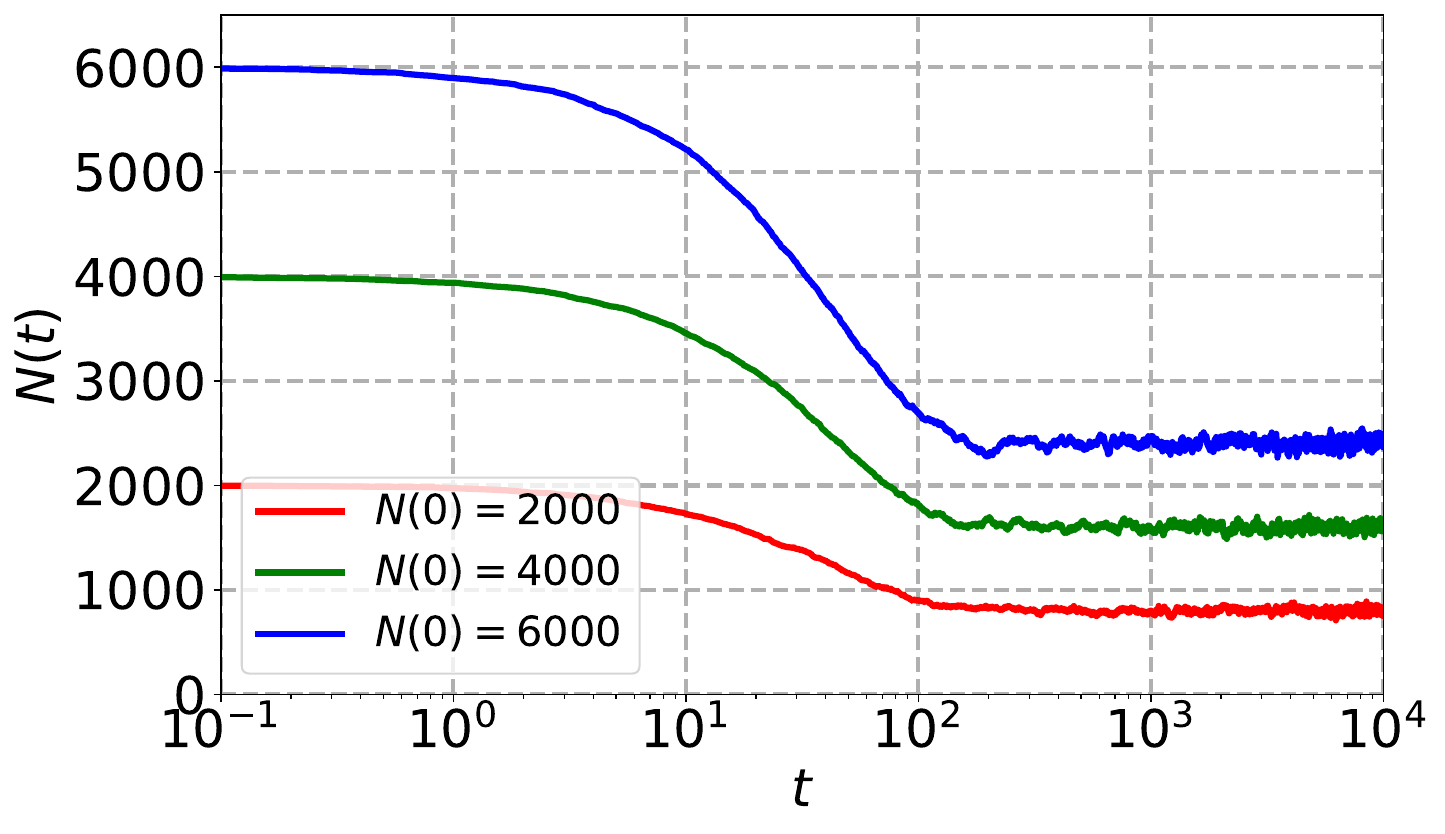}
    \label{fig:2/3}
  }
    \hspace{-5mm}
  \vspace{-3.3mm}

  \subfigure[$\lambda=0.020, \mu=0.005$]{
    \includegraphics[scale=0.21]{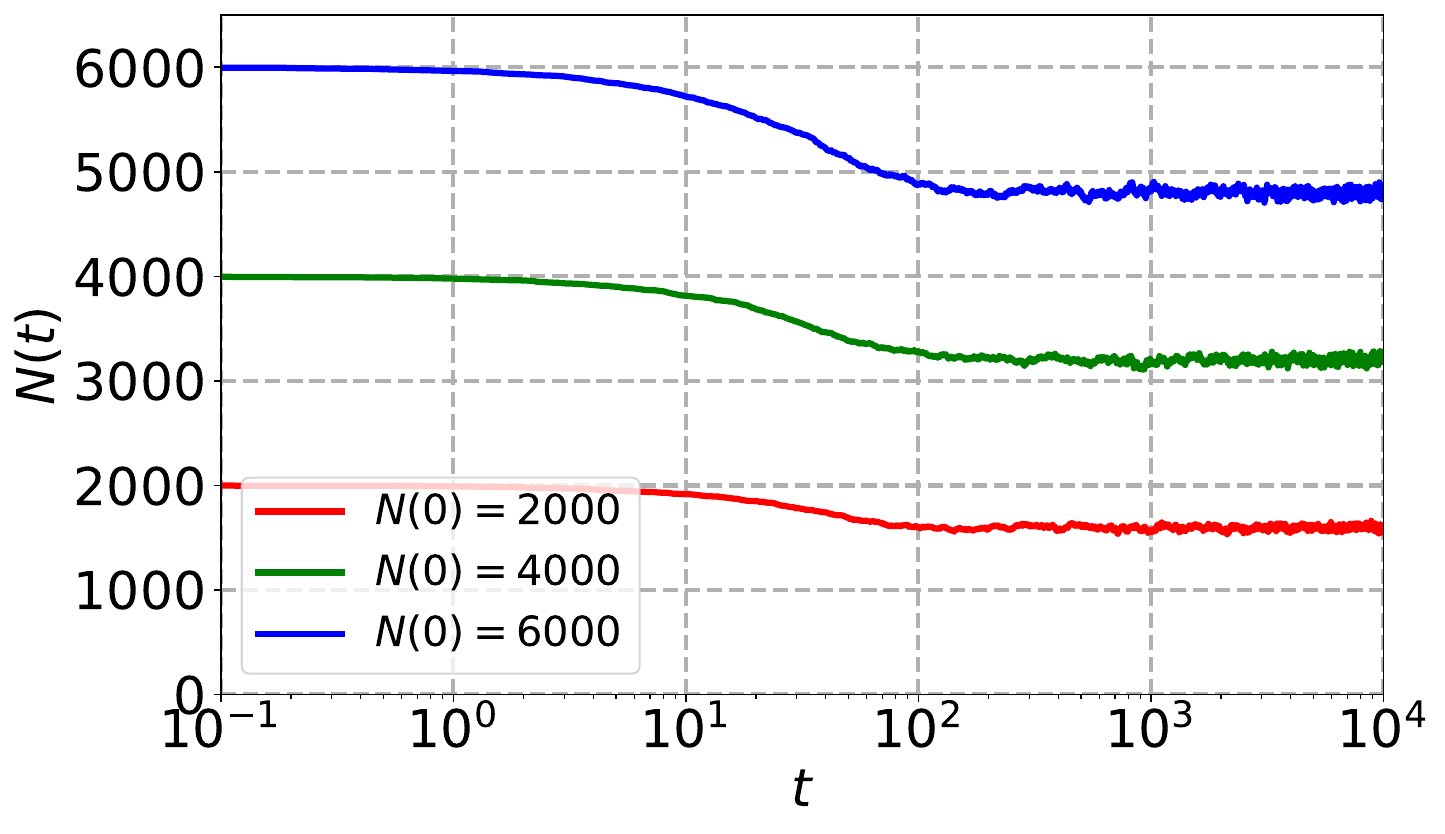}
    \label{fig:2/4}
  }
    \hspace{-5mm}
  \subfigure[$\lambda=0.020, \mu=0.010$]{
    \includegraphics[scale=0.21]{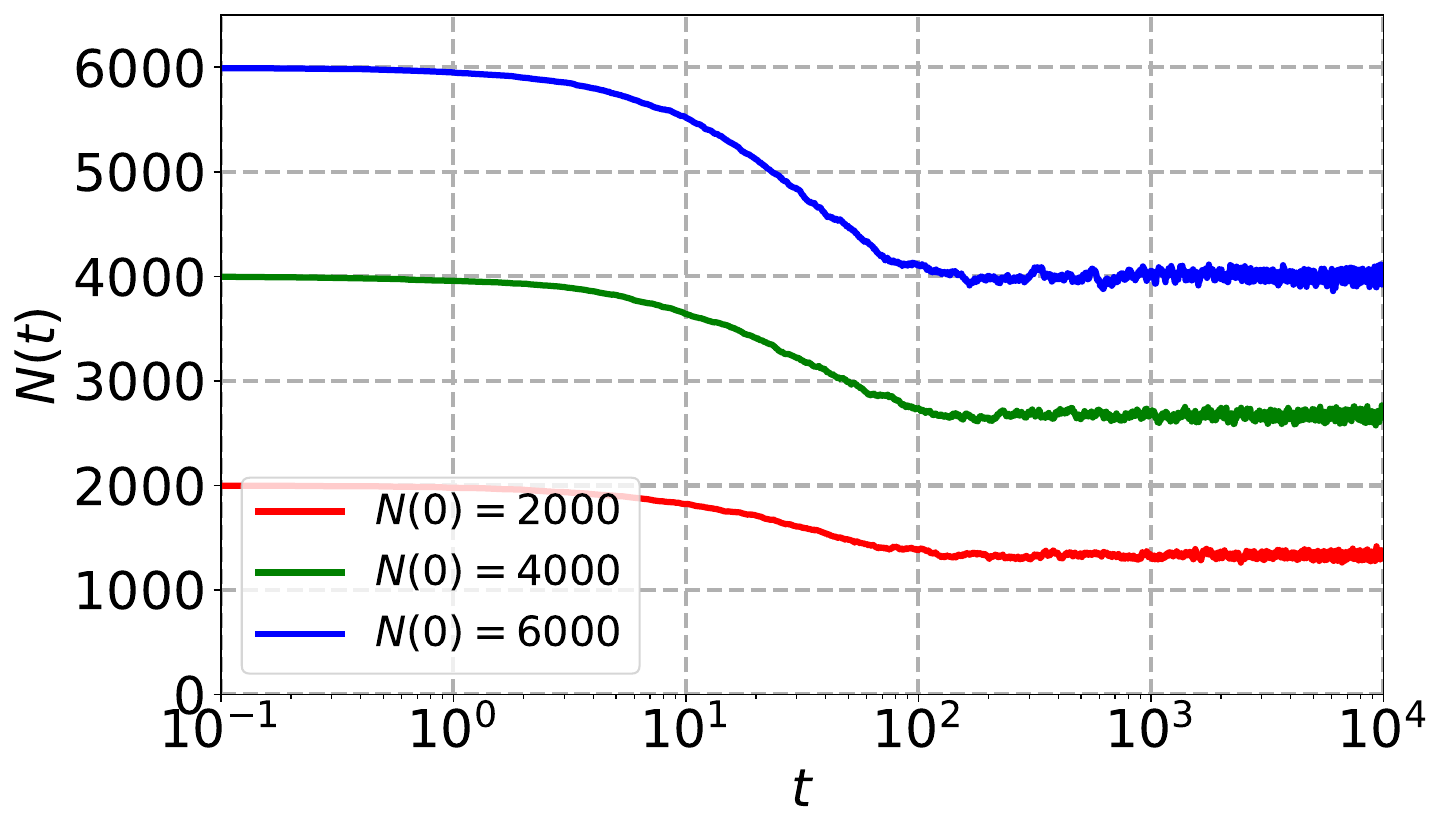}
    \label{fig:2/5}
  }
    \hspace{-5mm}
  \subfigure[$\lambda=0.020, \mu=0.015$]{
    \includegraphics[scale=0.21]{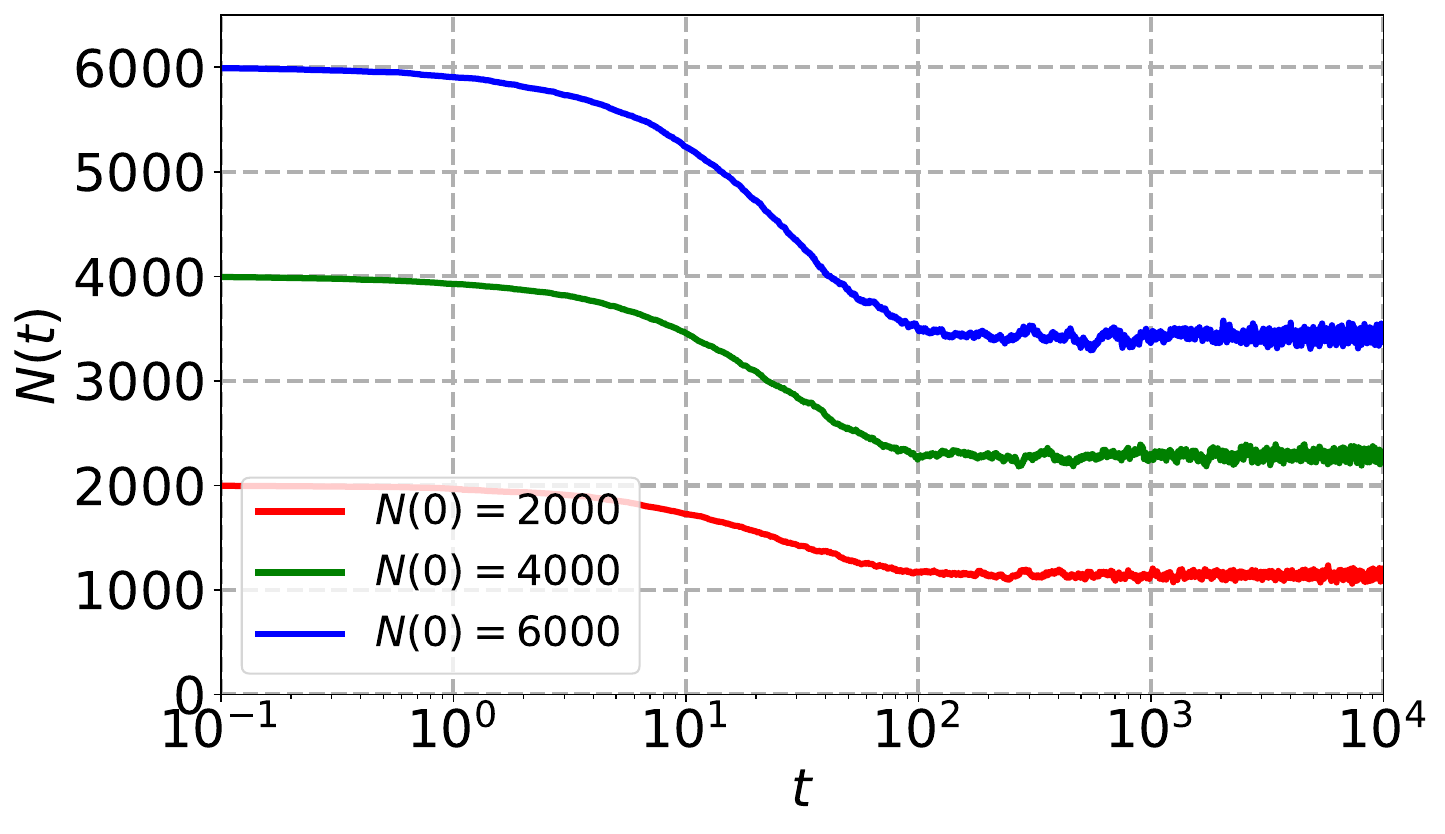}
    \label{fig:2/6}
  }
    \hspace{-5mm}
  \vspace{-3.3mm}
\caption{\textbf{Network sizes as functions of time. }The online network sizes are stable as time evolves and estimable by $N(0)$, $\lambda$, and $\mu$. The volatility is relatively acceptable. We show the evolution process of network sizes under different $\lambda$s, $\mu$s and $N(0)$s. We set $N(0)=2\times10^3$ (red plots), $4\times10^3$ (green plots), $6\times10^3$ (blue plots) and $\lambda$s and $\mu$s as \textbf{\ref{fig:2/1}} $\lambda=0.010$, $\mu=0.005$, \textbf{\ref{fig:2/2}} $\lambda=0.010$, $\mu=0.010$, \textbf{\ref{fig:2/3}} $\lambda=0.010$, $\mu=0.015$, \textbf{\ref{fig:2/4}} $\lambda=0.020$, $\mu=0.005$, \textbf{\ref{fig:2/5}} $\lambda=0.020$, $\mu=0.010$, \textbf{\ref{fig:2/6}} $\lambda=0.020$, $\mu=0.015$. We set the $y$-axis as the network size $N$ and its range as $[0, 6.5\times10^3]$. We observe the evolution process of each network's size in the time interval $[0, 10^4]$. We emphasize again that the network size is not affected by the initial network type. Therefore, we do not set the initial network type as a variable. (color online)}
\label{fig:2}
\vspace{-3mm}
\end{figure*}
\begin{figure*}[thbp]
\centering
\vspace{-1mm}
  \subfigure[$\lambda=0.005$]{
    \includegraphics[scale=0.21]{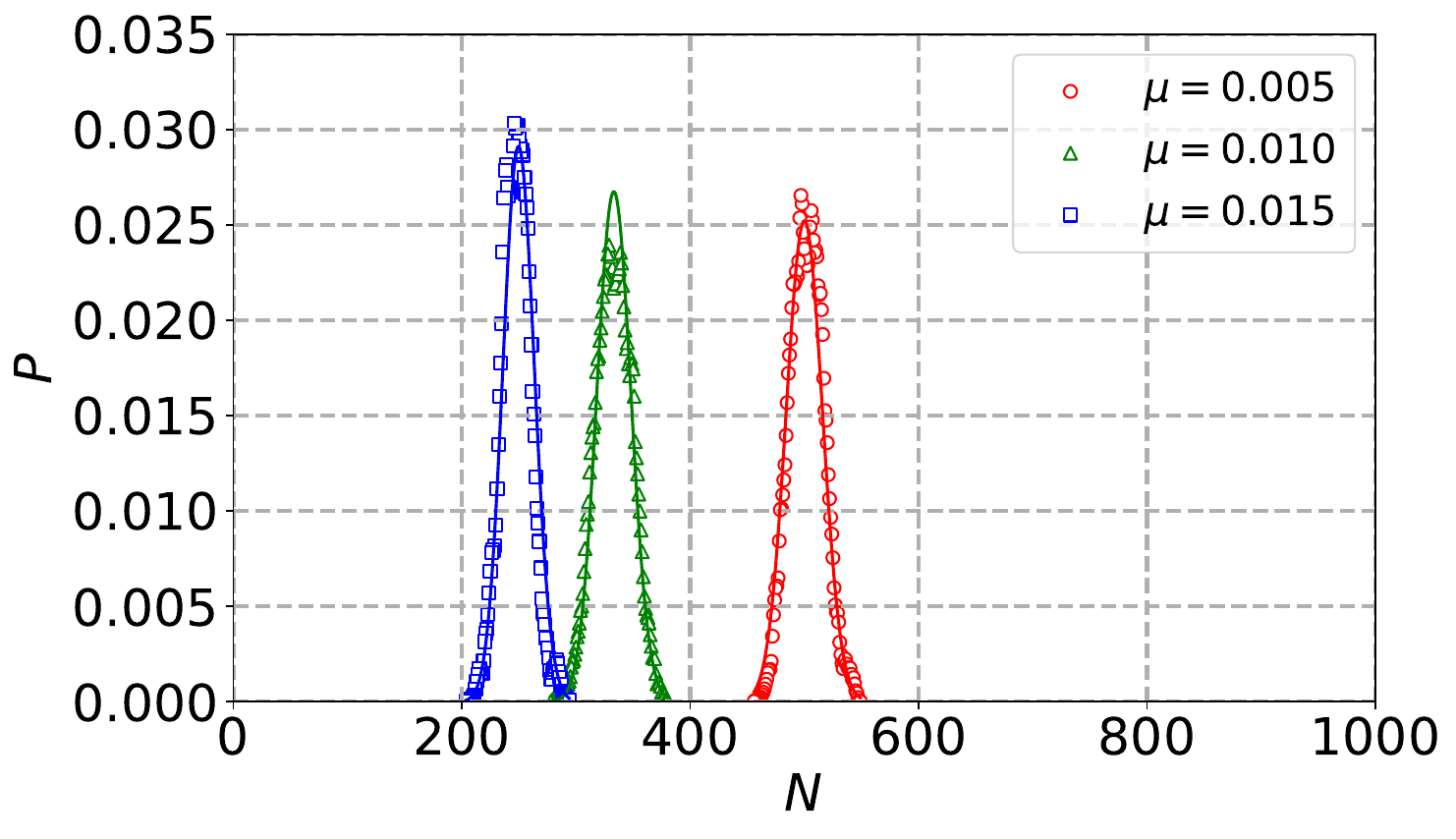}
    \label{fig:3/1}
  }
    \hspace{-5mm}
  \subfigure[$\lambda=0.010$]{
    \includegraphics[scale=0.21]{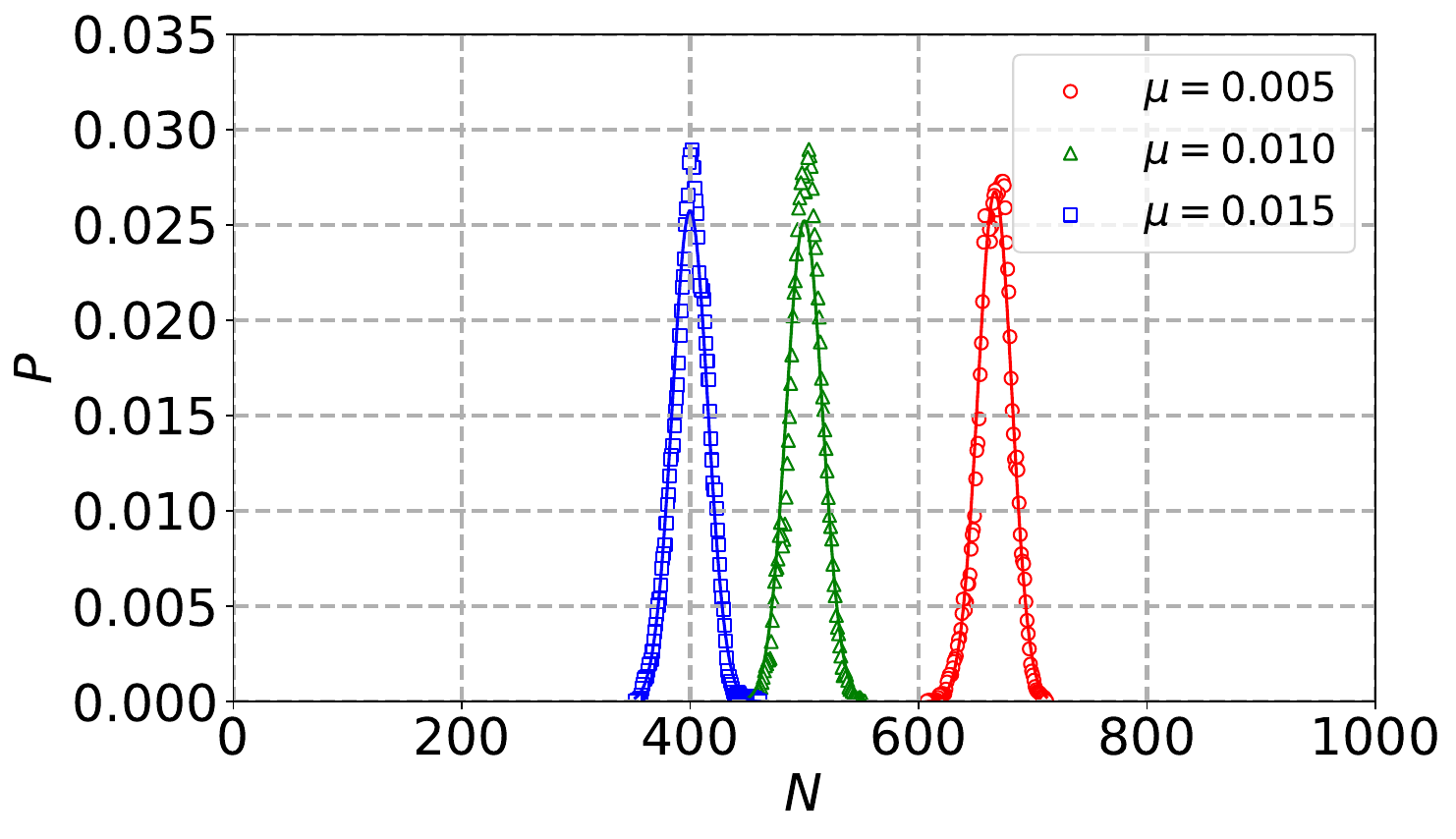}
    \label{fig:3/2}
  }
    \hspace{-5mm}
  \subfigure[$\lambda=0.015$]{
    \includegraphics[scale=0.21]{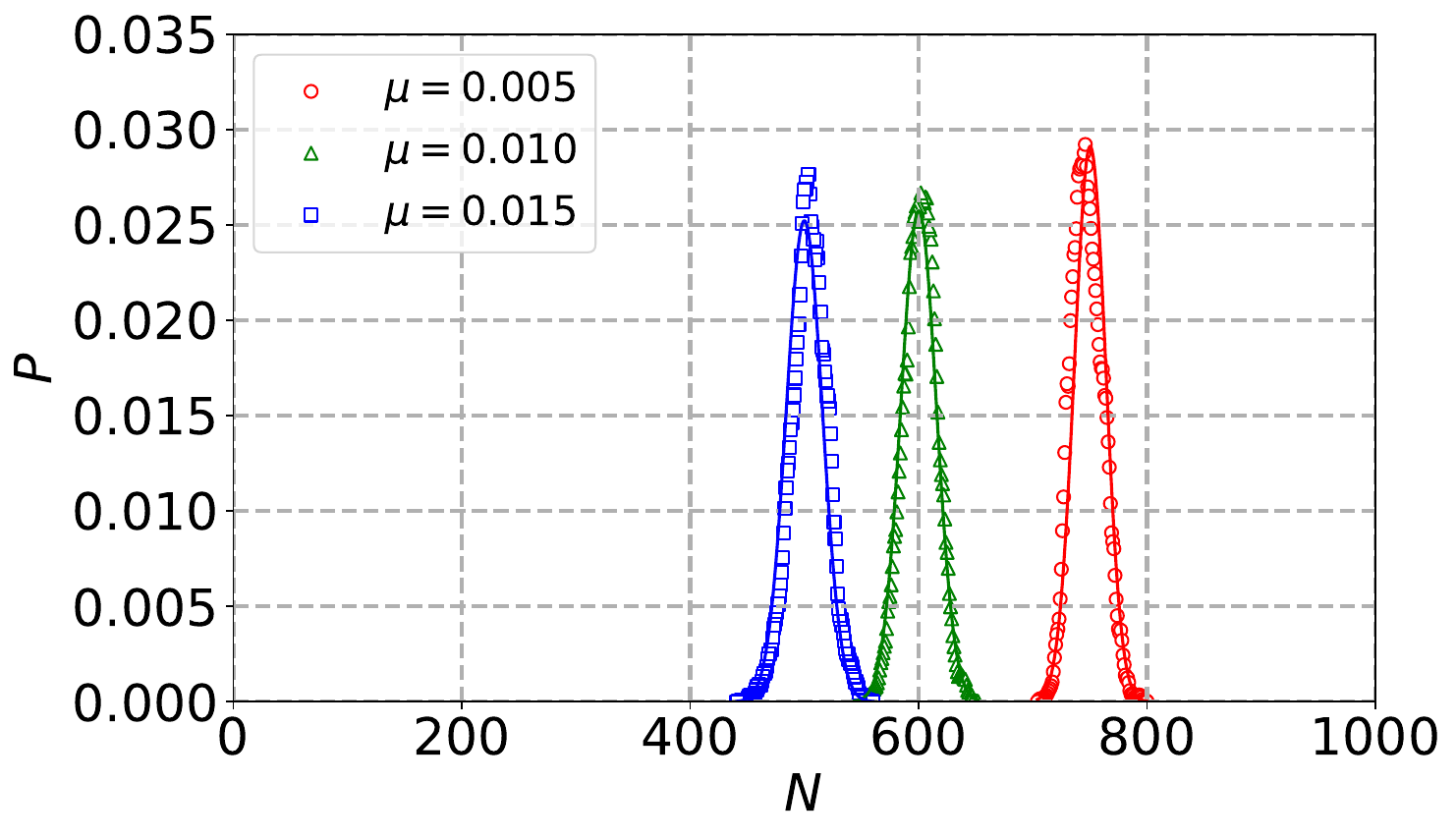}
    \label{fig:3/3}
  }
    \hspace{-5mm}
  \vspace{-3.3mm}
\caption{\textbf{Statistical distribution of network sizes. }The probability distribution of online network sizes is homogeneous and can be fitted by the proposed Theorem \ref{theorem:2}. We present the frequency that each network's size is under different $\lambda$s and $\mu$s. $\lambda$s, $\mu$s and initial sizes are set as \textbf{\ref{fig:3/1}} $\lambda=0.005$, $\mu$=$0.005$ (rounded scatters in red), $0.010$ (triangular scatters in green), $0.015$ (square scatters in blue), $N(0)=10^3$. \textbf{\ref{fig:3/2}} $\lambda=0.010$, $\mu$s and $N(0)$ are the same as above. \textbf{\ref{fig:3/3}} $\lambda=0.015$, $\mu$s and $N(0)$ are the same as above. To show our results more clearly, we set $x$-axis as the network size $N$ and its range as $[0, 10^3]$, $y$-axis as the probability $P$ and its range as $[0, 0.035]$. To guarantee the stationary state is reached and the sample size is large enough, we observe each network's size in the time interval $[10^4, 2\times10^4]$. Besides, the network size is not affected by the initial network type. Therefore, we do not set the initial network type as a variable. (color online)}
\label{fig:3}
\vspace{-5mm}
\end{figure*}
Next, we focus on the online network size under different parameters, which is interpreted as the stochastic process $N(t)$. As mentioned in Sec. \ref{sec:Complex Networks With Online and Hidden Vertices}, once a vertex becomes hidden, it is removed from the network temporarily. Besides, once it go online again, it reconnects to its original neighbors. We set the initial network size as $N(0)=2\times10^3, 4\times10^3, 6\times10^3$, the online rate as $\lambda=[0.010, 0.020]$ and the hidden rate as $\mu=[0.005, 0.010, 0.015]$. It is apparent that a online network size is not affected by the initial network types. Therefore, we do not consider whether the process is on SFs or SWs. In addition, to better illustrate the ascent stage and the stationary process of the plots, the $x$-axis is set as a log coordinate.

In Fig. \ref{fig:2}, we present the evolution process of the online network size under different $N(0)$, $\lambda$s and $\mu$s. Our results for $N(0)=2\times10^3, 4\times10^3, 6\times10^3$ are shown in blue, green and red plots respectively. It is obvious that each online network size declines from the beginning and becomes stable gradually, and the online network size is larger if $N(0)$ is larger. In addition, network sizes with different parameters are all stable when $t=10^3$, which provides a basis for subsequent numerical simulations. Besides, in Figs. \ref{fig:2/1}-\ref{fig:2/3}, where $\lambda=0.010$, each online network size declines more sharply and becomes stable at a lower value as $\mu$ values from $0.005$ to $0.015$, which corresponds to our theoretical results in Theorem \ref{theorem:3}. The same is true in Figs. \ref{fig:2/4}-\ref{fig:2/6}, where $\lambda=0.020$. In Fig. \ref{fig:2/1}, where $\lambda=0.010$ and $\mu=0.005$, online network sizes for $N(0)=2\times10^3, 4\times10^3, 6\times10^3$ reduce to about $1333,2666,4000$. Fixing $\lambda=0.010$, in Figs. \ref{fig:2/2} and \ref{fig:2/3}, where $\mu=0.010$ and $0.015$, the corresponding online network sizes reduce to approximately $1000,2000,3000$ and $800,1600,2400$ respectively. Besides, in Fig. \ref{fig:2/4}, where $\lambda=0.020$ and $\mu=0.005$, the online network sizes reduce from $6000,4000,2000$ to around $4800,3200,1600$. Remaining $\lambda=0.020$ unchanged, in Figs. \ref{fig:2/5} and \ref{fig:2/6}, where $\mu=0.010$ and $0.015$, the corresponding online network sizes reduce to about $4000,2666,1333$ and $3428,2285,1142$. Considering the same hidden rate $\mu$, setting $\lambda=0.020$ brings more online vertices to each network than for $\lambda=0.010$. Moreover, considering the same $\lambda$ and $\mu$, the change of network size is larger if the initial size is larger, where $N(0)=6\times10^3$ (blue plots) reduces the most, and $N(0)=4\times10^3, 2\times10^3$ comes second. Besides, the $95\%$ confident intervals of the online network sizes are no more than $7$, which also means that the online network sizes are stable and measurable. We note that the online network sizes seem unchanged at the beginning of the evolution process. The usage of the logarithmic time axis causes that in the first half of each subfigure, a small time interval (from $10^{-1}$ to $10^{1}$) is shown compared to the total $10^4$. Additionally, our employed $\lambda$s and $\mu$s are small, leading to the fact that the expected phase transition time cannot be tracked in $\Delta t=10$, but can be observed in the whole time axis. 
\begin{figure*}[htbp]
\centering
  \subfigure[$\lambda=0.005$]{
    \includegraphics[scale=0.21]{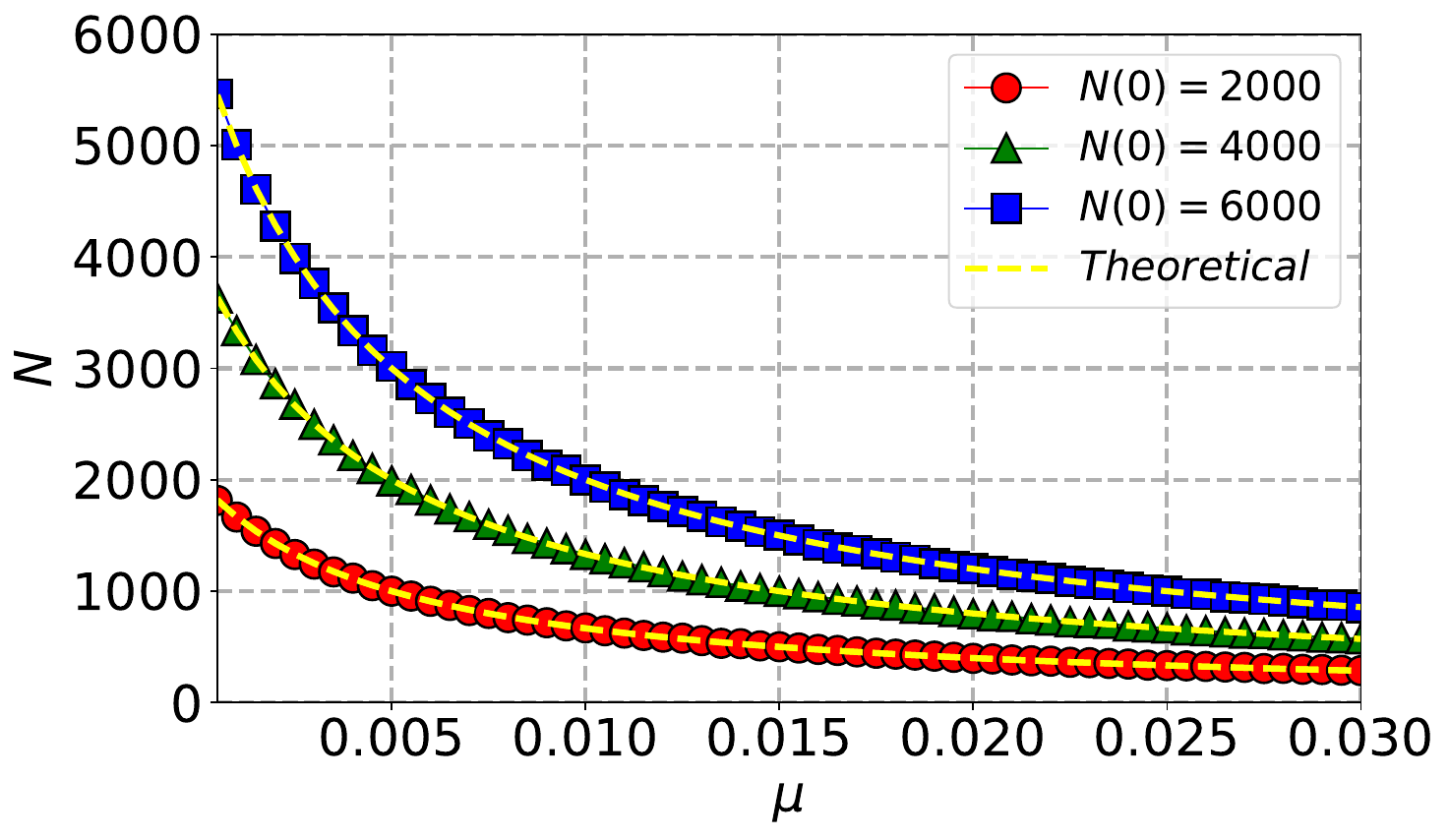}
    \label{fig:4/1}
  }
    \hspace{-5mm}
  \subfigure[$\lambda=0.010$]{
    \includegraphics[scale=0.21]{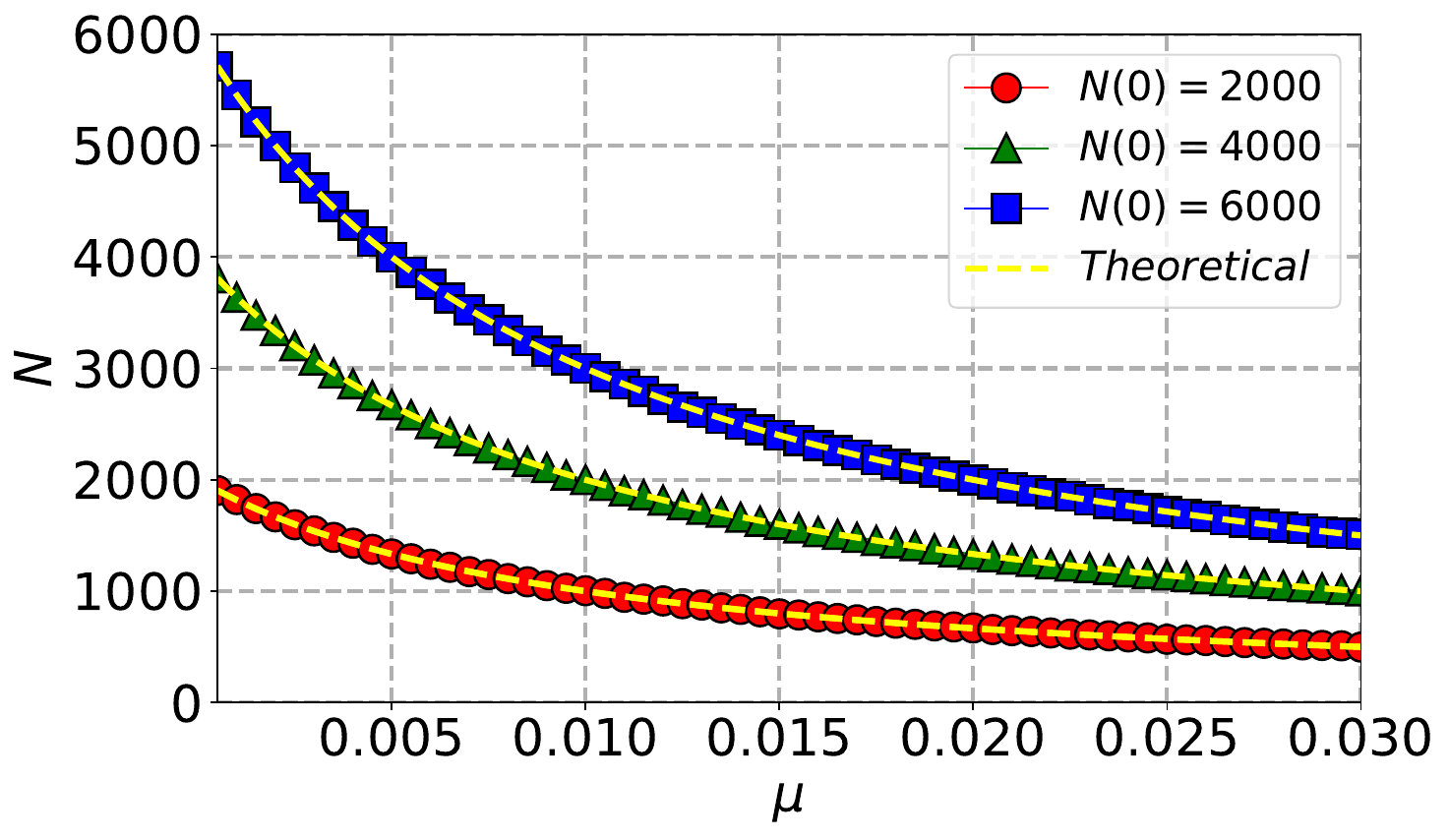}
    \label{fig:4/2}
  }
    \hspace{-5mm}
  \subfigure[$\lambda=0.015$]{
    \includegraphics[scale=0.21]{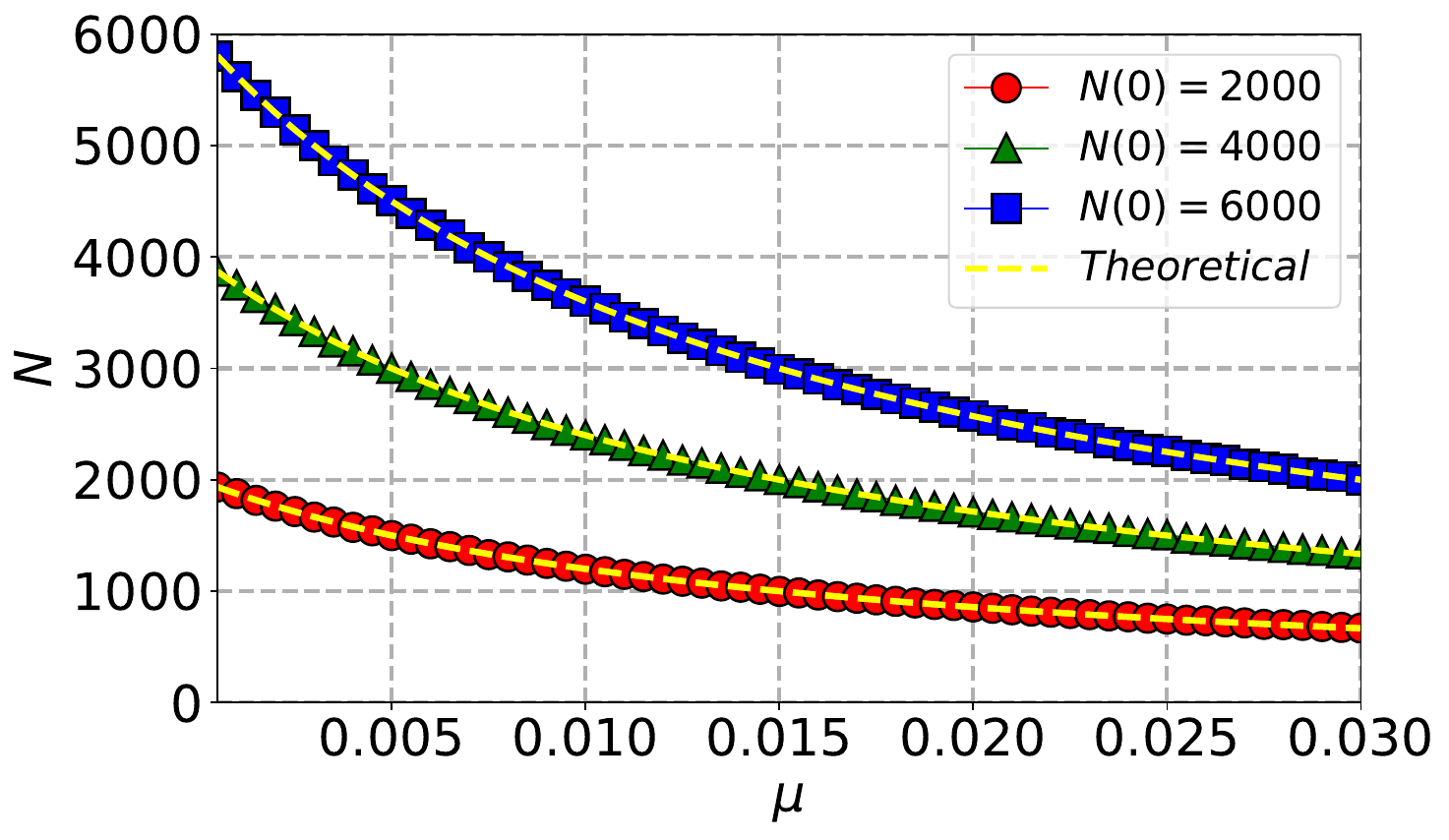}
    \label{fig:4/3}
  }
    \hspace{-5mm}
\vspace{-4mm}
\caption{\textbf{Network sizes as functions of hidden rates. }The expected online network sizes are predictable by Theorem \ref{theorem:3}. We report network sizes for different $\lambda$s, $\mu$s and $N(0)$s and theoretical results (dotted plots in purple). In each subfigure, we set $N(0)=2\times10^3$ (rounded scatters in red), $4\times10^3$ (triangular scatters in green), $6\times10^3$ (square scatters in blue), $\mu$s in the range $[0.05, 3.00]$ with the interval $0.05$. We set \textbf{\ref{fig:4/1}} $\lambda=0.50$, \textbf{\ref{fig:4/2}} $\lambda=1.00$, \textbf{\ref{fig:4/3}} $\lambda=1.50$. We let the $x$-axis be $\mu$ and its range be $[0, 3.00]$, the $y$-axis be the network size $N$ and its range be $[0, 6\times10^3]$. Each network size that we obtain is averaged by the network size in the time interval $[2\times10^4, 10^6]$ to guarantee the stationarity of the stochastic process. (color online)}
\label{fig:4}
\end{figure*}

From the simulations above, we show that each online network size becomes stable around a specific value if the initial size $N(0)$, the online rate $\lambda$ and the hidden rate $\mu$ are given, which corresponds to our theoretical analysis in Theorems \ref{theorem:2} and \ref{theorem:3}. Additionally, in Theorems \ref{theorem:2}, we reckon that during a long-range time, the online network size follows a probability distribution as in Eq. \ref{eq:stationary distribution Nt}. Therefore, in the following simulations, we focus on the frequency of the online network size values in a long-range time. For the reason that the online network size is not affected by the initial network type, we do not set the initial network type as SF or SW for comparison.

The results shown in Fig. \ref{fig:3} indicate that the NOH model guarantees that the online network size is constantly fluctuating in a range around the most probable network size, where we set $\lambda=0.005, 0.010, 0.015$, $\mu=0.005, 0.010, 0.015$ and $N(0)=10^3$. Figs. \ref{fig:3/1}, \ref{fig:3/2} and \ref{fig:3/3} present the online network size distribution provided $\lambda=0.005, 0.010, 0.015$ respectively. For this purpose, it is supposed to ensure that our results are carried out when the stochastic process is stable, and the samples are large enough according to the law of large numbers. As shown in Fig. \ref{fig:2} that each stochastic process of the online network size is stable at $t=10^4$, we count the frequency of the online network sizes during the time interval $[10^4, 2\times10^4]$. It is obvious that there is a peak value for each frequency distribution, which indicates the most probable online network size. In Figs. \ref{fig:3/1}-\ref{fig:3/3} with $\lambda=0.005, 0.010, 0.015$, for $\mu=(0.005, 0.010, 0.015)$, the most probable online network sizes are approximately $(500, 333, 250)$, $(666, 500, 400)$ and $(750, 600, 500)$ respectively, which correspond to the expected network size in Theorem \ref{theorem:3}. In this way, the online network size that emerges with the highest frequency is significantly larger if the hidden rate $\mu$ is smaller. Besides, the frequency distribution shown in Fig. \ref{fig:3} is homogeneous on both sides of the peak. Moreover, when the stable state is reached, in Figs. \ref{fig:3/1}-\ref{fig:3/3} with $\lambda=0.005, 0.010, 0.015$, for $\mu=(0.005, 0.010, 0.015)$, minimum online network sizes in shown results are around $(210, 390, 460)$, $(360, 450, 720)$ and $(450, 540, 690)$ respectively. Additionally, corresponding maximum network sizes are about $(290, 380, 560)$, $(430, 540, 700)$ and $(550, 640, 800)$ respectively. 

To verify our simulations based on the theoretical analysis in Eq. \ref{eq:stationary distribution Nt}, we apply the Kullback-Leibler divergence (KL) to compare the closeness of the theoretical and the experimental distribution because this indicator presents an intuitive result for the distance of two probability distributions. The KL in discrete form is denoted as
\begin{equation}\label{eq:KL}
KL(\pi \vert\vert Q)=\sum \pi_n\log\frac{\pi_n}{Q_n},
\end{equation}
where $\pi_n$ and $Q_n$ indicate the theoretical (Eq. \ref{eq:stationary distribution Nt}) and the experimental distribution (results in Fig. \ref{fig:3}) respectively. We compare the similarity of these two probability distributions by calculating Eq. \ref{eq:KL}. If the KL divergence of two probability distributions is small, their distance is small as well. It is worth noting that the theoretical results for the probability distribution may exceed the upper limit of the computer. Therefore, we first change Eq. \ref{eq:stationary distribution Nt} into logarithmic form to avoid the overflow as follow
\begin{equation}
    \ln{\pi}=\ln{C_{N(0)}^n}+n\ln{(\lambda\mu^{-1})}+N(0)\ln{(1+\lambda\mu^{-1})}. 
\end{equation}
After calculating the above result, we get the probability distribution $\pi_n=e^{\ln{\pi_n}}$. Besides, $Q_n$ denotes the frequency that the size $n$ emerges in the stochastic system. In our simulation, we obtain this distribution by investigating the probability that the network stays on a specific size during $[10^4, 2\times10^4]$. In Tab. \ref{tab:KL}, we display the aforementioned KL divergence. It is clear that all KL divergences are no more than $0.030$, which verifies that the theoretical distribution in Eq. \ref{eq:stationary distribution Nt} fits the simulation data well. 
\begin{table}[h]
\centering
\caption{\textbf{The Kullback-Leibler divergence of the theoretical and experimental distribution.}}
\begin{tabular}{c|c|c|c}
\hline
\diagbox{$\lambda$}{$\mu$}&0.005      &0.010        &0.015\\
\hline
0.005&0.027&0.014&0.018\\
\hline
0.010&0.006&0.015&0.011\\
\hline
0.015&0.012&0.030&0.010\\
\hline
\end{tabular}
\label{tab:KL}
\end{table}
\begin{figure*}
\centering
\vspace{-3.3mm}
  \subfigure[SF, $\lambda=0.010, \mu=0.005$]{
    \includegraphics[scale=0.21]{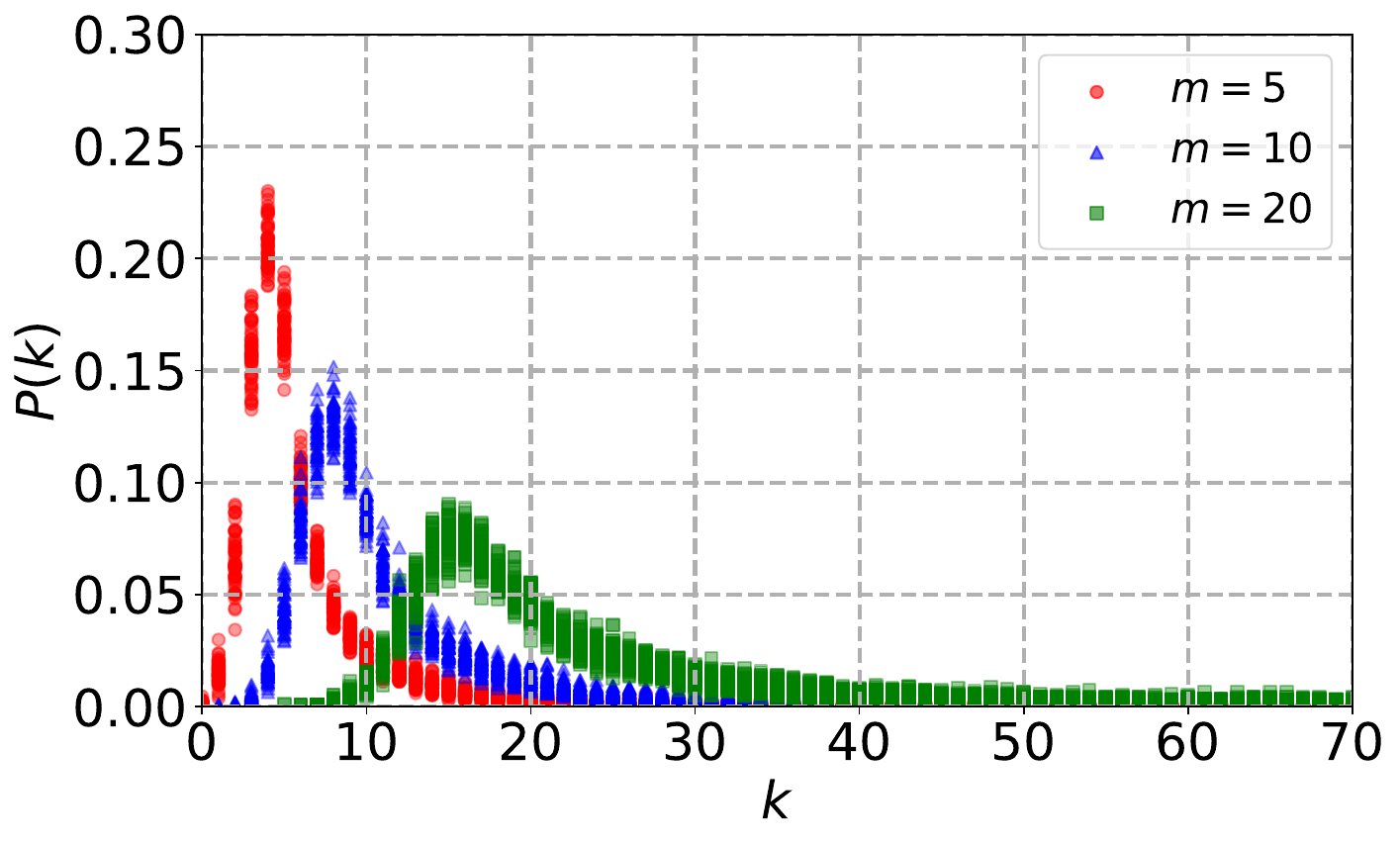}
    \label{fig:5/SF1}
  }
    \hspace{-5mm}
  \subfigure[SF, $\lambda=0.010, \mu=0.010$]{
    \includegraphics[scale=0.21]{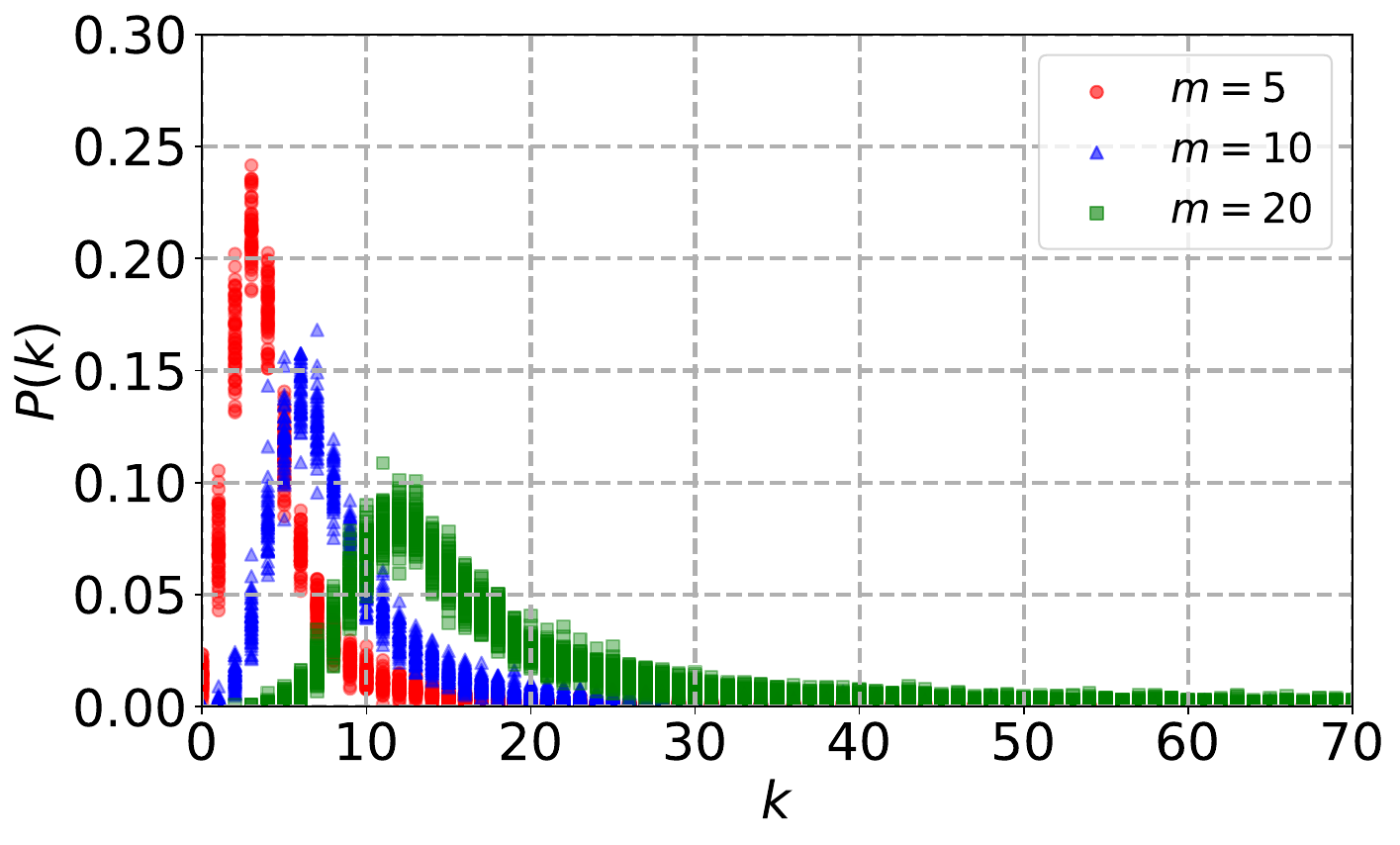}
    \label{fig:5/SF2}
  }
    \hspace{-5mm}
  \subfigure[SF, $\lambda=0.010, \mu=0.015$]{
    \includegraphics[scale=0.21]{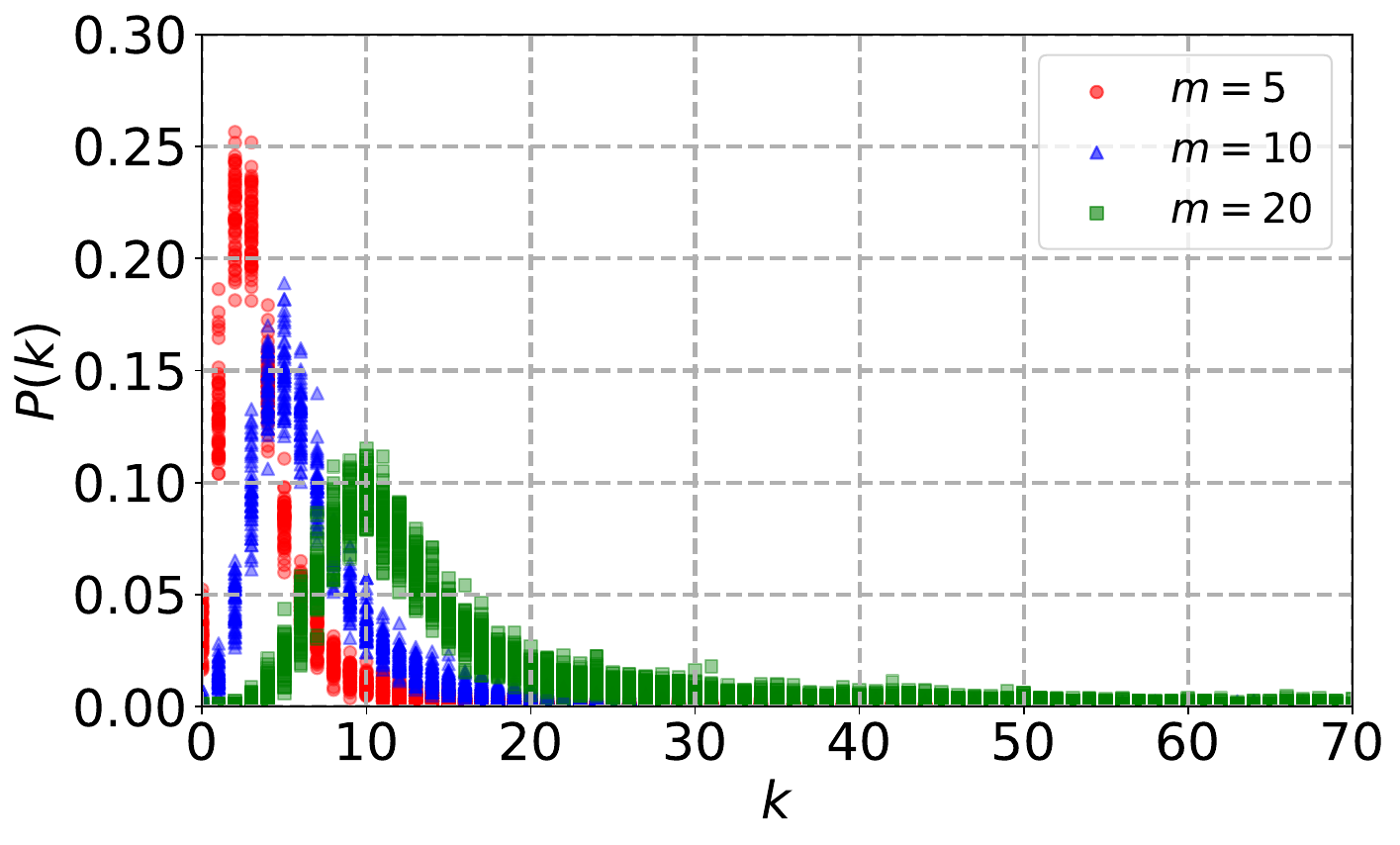}
    \label{fig:5/SF3}
  }
    \hspace{-5mm}
\vspace{-3.3mm}

  \subfigure[SF, $\lambda=0.020, \mu=0.005$]{
    \includegraphics[scale=0.21]{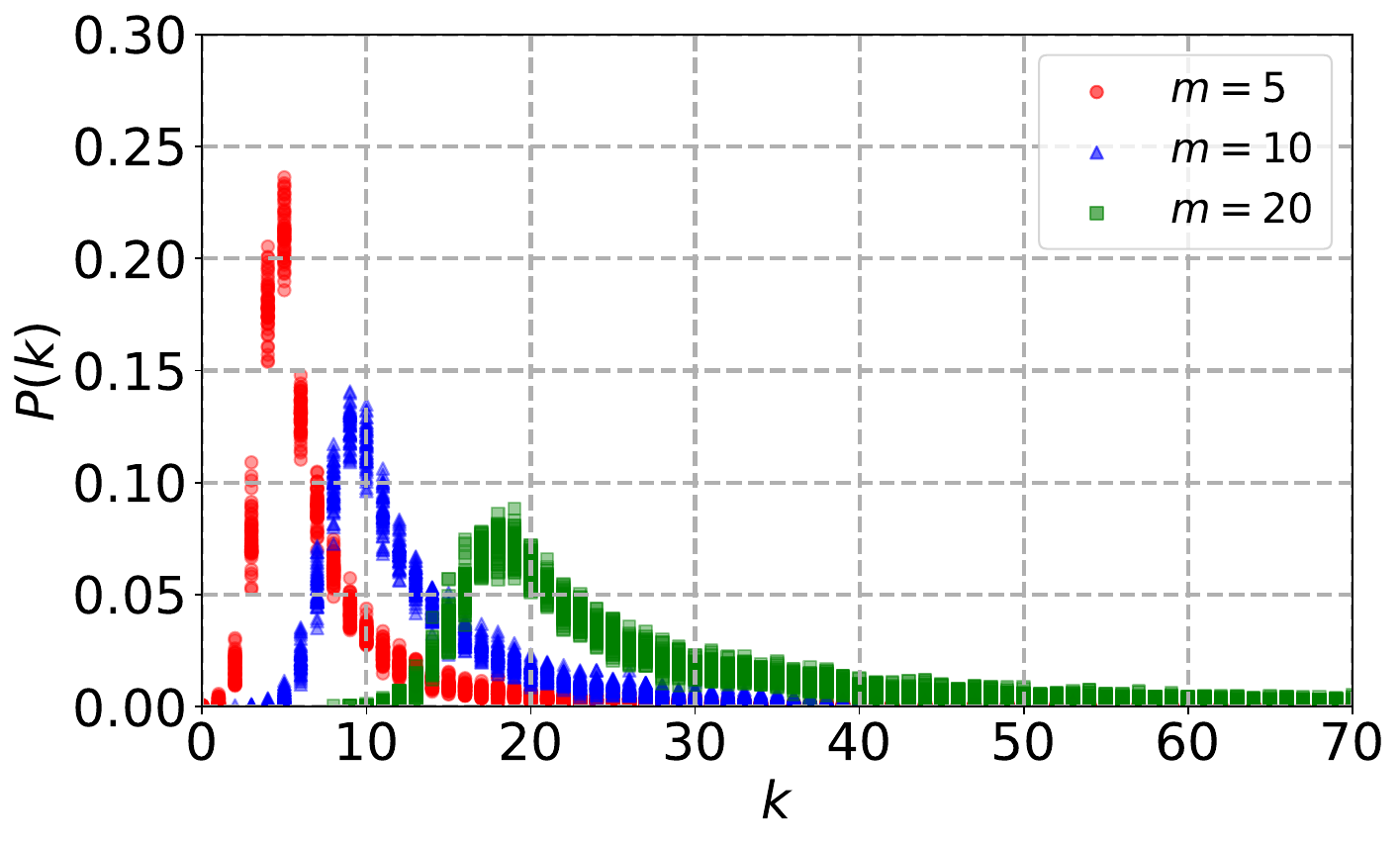}
    \label{fig:5/SF4}
  }
    \hspace{-5mm}
  \subfigure[SF, $\lambda=0.020, \mu=0.010$]{
    \includegraphics[scale=0.21]{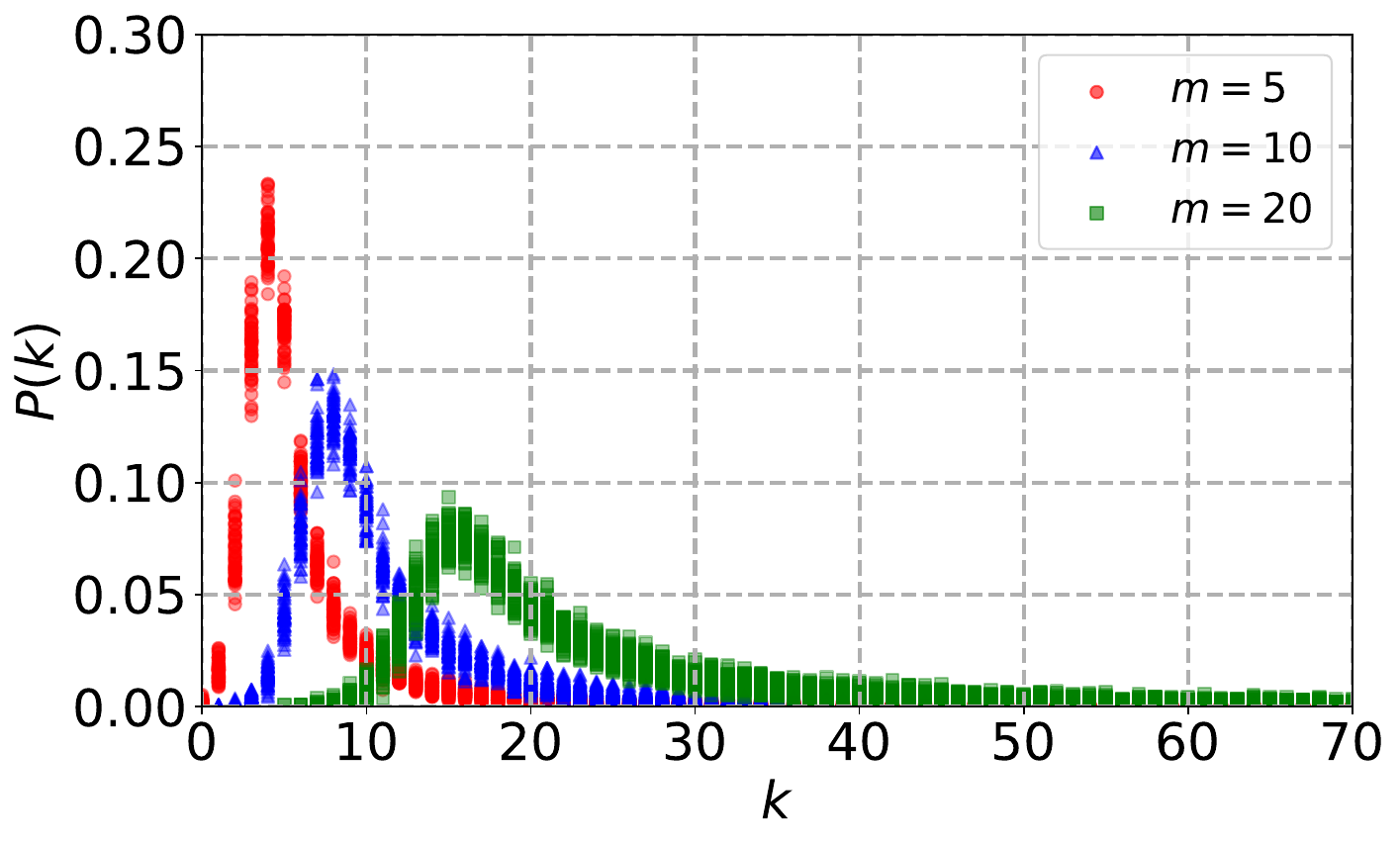}
    \label{fig:5/SF5}
  }
    \hspace{-5mm}
  \subfigure[SF, $\lambda=0.020, \mu=0.015$]{
    \includegraphics[scale=0.21]{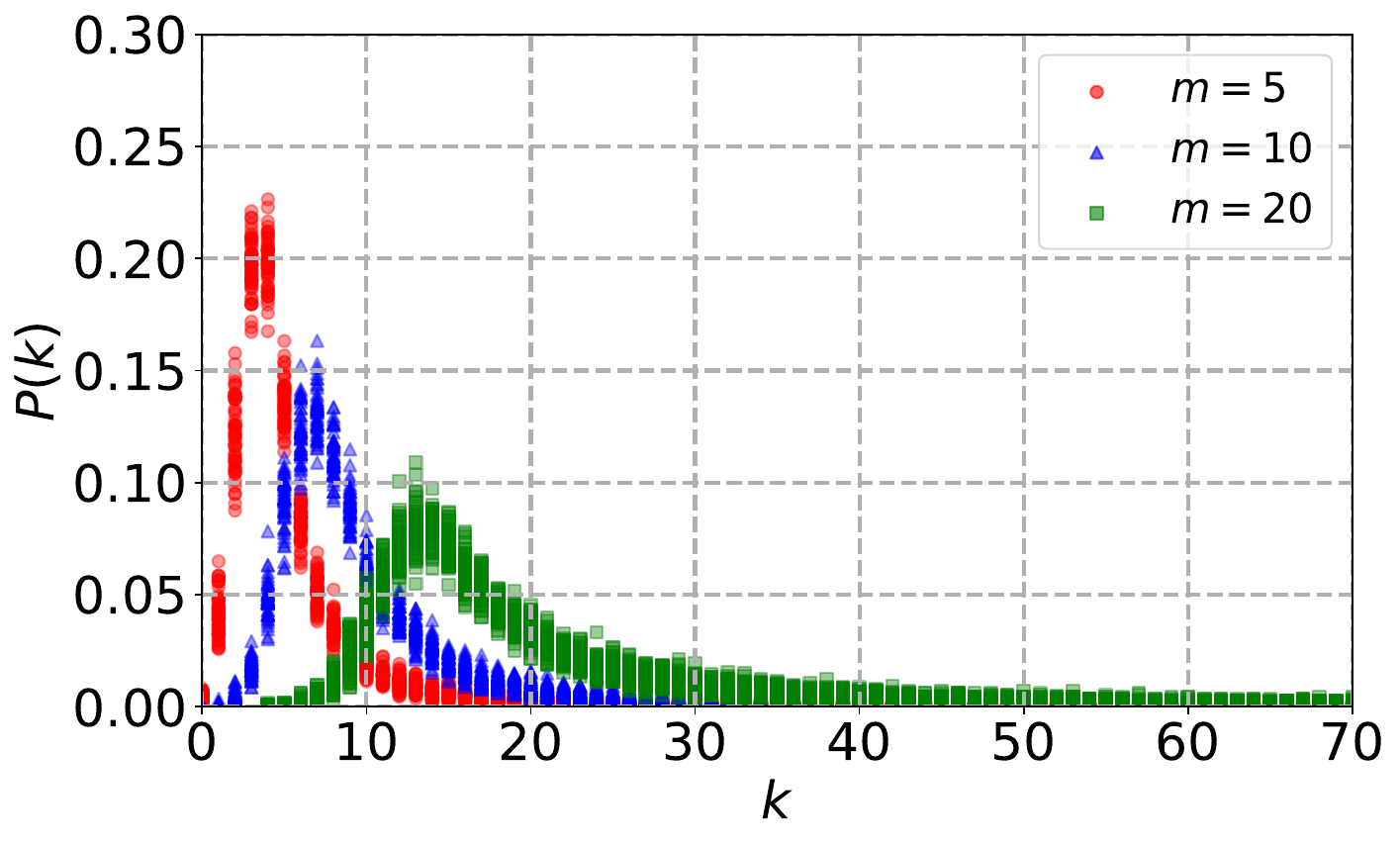}
    \label{fig:5/SF6}
  }
    \hspace{-5mm}
  \vspace{-3.3mm}

  \subfigure[SW, $\lambda=0.010, \mu=0.005$]{
    \includegraphics[scale=0.21]{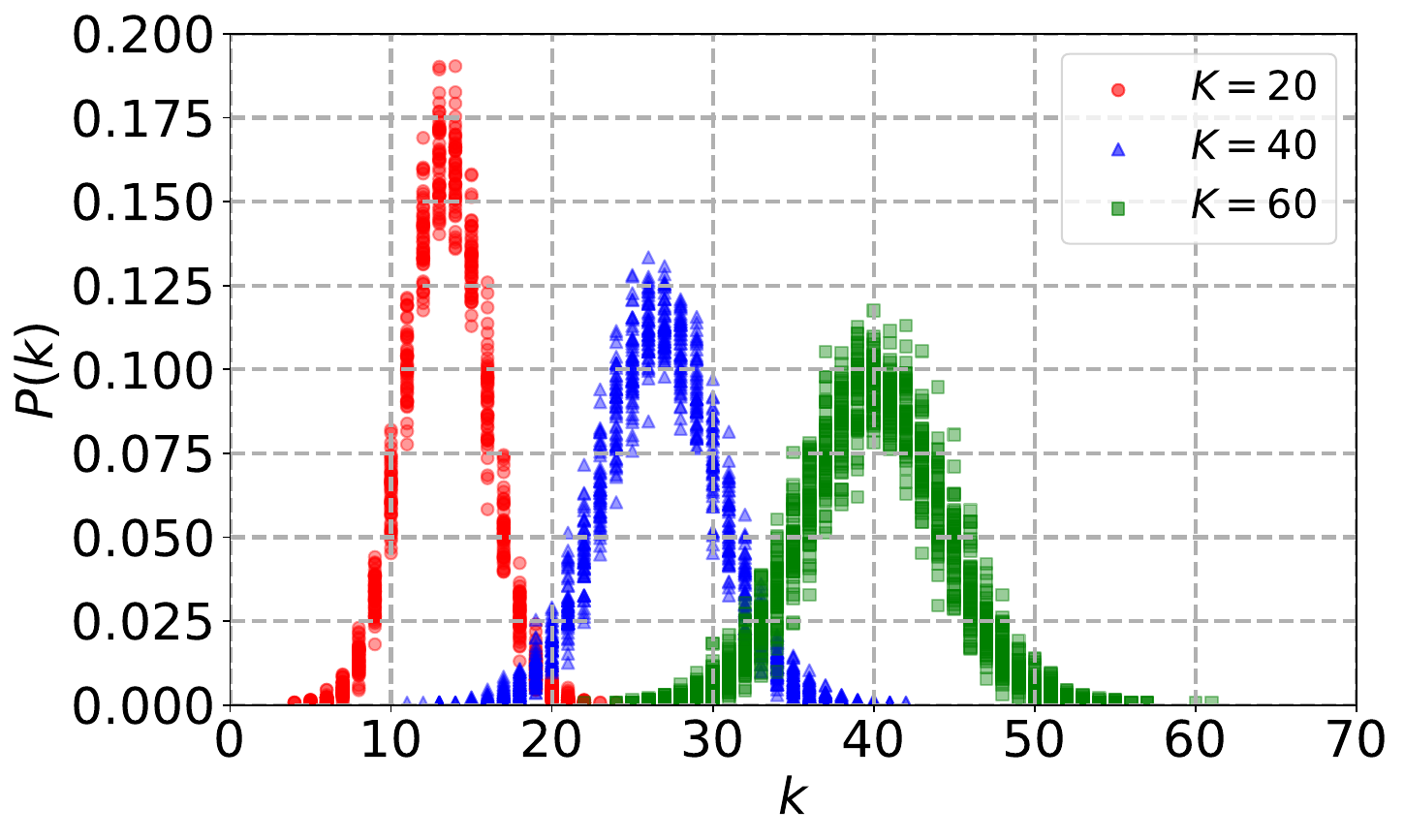}
    \label{fig:5/SW1}
  }
    \hspace{-5mm}
  \subfigure[SW, $\lambda=0.010, \mu=0.010$]{
    \includegraphics[scale=0.21]{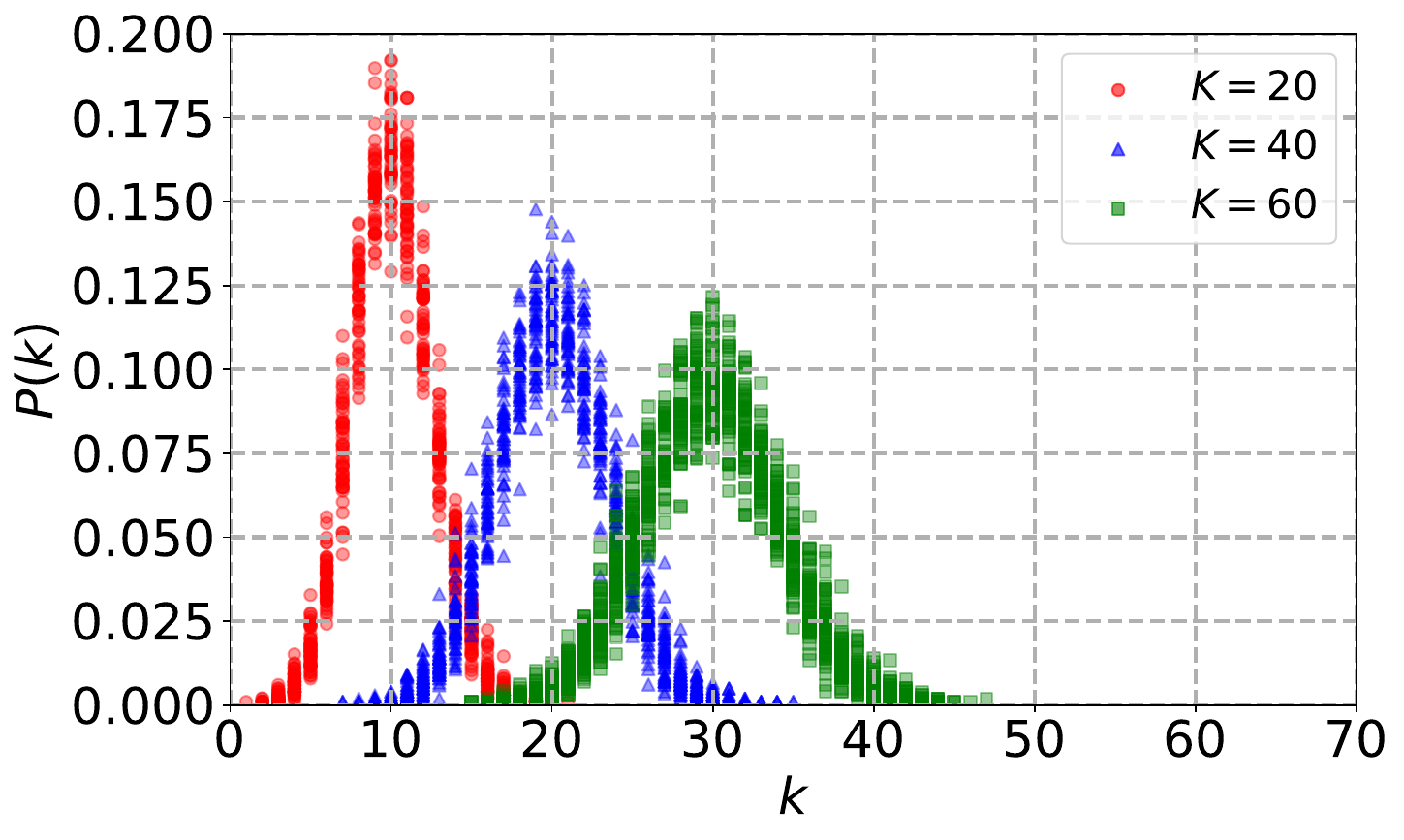}
    \label{fig:5/SW2}
  }
    \hspace{-5mm}
  \subfigure[SW, $\lambda=0.010, \mu=0.015$]{
    \includegraphics[scale=0.21]{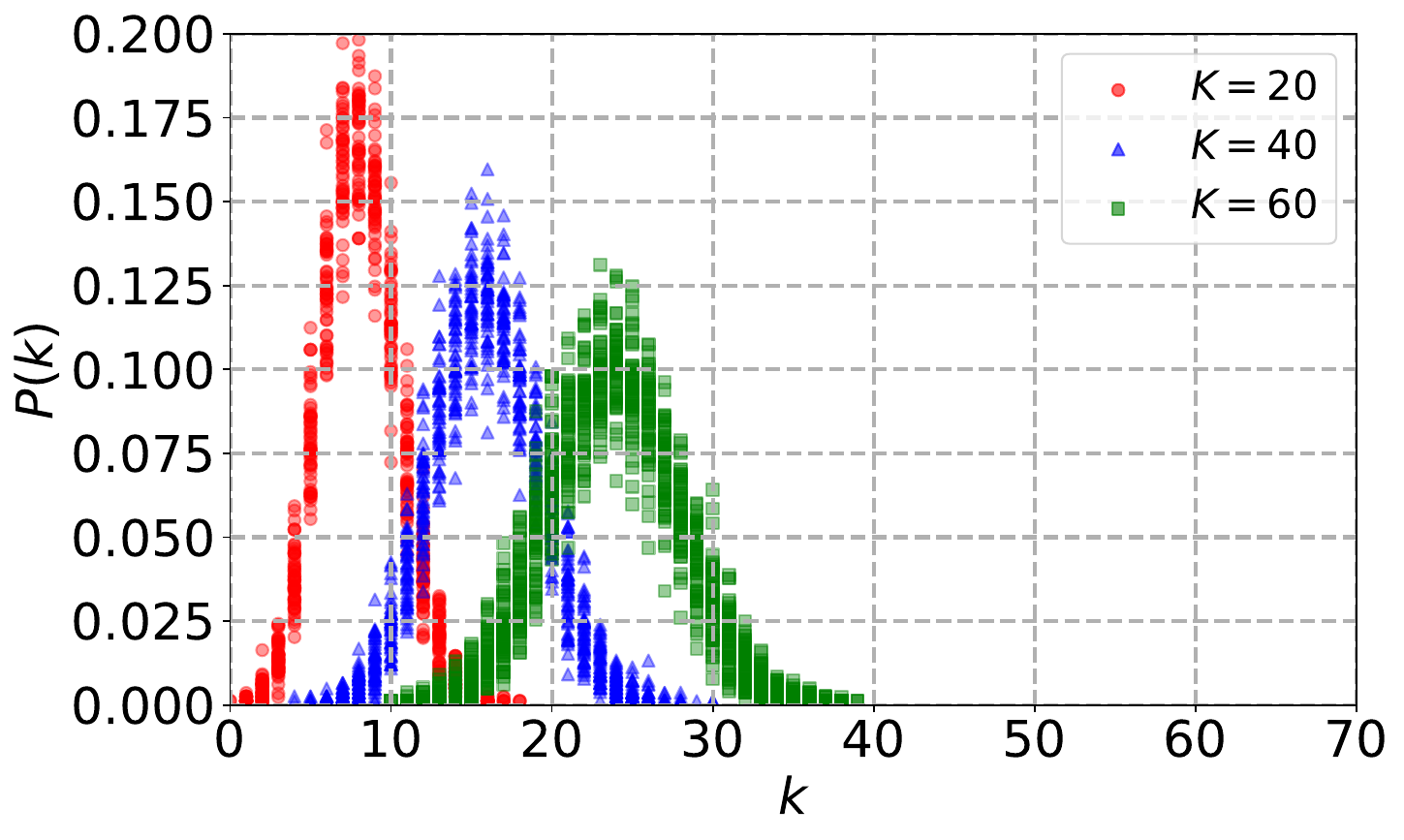}
    \label{fig:5/SW3}
  }
    \hspace{-5mm}
  \vspace{-3.3mm}

  \subfigure[SW, $\lambda=0.020, \mu=0.005$]{
    \includegraphics[scale=0.21]{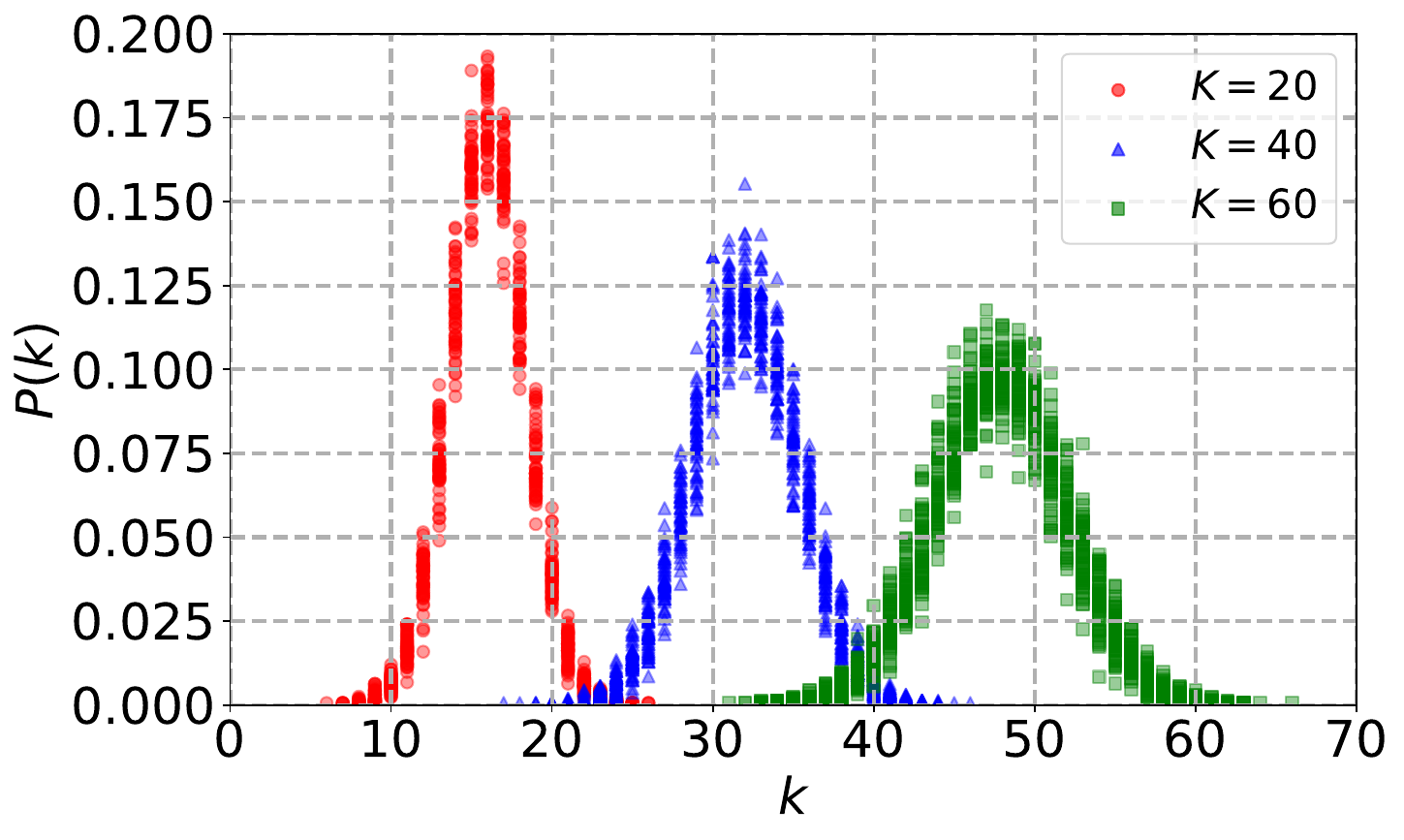}
    \label{fig:5/SW4}
  }
    \hspace{-5mm}
  \subfigure[SW, $\lambda=0.020, \mu=0.010$]{
    \includegraphics[scale=0.21]{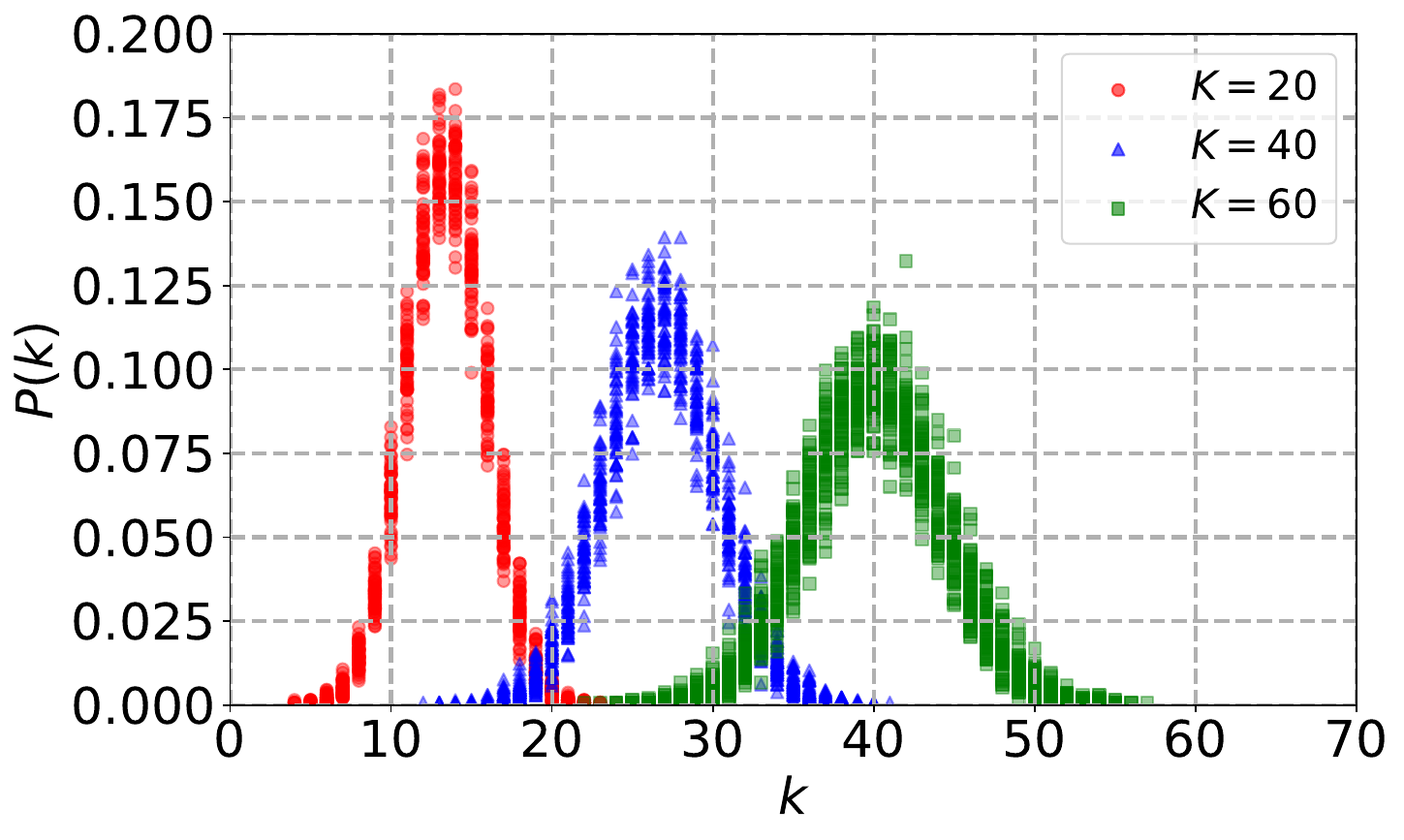}
    \label{fig:5/SW5}
  }
    \hspace{-5mm}
  \subfigure[SW, $\lambda=0.020, \mu=0.015$]{
    \includegraphics[scale=0.21]{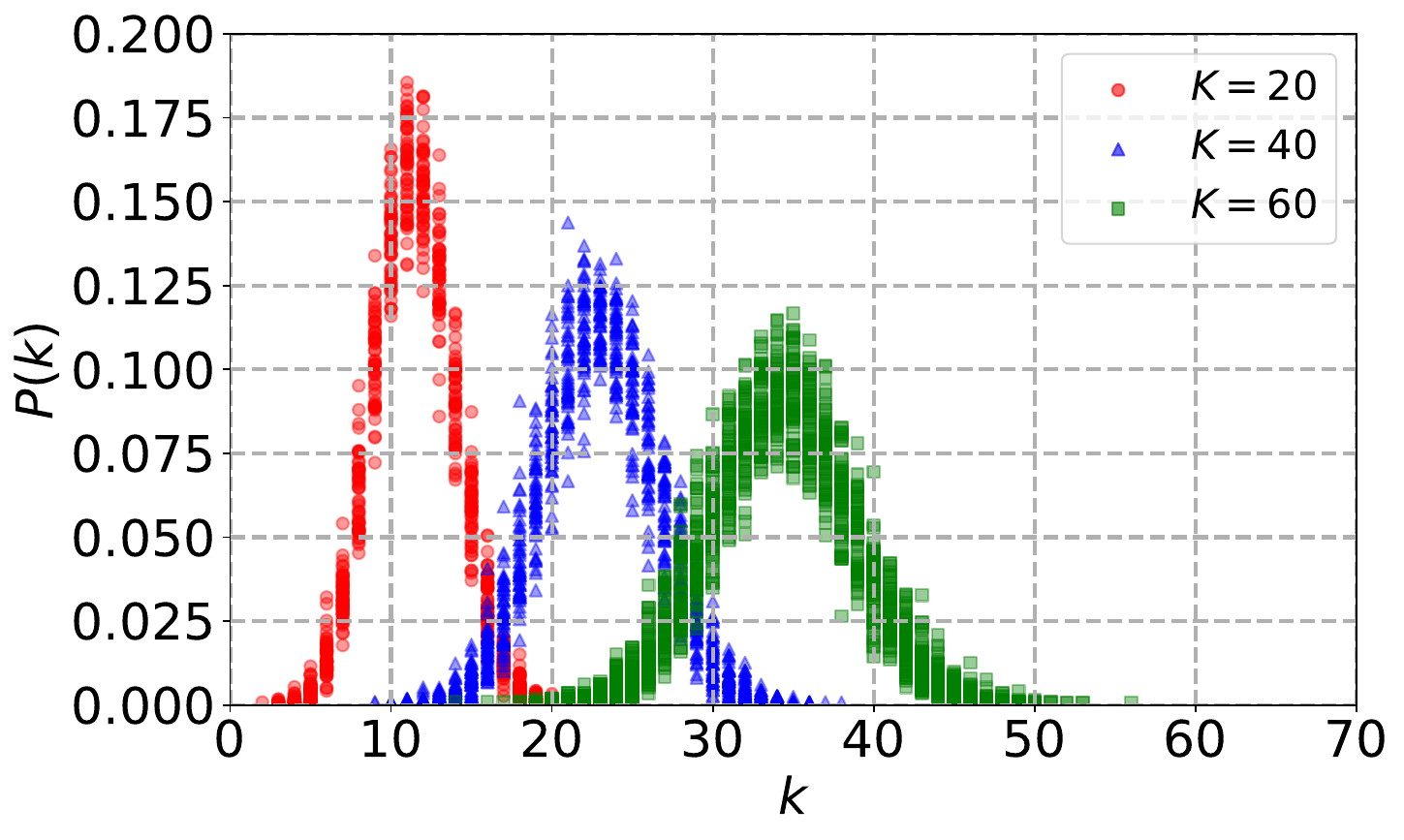}
    \label{fig:5/SW6}
  }
    \hspace{-5mm}
  \vspace{-3.3mm}
\caption{\textbf{Degree distributions of NOHs. }The switching of online and hidden states reduces the heterogeneity of SFs but maintains the homogeneity of SWs. This figure presents the degree distribution of NOHs. We consider the initial network as SFs with $m=5,10,20$ and SWs with $K=20,40,60$, $p=0.200$, where the initial size is set as $N(0)=2\times10^3$. In each subfigure, we show results by observing each network when $t>10^4$ every 200 unit times until $t=2\times10^4$, where the network size is stable when $t>10^4$ as aforementioned. In Fig. \ref{fig:5/SF1}-\ref{fig:5/SF6}, results are shown in red circles for $m=5$, in blue triangles for $m=10$ and in green squares for $m=20$. In Fig. \ref{fig:5/SW1}-\ref{fig:5/SW6}, results are shown in red circles for $K=20$, in blue triangles for $K=40$ and in green squares for $K=60$. Provided the network type is SF, we present the degree distribution in \textbf{\ref{fig:5/SF1}} $\lambda=0.010$, $\mu=0.005$, \textbf{\ref{fig:5/SF2}} $\lambda=0.010$, $\mu=0.010$, \textbf{\ref{fig:5/SF3}} $\lambda=0.010$, $\mu=0.015$, \textbf{\ref{fig:5/SF4}} $\lambda=0.020$, $\mu=0.005$, \textbf{\ref{fig:5/SF5}} $\lambda=0.020$, $\mu=0.010$, \textbf{\ref{fig:5/SF6}} $\lambda=0.020$, $\mu=0.015$. Provided and network type is SW, we present the degree distribution in \textbf{\ref{fig:5/SW1}} $\lambda=0.010$, $\mu=0.005$, \textbf{\ref{fig:5/SW2}} $\lambda=0.010$, $\mu=0.010$, \textbf{\ref{fig:5/SW3}} $\lambda=0.010$, $\mu=0.015$, \textbf{\ref{fig:5/SW4}} $\lambda=0.020$, $\mu=0.005$, \textbf{\ref{fig:5/SW5}} $\lambda=0.020$, $\mu=0.010$, \textbf{\ref{fig:5/SW6}} $\lambda=0.020$, $\mu=0.015$. Besides, we let $x$-axis be the degree with the range $[0,70]$, the $y$-axis be the degree frequency with the range $[0,0.300]$ in Fig. \ref{fig:5/SF1}-\ref{fig:5/SF6} and $[0,0.200]$ in Fig. \ref{fig:5/SW1}-\ref{fig:5/SW6}. (color online)}
\label{fig:5}
\vspace{-3.3mm}
\end{figure*}

Based on the distribution of the network sizes, in Theorem \ref{theorem:3}, we can measure the expected network sizes by Eq. \ref{eq:EN}, which is determined by $N(0)$, $\lambda$ and $\mu$. In the following simulations, we verify the theorem by setting $N(0)=2\times10^3, 4\times10^3, 6\times10^3$, $\lambda=0.005, 0.010, 0.015$ and $\mu\in[5\times10^{-4}, 0.03]$ with the interval $5\times10^{-4}$. The results in Fig. \ref{fig:4} presents the average network sizes as functions of the hidden rates $\mu$s, where the theoretical solutions (Eq. \ref{eq:EN}) are in purple, and $N(0)=2\times10^3, 4\times10^3, 6\times10^3$ are shown in rounded red, triangular blue and square blue scatters respectively. Each data point in our simulation results is obtained by averaging the network sizes in the interval $t\in[10^4, 1.5\times10^4]$. It is evident that our theoretical solutions fit well with our simulation results. For each parameter pair we adopt in Fig. \ref{fig:4}, it is apparent that the average network size reduces as the hidden rate $\mu$ grows. If $\mu$ is large, the reduction rate is relatively low, i.e., the sensitivity of the network sizes to hidden rates is weakened. Furthermore, each blue curve ($N(0)=6\times10^3$) is above each green curve ($N(0)=4\times10^3$), and each green curve is above each red curve ($N(0)=2\times10^3$). Therefore, provided the same pair of $\lambda$ and $\mu$ is given, the online network size is larger if $N(0)$ is larger. In Fig. \ref{fig:4/1}, the average online network sizes for $N(0)=2\times10^3, 4\times10^3, 6\times10^3$ decline to approximately $286$, $571$, $857$, and in Figs. \ref{fig:4/2}, \ref{fig:4/3}, corresponding average online network sizes for each initial size decline to about $500$, $1000$, $1500$ and $666$, $1333$, $2000$ respectively. Additionally, the mean absolute errors between theoretical and numerical results are no more than 5 and based on the initial network sizes. Given a high initial network size, the mean absolute error is high as well. As introduced in Sec. \ref{sec:Complex Networks With Online and Hidden Vertices}, the analysis of the online network size has the same principle as each vertex's online neighbors. Therefore, to simplify the simulation section, we omit the simulation on the number of each vertex's online neighbors.
\subsection{Degree Distributions}
\label{sec:degree distribution}
Next, we focus on the degree distribution of the NOH model, which describes the probability or frequency of a vertex's degree values. Our results are carried out on the networks initialized as SFs and SWs. 

By setting $\lambda=0.010,0.020$, $\mu=0.005,0.010,0.015$, we observe degree distributions on SFs with $m=5,10,20$ and SWs with $K=20,40,60$, $p=0.20$ as Fig. \ref{fig:5} presents. We record the degree distributions when $10^4<t<2\times10^4$ every 200 unit times and plot them in the same panels to show that the distributions are relatively stable.

In Fig. \ref{fig:5/SF1}-\ref{fig:5/SF6}, we present degree distributions on SFs with $m=5$ in red circles, $m=10$ in blue triangles, and $m=20$ in green squares. It is shown that the NOHs do not show power-law characteristics provided the initial networks are SFs. Besides, remaining the online rate $\lambda$ and the hidden rate $\mu$ unchanged, a lower $m$ brings a higher peak value for the degree distribution and a lower degree value with the highest probability. In addition, with a larger $m$, degree distributions are shown to be wider, which indicates that the deviations are higher. Remaining $m$ and $\lambda$ unchanged, if a higher hidden rate $\mu$ is given, the degree distribution deviates more to the left and becomes narrower. On the contrary, fixing $m$ and $\mu$, if a higher online rate $\lambda=0.020$ is given, the degree distribution deviates more to the right than for $\lambda=0.010$. 

It is worth noting that the SFs follow the power-law degree distribution. However, with our proposed online and hidden mechanism, the degree distributions of SFs become skewed. That is to say, the complex network systems with the preferential attachment mechanism may not show the power-law degree distribution if there are both online and hidden vertices, but present a skewed degree distribution with a peak. Additionally, the corresponding skewnesses are presented in Tab. \ref{tab: degree distribution SF}. Each skewness value is obtained by averaging the skewnesses during the same time as how we obtained the simulations of degree distributions. The degree distributions are all positively skewed for the parameters we presume. With a high $m$, the skewness value is low, resulting in a degree distribution that is close to a normal distribution. 

In Figs. \ref{fig:5/SW1}-\ref{fig:5/SW6}, setting $p=0.20$, we present degree distributions on SWs with $K=20$ in red circles, $K=40$ in blue triangles and $K=60$ in green squares. It is apparent that NOHs initialized by SWs show homogeneous degree distributions. Remaining the online rate $\lambda$ and the hidden rate $\mu$ unchanged, the degree value with the highest probability is lower and the degree distribution deviates more to the left if $K$ is lower. Remaining $\lambda$ and $K$ unchanged, the degree distribution deviates more to the left and becomes more narrow provided a higher $\mu$ is given. Besides, fixing $K$ and $\mu$, the degree distribution deviates more to the right with higher $\lambda=0.020$. Additionally, in Tab. \ref{tab: degree distribution SW}, we present the skewnesses of degree distributions of NOHs initialized by SWs. Comparing Tabs. \ref{tab: degree distribution SF} and \ref{tab: degree distribution SW}, we find that the skewnesses on SWs with online and hidden vertices are much smaller than SFs, most of which are positive for the parameters we presume. Therefore, the homogeneity of NOHs initialized by SWs is almost not influenced.
\begin{table}[h]
\centering
\caption{\textbf{The skewness of degree distributions of NOHs on SFs. }}\label{tab: degree distribution SF}
\begin{tabular}{c|ccc|ccc}
\toprule[2pt]
\midrule
$\lambda$s & \multicolumn{3}{c|}{$0.01$}                                                                        & \multicolumn{3}{c}{$0.02$}                                                                        \\ \midrule
$\mu$s     & \multicolumn{1}{c|}{$0.005$} & \multicolumn{1}{c|}{$0.01$} & $0.015$ & \multicolumn{1}{c|}{$0.005$} & \multicolumn{1}{c|}{$0.01$} & $0.015$ \\ \midrule
$m=5$                      & \multicolumn{1}{c}{5.99}                     & \multicolumn{1}{c}{5.18}                    & 5.32                     & \multicolumn{1}{c}{6.32}                     & \multicolumn{1}{c}{6.10}                    & 6.77                     \\ 
$m=10$                     & \multicolumn{1}{c}{5.59}                     & \multicolumn{1}{c}{4.94}                    & 5.15                     & \multicolumn{1}{c}{5.83}                     & \multicolumn{1}{c}{5.40}                    & 5.68                     \\ 
$m=20$                     & \multicolumn{1}{c}{5.04}                     & \multicolumn{1}{c}{4.56}                    & 4.64                     & \multicolumn{1}{c}{5.24}                     & \multicolumn{1}{c}{4.93}                    & 5.08                     \\ \midrule\bottomrule[2pt]
\end{tabular}
\end{table}
\begin{table}[h]
\centering
\caption{\textbf{The skewness of degree distributions of NOHs on SWs. }}\label{tab: degree distribution SW}
\begin{tabular}{c|ccc|ccc}
\toprule[2pt]
\midrule
$\lambda$s & \multicolumn{3}{c|}{$0.01$}                                                                        & \multicolumn{3}{c}{$0.02$}                                                                        \\ \midrule
$\mu$s     & \multicolumn{1}{c|}{$0.005$} & \multicolumn{1}{c|}{$0.01$} & $0.015$ & \multicolumn{1}{c|}{$0.005$} & \multicolumn{1}{c|}{$0.01$} & $0.015$ \\ \midrule
$K=20$                      & \multicolumn{1}{c}{0.09}                     & \multicolumn{1}{c}{0.22}                    & 0.37                     & \multicolumn{1}{c}{0.01}                     & \multicolumn{1}{c}{0.08}                    & 0.17                     \\ \midrule
$K=40$                     & \multicolumn{1}{c}{0.06}                     & \multicolumn{1}{c}{0.20}                    & 0.30                     & \multicolumn{1}{c}{0}                     & \multicolumn{1}{c}{0.06}                    & 0.14                     \\ \midrule
$K=60$                     & \multicolumn{1}{c}{0.06}                     & \multicolumn{1}{c}{0.17}                    & 0.25                     & \multicolumn{1}{c}{-0.01}                     & \multicolumn{1}{c}{0.05}                    & 0.11                     \\ \midrule\bottomrule[2pt]
\end{tabular}
\end{table}

Note that we can bring the concept of online and offline behaviors of individuals to any static network model by setting this static architecture as the initial network as we have done. In this case, all individuals are assumed to be online initially, and the phase transitions of the networked population become stable with sufficient time. Additionally, one can set an arbitrary initial number of online individuals as required. 

\subsection{Using NOHs to Fit the Real Network}
\label{sec: real}
We now present the model efficiency by fitting the NOH model with a real network data set and comparing our proposed model with current network models. The data set is a snapshot of the Amazon co-purchase network \cite{kunegis2013konect}\cite{yang2012defining}, containing $334,863$ vertices and $925,872$ undirected and unweighted edges. The average path length and the clustering coefficient are $11.7253$ and $0.205224$ separately. Therefore, we set the SF as the initial network. Additionally, the online and offline rates are set as $\lambda=0.01$ and $\mu=0.013$ respectively in our comparison. 
\begin{figure}
    \centering
    \includegraphics[scale=0.32]{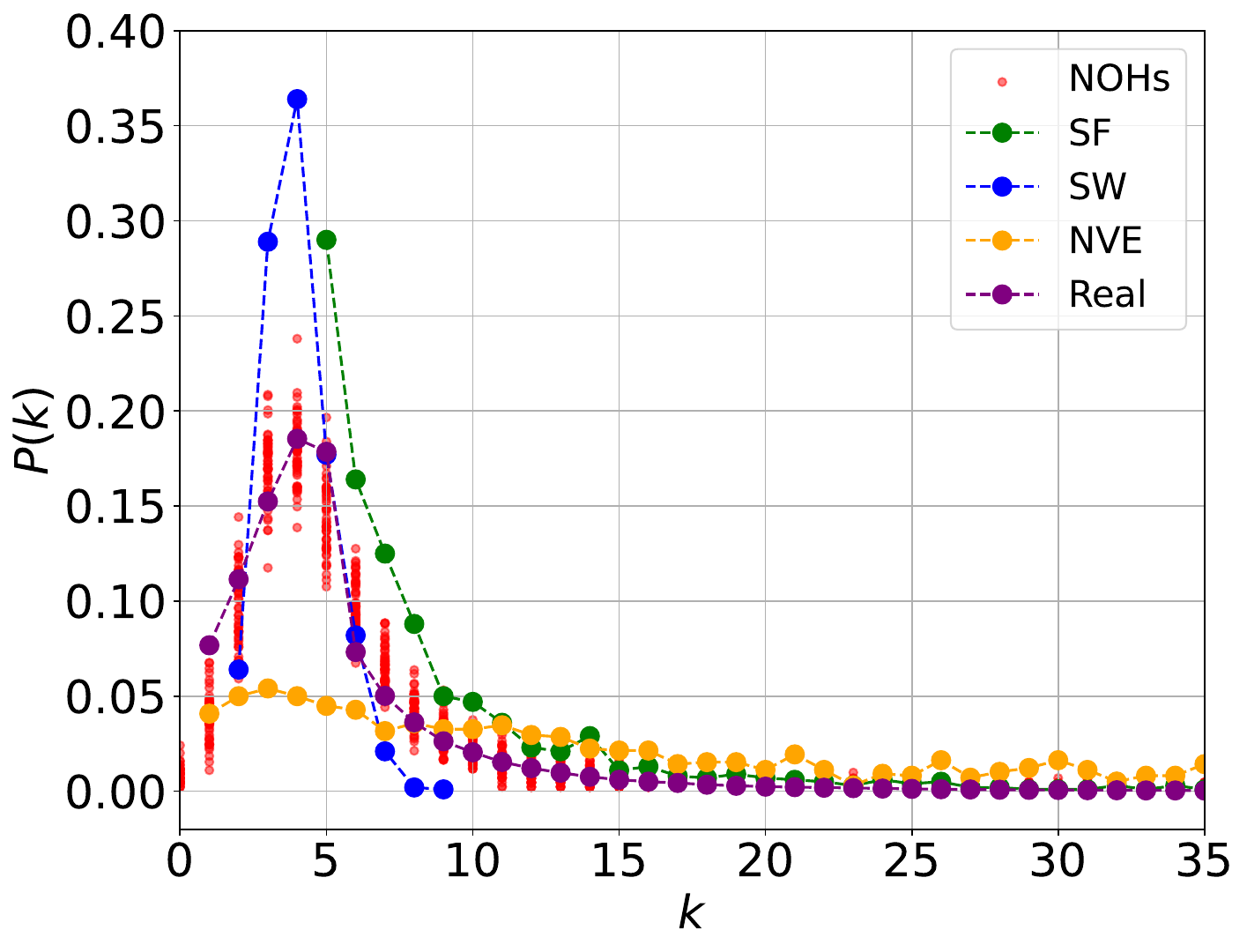}
    \caption{\textbf{Fitting the degree distribution of Amazon co-purchase network data and comparison.} The proposed NOH model is effective in fitting real-world network data sets. The green plot shows the SF with $m=5$. The blue plot is the SW with $K=5$ and $p=0.4$. The orange plot denotes the NVE with increments following the lognormal distribution with $\mu_{NVE}=2$ and $\sigma_{NVE}=0.5$. The red scatters are the NOHs with $\lambda=0.01$ and $\mu=0.013$. The purple plot indicates the degree distribution of the real network data. (Color online)}
    \label{fig: 9}
\end{figure}

To demonstrate the efficiency in comparison to other models, we select three current network models and compare their results with our proposed model, including the SF ($m=5$) \cite{barabasi1999emergence}, SW ($K=5$ and $p=0.4$) \cite{watts1998collective}, and the network model proposed by Ref. \cite{feng2020accumulative} (NVE) (increments follow the lognormal distribution with $\mu_{NVE}=2$ and $\sigma_{NVE}=0.5$). The employed parameters are obtained through traversal. Note that in Sec. \ref{sec:Complex Networks With Online and Hidden Vertices}, we briefly introduce the concepts of the network models above. Additionally, we present the degree distribution results on the proposed NOH model from $t=10^4$ to $2\times10^4$ with the interval $200$. In Fig. \ref{fig: 9}, we show the fitting and comparison result of the real network data and the network models. The degree distribution of real network data is presented in purple. In the network models we presume to compare, the fitting effect between the SF and real data seems not very good because of the heterogeneity provided by the SF. The SW and NVE can show the peak positions similar to the degree distribution of real network data. However, the peak values of probability are quite different from the real data. That is, although the positions of peak degree values can be well quantified by the SW and NVE, these two network models may not describe the degree frequencies of the Amazon co-purchase network very precisely. Compared to these three network models, our model is highly consistent with the actual data as shown in Fig. \ref{fig: 9}. 

Additionally, to quantify this efficiency, we employ the KL divergence in Eq. \ref{eq:KL} to measure the distance between two degree distributions. Tab. \ref{tab: real} presents the KL divergence between real data and each model. Obviously, the proposed NOH model provides the smallest KL divergence ($0.04$), leading to a minimum degree distribution difference. The KL divergences of the SF, the SW, and the NVE are $0.34$, $0.10$, and $0.60$ separately. The real data does not show any heterogeneity, resulting in a low KL divergence for the SF. For the SW, the degree frequencies around the peak values are higher than the real network data. On the contrary, the NVE provides lower frequencies around the peak values than the real data. Additionally, the mean degree and assortativity of the real network can be described in our proposed NOH. Therefore, we can conclude that the NOH model is efficient in fitting the real data and finding some regular patterns in real systems. However, the employed real network has a high clustering characteristic, but none of these mentioned models can fit the clustering coefficient and skewness of the degree distribution. Moreover, we discuss the complexity of each network generation algorithm in Tab. \ref{tab: real}. The complexities of SF, SW, and NVE are $O(N^2)$, $O(N)$, and $O(N^2)$ respectively. Our proposed NOH model calculates the transition time of each vertex, leading to the complexity $O(N)$. 
\begin{table}[h]
\centering
\caption{\textbf{The comparison of models in fitting real networks.}}
\begin{tabular}{c|c|c|c|c|c|c}
\toprule[2pt]
\hline
&KL& Clustering  & Mean degree & Assortativity& Skewness&Complexity\\
\midrule
SF&0.34&0.01&10.00&-0.03&37.72&$O(N^2)$\\
SW&0.10&0.12&4.00&-0.07&0.48&$O(N)$\\
NVE&0.60&0.03&23.41&-0.01&5.25&$O(N^2)$\\
NOH&0.04&0.01&6.02&-0.03&3.72&$O(N)$\\
Real& &0.40&5.52&-0.05&10.56&\\
\midrule
\bottomrule[2pt]
\end{tabular}
\label{tab: real}
\end{table}
\section{Conclusion and outlook}
\label{sec:conclusion}
In this paper, we introduce an innovative network model with both hidden and online vertices based on a birth and death process. In our model, a hidden vertex becomes online with an online rate, and an online vertex becomes hidden with a hidden rate, where each vertex's online and hidden duration on each state follows an exponential distribution with two independent parameters. By defining corresponding stochastic processes, we perform a theoretical analysis on the number of each vertex's online neighbors and the online network size, which are found to be stable and follow a probability distribution in the same form. Furthermore, we calculate the expected online network size and the variance. In simulations, considering initial networks as SFs and SWs, we analyze the numbers of each vertex's online neighbors, online network sizes, and degree distributions. Besides, we find that our theoretical analysis fits well with the simulation results. 

However, the NOH model can be further studied. For example, degree distributions are shown to be stable as results shown in simulations. Nevertheless, we did not solve the theoretical degree distribution in Sec. \ref{sec:Complex Networks With Online and Hidden Vertices}. Moreover, more topological characteristics, including the mean degree, the average clustering coefficient, and the average path length, are not given in this paper. Besides, each vertex's duration can follow other probability distributions, like the uniform distribution and the normal distribution. In this way, each vertex's state transition can be regarded as an alternating renewal process, which may lead to different theoretical results. Additionally, a real network usually allows new vertices to join in and quit from it, i.e., the total number of vertices is time-varying instead of static. From the perspective of the application, it is worth studying how to apply the NOH model to simulate real networks with online and hidden individuals. Moreover, our proposed framework of vertex phase transition between the online and hidden states may help the study of epidemic propagation in social networks. All these issues require further work.

\appendix
\section{Notations}
\label{sec: appendix-notations}
The appendix includes some notations (Tab. \ref{tab: notation}) for readers' convenience and the proofs for the mentioned theorem in the main text. 
\section{Proof for Theorem 1}
\begin{proof}
According to Eqs. \ref{eq:transition rate of degree} and \ref{eq:transition rate of size} in the main body text, the elements of the probability transition rate matrix of $k_i$ ($\boldsymbol{Q_i}$) can be denoted as
\begin{equation}
Q_{i(m,n)}=
\left\{
\begin{array}{ll}
[k_i(0)-m]\lambda ,& n=m+1 \\
m\mu ,& n=m-1 \\
1-[k_i(0)-m]\lambda- m\mu,& n=m \\
0,& \vert n-m\vert\geq 2\\
\end{array}.
\right.
\end{equation}
According to Lemma \ref{lem:PQ=0}, solving the system of equations $\boldsymbol{\Pi Q=0}$, we obtain the stationary distribution of the number of the vertex $i$'s online neighbors
\begin{equation}
\begin{aligned}
\pi_{i(n)}&=\frac{C_{k_i(0)}^{n}(\lambda\mu^{-1})^n}{\sum_{j=0}^{k_i(0)}C_{k_i(0)}^{j}(\lambda\mu^{-1})^j}=\frac{C_{k_i(0)}^{n}(\lambda\mu^{-1})^n}{(1+\lambda\mu^{-1})^{k_i(0)}}.
\end{aligned}
\end{equation}
For $N(t)$, the elements of the probability transition rate matrix $\boldsymbol{Q}$ are
\begin{equation}
Q_{m,n}=
\left\{
\begin{array}{ll}
[N(0)-m]\lambda ,& n=m+1 \\
m\mu ,& n=m-1 \\
1-[N(0)-m]\lambda- m\mu,& n=m \\
0,& \vert n-m\vert\geq 2\\
\end{array}.
\right.
\end{equation}
Similarly, we have the stationary distribution of the network size
\begin{equation}
\pi_{n}=\frac{C_{N(0)}^{n}(\lambda\mu^{-1})^n}{(1+\lambda\mu^{-1})^{N(0)}}.
\end{equation}
The result follows.
\end{proof}
\label{sec: appendix-proof1}
\begin{table}
\centering
\caption{Notations}\label{tab: notation}
\begin{tabular}{cc}
   \toprule
   Symbol & Definition\\
   \midrule
   $\lambda$ & The online rate\\
   $\mu$ & The offline rate\\
   $k_i(t)$ & The vertex $i$'s online neighbor number at $t$ \\
   $N(t)$ & The online network size at $t$ \\
   $p_{i(m,n)}(\Delta t)$ & The transition probability of $k_{i}(t)$\\
   $p_{m,n}(\Delta t)$ & The transition probability of $N(t)$\\
   $q_{i(m,n)}$ & The transition rate of $k_{i}(t)$ \\
   $q_{m,n}$ & The transition rate of $N(t)$\\
   $\pi_{i(n)}$ & The limit distribution of $k_{i}(t)$ \\
   $\pi_n$ & The limit distribution of $N(t)$ \\
   \bottomrule
\end{tabular}
\end{table}
\section{Proof for Theorem 2}
\label{sec: appendix-proof2}
\begin{proof}
Applying the binomial formula
\begin{equation}\label{eq:binomial formula}
(1+\lambda\mu^{-1})^{k_i(0)}=\sum_{j=0}^{k_i(0)}C_{k_i(0)}^{j}(\lambda\mu^{-1})^j.
\end{equation}
Finding the first derivative of $\lambda\mu^{-1}$ at both ends of Eq. \ref{eq:binomial formula}, we have
\begin{equation}\label{eq:first derivative}
k_i(0)(1+\lambda\mu^{-1})^{k_i(0)-1}=\sum_{j=0}^{k_i(0)}jC_{k_i(0)}^j(\lambda\mu^{-1})^{j-1}.
\end{equation}
Multiplying $\lambda\mu^{-1}$ at both ends of the Eq. \ref{eq:first derivative}, we obtain
\begin{equation}
k_i(0)\lambda\mu^{-1}(1+\lambda\mu^{-1})^{k_i(0)-1}=\sum_{j=0}^{k_i(0)}jC_{k_i(0)}^{j}(\lambda\mu^{-1})^j.
\end{equation}
Therefore, the expected number of online vertices in the vertex $i$'s neighbors is
\begin{equation}
\begin{aligned}
E[k_i]&=\sum_{j=0}^{k_i(0)}j\pi_{i(j)}=\frac{k_i(0)\lambda\mu^{-1}}{1+\lambda\mu^{-1}}.
\end{aligned}
\end{equation}
Finding the second derivative of $\lambda\mu^{-1}$ at both ends of Eq. \ref{eq:binomial formula}, we get
\begin{equation}\label{eq:second derivative}
k_i(0)[k_i(0)-1](1+\lambda\mu^{-1})^{k_i(0)-2}=\sum_{j=0}^{k_i(0)}j(j-1)C_{k_i(0)}^{j}(\lambda\mu^{-1})^{j-2}.
\end{equation}
Multiplying $(\lambda\mu^{-1})^2$ at both ends of the Eq. \ref{eq:second derivative}, we obtain
\begin{equation}
\begin{aligned}
&\sum_{j=0}^{k_i(0)}j^2C_{k_i(0)}^{j}(\lambda\mu^{-1})^j\\
&=(\lambda\mu^{-1})^2k_i(0)[k_i(0)-1](1+\lambda\mu^{-1})^{k_i(0)-2}\\
&+k_i(0)\lambda\mu^{-1}(1+\lambda\mu^{-1})^{k_i(0)-1}.\\
\end{aligned}
\end{equation}
Therefore, the corresponding variance is
\begin{equation}
\begin{aligned}
D[k_i]=E[k_i^2]-E[k_i]^2=\frac{k_i(0)\lambda\mu^{-1}}{(1+\lambda\mu^{-1})^2}.
\end{aligned}
\end{equation}
Similarly, we have the expected online network size and its corresponding variance
\begin{equation}
E[N]=\frac{N(0)\lambda\mu^{-1}}{1+\lambda\mu^{-1}},
\end{equation}
and
\begin{equation}
D[N]=\frac{N(0)\lambda\mu^{-1}}{(1+\lambda\mu^{-1})^2}.
\end{equation}
The result follows.
\end{proof}
\printcredits
\bibliographystyle{elsarticle-num}
\bibliography{refs}
\end{document}